\documentstyle[epsfig]{mn}
\begin{document}

\title[The first stars]{The first generation of stars in the $\Lambda$CDM cosmology}
\author[Gao L. et al]{L. Gao$^{1,2}$\thanks{Email: liang.gao@durham.ac.uk},
                     N. Yoshida$^3$,
                     T. Abel$^4$,
                     C. S. Frenk$^1$,
                     A. Jenkins$^1$,
                     V. Springel$^2$ \\
$^1$ Institute for Computational Cosmology, Department of Physics,
University of Durham, South Road, Durham, DH1 3LE \\
$^2$ Max--Planck--Institut f\"ur Astrophysik, D-85748, Garching,
Germany \\
$^3$ Department of Physics, Nagoya University, Furocho, Nagoya, Aichi
464-8602, Japan \\
$^4$ Kavli Institue for Particle Astrophysics and Cosmology, Standford
University, 2575 Sand Hill Road, MS 29, Menlo Park, CA 94205, USA}
\maketitle

\begin{abstract}
We have performed a large set of high-resolution cosmological
simulations using smoothed particle hydrodynamics ({\small SPH}) to
study the formation of the first luminous objects in the $\Lambda$CDM
cosmology. We follow the collapse of primordial gas clouds in eight
early structures and document the scatter in the properties of the
first star-forming clouds. Our first objects span formation redshifts
from $z \sim 10$ to $z \sim 50$ and cover an order of magnitude in
halo mass. We find that the physical properties of the central
star-forming clouds are very similar in all of the simulated objects
despite significant differences in formation redshift and
environment. This suggests that the formation path of the first stars
is largely independent of the collapse redshift; the physical
properties of the clouds have little correlation with spin, mass, or
assembly history of the host halo.  The collapse of proto-stellar
objects at higher redshifts progresses much more rapidly due to the
higher densities, which accelerates the formation of molecular
hydrogen, enhances initial cooling and shortens the dynamical
timescales. The mass of the star-forming clouds cover a broad range,
from a few hundred to a few thousand solar masses, and exhibit various
morphologies: some have disk-like structures which are nearly
rotational supported; others form flattened spheroids; still others
form bars. All of them develop a single protostellar `seed' which
does not fragment into multiple objects up to the moment that the
central gas becomes optically thick to H$_2$ cooling lines. At this
time, the instantaneous mass accretion rate onto the centre varies
significantly from object to object, with disk-like structures having
the smallest mass accretion rates. The formation epoch and properties
of the star-forming clouds are sensitive to the values of cosmological
parameters.
\end{abstract}

\label{firstpage}


\section{Introduction}
The first generation stars, often referred to as ``Population-III''
stars, are thought to be the first sources of light in our Universe
and the origin of the heavy elements required for the subsequent
formation of ordinary stellar populations. Their remnant black holes
may be the seeds from which supermassive black holes grow, including
those that power quasars at redshift $\sim 6$ (e.g. Fan et al. 2003),
when the Universe was just $\sim 800$ million years old. In the
standard $\Lambda$CDM cosmology with a high amplitude of initial
density perturbations, $\sigma_8=0.9$, the first stars are predicted
to form at redshifts of about $20$ and higher, based on the argument
that gas in $3\sigma$-halos at such redshifts should be able to cool
and condense by molecular hydrogen cooling. Recent
determinations of the cosmological parameters, obtained by combining
the 2dFGRS galaxy power spectrum with the first and/or third year
{\small WMAP} data (Sanchez et al 2005, Spergel et al. 2006), give a
lower value of $\sigma_8\sim 0.75$, implying that the formation epoch
of the first stars could be as late as $z\sim 10$, opening the
possibility that these stars may be detected directly by the next
generation of space-borne and ground-based telescopes.

From a theoretical point of view, the formation of the first stars is
a well-defined problem. Quantum fluctuations imprinted during the
inflationary epoch provide the initial seeds for the formation and
growth of the dark matter halos that host the first stars. In the
absence of heavy elements, the only efficient coolant that can promote
the formation of primordial star-forming gas clouds is molecular
hydrogen (Peebles \& Dicke 1968; Matsuda, Sato \& Takeda 1969). Thus
once the cosmological parameters and the fluctuation spectrum are
specified, the formation of the first luminous objects in the universe
becomes, in principle, a straightforward physics problem.

Over the past decades, much progress has been achieved in following
this program. Early studies used semi-analytical methods (e.g.
Doroshkevich, Zel'Dovich \& Novikov 1967; Peebles \& Dicke 1968;
Couchman \& Rees 1986; Tegmark et al. 1997; Silk \& Langer 2006),
while others utilised direct numerical simulations, either in one
dimension (Haiman \& Thoul \& Loeb 1996; Omukai \& Nishi 1998;
Nakamura \& Umemura 2001; Ahn \& Shapiro 2007) or in three dimensions
(Abel, Bryan \& Norman 2000, 2002, hereafter ABN02; Bromm et al. 1999,
2002; Fuller \& Couchman 2000; Yoshida et al. 2003, 2006; O'Shea \& Norman
2006a,b). While the semi-analytical calculations have been
instrumental in sketching out the relevant physics, the numerical
simulations have been able to deliver detailed predictions for the
outcome of the coupling between the chemo-thermal evolution and
gravitational collapse.  Comprehensive reviews of recent progress on
the formation of the first luminous objects and associated radiative
feedback processes may be found in Barkana \& Loeb (2001), Bromm \&
Larson (2004), Glover (2005) and Ciardi \& Ferrara (2005), and
Ripamonti \& Abel (2005).

Pioneering  three-dimensional simulations of the formation of the first
objects including non-equilibrium chemistry suggested that the first
stars could be more massive than $100\,{\rm M_\odot}$ (Abel, Bran
\& Norman 2000; 2001; Bromm et al. 1999, 2002;). ABN02 employed 
cosmological initial conditions and adaptive mesh refinement ({\small
AMR}) techniques to follow a star-forming gas cloud until it reached a
central hydrogen number density, $n_{\rm H}\sim 10^{10}{\rm cm^{-3}}$,
the point at which the assumption that the gas is everywhere optically
thin breaks down and a radiative transfer calculation is
required. Recently, Yoshida et al. (2006) included the effect of
molecular line opacities and followed the further collapse of the
central core to ${\rm n_{H} \sim 10^{16}cm^{-3}}$. Bromm et al. (2002)
(hereafter BCL02) and Bromm \& Loeb (2004) used constrained initial
conditions and smooth particle hydrodynamics ({\small SPH}) to study
the same problem. Their simulations agree with those of ABN02 on many
of the properties of the resulting star-forming cloud, but they also
differ in a number of details.

A major disagreement is that ABN02 proposed that only one star forms
per halo, whilst BCL02 argued that multiple stellar systems can form
if the gas has sufficient initial rotation. In the latter case, the
gas first collapses into a disk-like structure which fragments into
several clumps very quickly. These clumps will then make their own
stars if they can survive negative feedback effects from more advanced
neighbouring clumps. Since ABN02 only simulated one particular object
and the initial set-up of BCL was not realistic, there is hope that
progress can be made by carrying out a number of simulations with
realistic initial conditions in order to establish the typical
properties of star-forming clouds and their variation.

One important caveat for simulations of the first stars has been noted
by White \& Springel (2000) and Gao et al.~(2005) (hereafter
G05). They argue that the first halos of any given mass do not form in
`typical' regions, but rather in `protocluster' regions corresponding
to peaks of the large-scale density field in the early universe. Since
a significant contribution to the local overdensity at the sites of
first star formation comes from megaparsec and larger wavelength
fluctuations, some conclusions about the first objects reached from
simulations of small volumes (which represent `typical' regions) may
be misleading. In particular, the formation redshift of the first
halos of any given mass or temperature will be systematically
underestimated. 

G05 carried out a sequence of $N$-body re-simulations to follow the
growth of one of the most massive progenitors of a supercluster
region, from redshift $z \sim 80$, when its mass was just $10\, {\rm
M_\odot}$, to the present epoch. They found that the mass of the first
object reaches $10^5{\rm M_\odot}$ already at $z\sim 50$.  A follow-up
study of Reed et al. (2005) grafted a semi-analytical model developed
by Yoshida et al. (2003) onto the dark matter merger trees of 
the G05 simulations and concluded that the first halo in this
sequence of simulations would make one or more stars already at $z
\sim 47$. Very rare objects like the one simulated by G05 are 
different from those formed in typical regions. These rare objects
reside in extremely dense environments and experience particularly
rapid growth. Their higher collapse redshift results in a higher
initial density than in a typical region and this could result in
significantly different evolution later on because the dynamical time,
the cooling time, and the reaction rates for the formation of
molecular hydrogen all depend on gas density. It is therefore an
interesting question whether the formation paths and properties of
star-forming clouds in these very rare objects are at all similar to
those forming later on in more typical environments.

In this paper, we address this issue by asking two important
questions: (1) {\it Is there a significant scatter in the properties
of primordial star-forming clouds?} (2) {\it Do the properties of
star-forming clouds depend on formation redshift?} We perform
simulations of a sample of 8 first objects. By following evolution
from different sets of initial conditions we can simulate a variety of
formation redshifts and investigate the redshift dependence of the
formation process. 

This paper is structured as follows. In Section~2, we briefly
introduce the chemical processes that are important for primordial
gas. In Section 3, we describe our simulations. In Section 4, we take
three of them as examples to sketch a general picture for the
formation path of the first stars and then focus on one of them in
greater detail. In Section 5, we study whether the properties of the
star-forming clouds depend on collapse redshift and, in Section~6, we
investigate how the baryon fraction influences the formation of the
first stars. In Section~7, we discuss the abundance of star-forming
halos. Finally, we give a summary of our results and set out our
conclusions in Section~8.

\section{Pristine gas chemistry}
In this section, we briefly summarize the basic gas processes that are
important for primordial gas unpolluted by metals.  Full details are
given by Abel et al. (1997), Galli \& Palla (1998) and Yoshida et
al. (2006). For simulations that start at very high z, we now also
include photo-reactions with cosmic microwave background (CMB)
photons.  At high redshift, in the absence of metals and when the
temperature of gas in halos is lower than $10^4\,{\rm K}$ (above which
atomic hydrogen line cooling is dominant), the principal coolant that
can enable gas to condense into stars is the ${\rm H_2}$ molecule.

The formation of molecular hydrogen has 3 main channels.  At high
redshifts, $z>200$, the ${\rm H_2^+}$ channel given by
\begin{eqnarray}
\rm H^+~+~H~   &\rightarrow &\rm ~H_2^+~+~\gamma \nonumber \\
\rm H_2^+~+~H~ &\rightarrow &\rm ~H_2~+\rm~H^+ \nonumber
\end{eqnarray}
is dominant (but see Hirata \& Padmanabhan (2006) for a different
viewpoint). At these early times, there are sufficient numbers of
energetic CMB photons to destroy ${\rm H^-}$ ions rapidly, so that a 
second possible channel, 
\begin{eqnarray}
{\rm H}~+~e^-&\rightarrow&~{\rm H^-~+~\gamma}  \nonumber \\
~~{\rm H^-~+~H}&\rightarrow&~{\rm H_2}~+~e^-   \nonumber
\end{eqnarray}
is initially strongly suppressed.  However, as the Universe expands
and the CMB photons become less energetic, the ${\rm H^{-}}$ channel
eventually becomes the main mechanism for creating molecular
hydrogen. In this process, the residual electrons left unattached
after recombination act as a catalyst.

When the density of gas exceeds ${n_{\rm H}} \sim 10^8 {\rm cm^{-3}}$,
a third channel becomes important, through the three-body reaction
(Palla, Salpeter \& Stahler 1983):
\begin{eqnarray}
&{\rm H~+~H~+~H} &\rightarrow{\rm~H_2~+~H},  \nonumber \\
&{\rm H~+~H~+~H_2} &\rightarrow{\rm~H_2~+~H_2}  \nonumber .
\end{eqnarray}
At sufficiently high densities, this reaction can, in fact, convert
almost all atomic hydrogen into molecular form in a very short
time. Our simulation code follows reactions for $9$ species $(e^{-}$,
${\rm H}$, ${\rm H^{+}}$, ${\rm He}$, ${\rm He^{+}}$, ${\rm He^{++}}$,
${\rm H_2}$, ${\rm H_2^+}$, ${\rm H^-}$), including these H2 formation
processes.

The cooling function for molecular hydrogen can be expressed as
\begin{equation}
\Lambda (\rm {n_H},T)=\frac{\Lambda _{\rm{LTE}}(T)}{1+\frac{
\Lambda _{\rm {LTE}}(T)}{\rm {n}_{\rm{H}}\Lambda _{\rm{n}_{\rm{H}
}\rightarrow 0}(T)}}, \nonumber
\end{equation}
where $\Lambda_{\rm LTE}$ and $\Lambda_{\rm n_H \rightarrow 0}$ are
the cooling functions for high and low density limits. We adopt the
high density limit given by Hollenbach \& McKee (1979), assuming local
thermal equilibrium ({\small LTE}), and the low density limit of Galli
\& Palla (1998). The transition between these two regimes occurs
approximately at a gas density of $n_{\rm H} \sim 10^4 {\rm
cm^{-3}}$. Above this value, the cooling rate per hydrogen molecule
is independent of density, while below it, it is proportional to
density.

\section{Simulations}

We have carried out a suite of simulations designed to study the
formation of the first stars. For the first series, a standard
$\Lambda$CDM cosmology, with matter, baryon and dark energy density
parameters, $\Omega_m=0.3$, $\Omega_b=0.04$, $ \Omega_{\Lambda}=0.7$
respectively, and a Hubble constant $h_{100}=0.7$, was adopted. The
initial power spectrum was generated with {\small CMBFAST} and
normalised to give an {\it rms} linear fluctuation amplitude of
$\sigma_8=0.9$ in $8\,h^{-1}{\rm Mpc}$ tophat spheres at $z=0$.  The
initial ionisation fraction was computed using {\small RECFAST}
(Seager, Sasselov \& Scott 2000).

One of the simulations used in this study has been previously
analysed extensively by G05 and Reed et al.~(2005). This
simulation was constructed through a sequence of
re-simulations of the most massive progenitors of a supercluster
region. Here we briefly describe the multi-grid procedure devised
by G05:

\begin{description}
\item[(1)] Identify a rich cluster size halo at $z=0$
in a cosmological simulation of a very large volume ($\sim 1 {\rm
  Gpc}^3$).

\item[(2)] Resimulate the evolution of this rich cluster and its
environment with higher mass resolution.

\item[(3)] Determine the most massive object in the high
  resolution region of the re-simulation as a function of redshift,
  and identify the time  when it first contains more than $10000$
  particles inside the viral radius.

\item[(4)] Resimulate the evolution of this progenitor object
and its immediate surroundings with a further improvement in mass
and force resolution.

\item[(5)] Iterate steps $3$ and $4$ until the desired redshift
and progenitor mass are reached. 
\end{description}

Following this procedure through $5$ levels of re-simulation, G05 were
able to resolve a massive progenitor of the supercluster region
capable of efficient ${\rm H}_2$-molecular cooling at redshift $z
\sim 50$. Their final, highest-resolution simulation had a particle mass
of $1.08{\rm M_\odot}$ and is labelled ``R5.'' Here we have re-ran
``R5,'' this time including radiative cooling by molecular hydrogen
and non-equilibrium chemistry. This is one of the main simulations
analysed in this paper.

We carried out two additional simulations drawn from a periodic box
of $300\, h^{-1}{\rm kpc}$ on a side. As noted by G05, the use of
periodic boundary conditions suppresses fluctuations with wavelength
longer than the side of the computational box. Therefore, if the box
size is small, the appearance of a halo that could host first stars
will be significantly delayed compared to a simulation of a larger
volume.  Our main motivation for simulating these two small periodic
boxes is to enable us to compare our results to earlier studies
carried out in this way (e.g. ABN02). In addition, however, since the
formation of the first stars could plausibly span a wide range in
redshift, our two small periodic box simulations are still interesting
for studying first star formation at comparatively low redshifts.

In order to achieve sufficient resolution, we again utilised a
resimulation technique similar to that of G05. First, we performed a
simulation of a $300\,h^{-1}{\rm kpc}$ cubic region using $256^3$
particles. We then identified the most massive objects at redshift $z
\sim 15$ which typically have mass of $\sim 10^6{\rm M_\odot}$ and
resimulated them with a resolution comparable to the R5
simulation. Note that since the dynamic range of this simulation is
much smaller than that of R5 (the initial volume is a factor $4 \times
10^9$ smaller), only one level of refinement, rather than five, is
required to create the initial conditions of the final target
simulation.  Table~1 summarizes the principal numerical parameters of
these simulations.

\begin{table}
\begin{center}
\begin{tabular}{l c c c}
\hline
& R5 & Z1 & Z2 \\ \hline
$m_{\rm dm}[{\rm M_\odot}]$ & $1.08$ & $3.1$ & $2.12$ \\
$m_{\rm gas}[{\rm M_\odot}]$ & $0.166$ & $0.48$ & $0.326$ \\
$M_{\rm 200}[{\rm M_\odot}]$ & $2.21\times 10^{5}$ & $3.75\times 10^{5}$ & $%
6.18\times 10^{5}$ \\
$z_{\rm end}$ & $47.91$ & $25.9$ & $21.0$ \\ \hline

\end{tabular}
\end{center}
\caption{Numerical parameters of our first set of simulations. Here
$m_{\rm dm}$ is the mass of each dark matter particle, while $m_{\rm
gas}$ is the mass of each gas particle; $z_{\rm end}$ is the time when
the central density of the gas cloud reaches $\sim 10^{10}$ when the
optically thin assumption breaks down; $M_{200}$ is the halo mass
defined as the enclosed mass in a sphere of mean interior overdensity
$200$ times the cosmic mean density at the final time. This simulation
series was performed using the standard cosmological parameters,
$\Omega_m=0.3$, $\Omega_\Lambda=0.7$, $\Omega_b=0.04$, $h_{100}=0.7$,
$\sigma_8=0.9$, with a primordial power spectrum index of $n_s=1$.} 
\label{Table:t1}
\end{table}

Since uncertaintes remain on the values of the cosmological parameters
which describe our Universe, we have also rerun all the simulations
listed in Table~\ref{Table:t1} adopting the parameters given in
Table~5 of Spergel et al. (2006), derived from the combination of $3$
years {\small WMAP} microwave data with the 2dFGRS galaxy power
spectrum (Cole et al. 2005). We retain the same phases in the initial
conditions but adjust the amplitudes of the modes according to the new
power spectrum shape and normalisation.  These cosmological parameters
are significantly different from the values adopted in our other
simulations: $\Omega_m=0.236$, $\Omega_{\Lambda}=0.764$,
$\Omega_b=0.042$ and $h_{100}=0.73$. The primordial power spectrum
slope index is $ n_s=0.95$, and the normalisation is
$\sigma_8=0.74$. Further details of this set of simulations are given
in Table~\ref{Table:t2}.

\begin{table}
\begin{center}
\begin{tabular}{lccc}
\hline
& R5wt & Z1wt & Z2wt \\ \hline
$m_{\rm dm}[{\rm M_\odot}]$ & $0.77 $ & $2.22$ & $1.52$ \\
$m_{\rm gas}[{\rm M_\odot}]$ & $0.167 $ & $0.48$ & $0.328$ \\
$M_{200}[{\rm M_\odot}]$ & $4.42\times 10^{5}$ & $1.16\times 10^{6}$ & $%
2.55\times 10^{6}$ \\
$z_{\rm end}$ & $30.04$ & $13.43$ & $10.33$ \\ \hline
\end{tabular}
\end{center}
\caption{Numerical parameters of our second set of simulations.  
The values of the cosmological parameters were 
taken from Table~5 of Spergel et al.~(2006): $\Omega_m=0.234$,
$\Omega_\Lambda=0.764$, $\Omega_b=0.042$, $h_{100}=0.73$,
$\sigma_8=0.74$, and $n_s=0.95$. The meaning of the tabulated 
quantities is the same as in Table~\ref{Table:t1}.}
\label{Table:t2}
\end{table}

We performed a third set of controlled simulations to complement the
other two.  By design, these simulations form stars at redshifts
intermediate between those of the first and the second sets. To
achieve this, we ran additional versions of Z1 and Z2, with a lower
value of $\sigma_8=0.74$, but employing the standard power spectrum,
in order to increase the power on small scales, whilst the other
cosmological parameters were set according to the 3rd year {\small
WMAP} results. These simulations are detailed in Table~\ref{Table:t3}.

\begin{table}
\begin{center}
\begin{tabular}{lcc}
\hline
& Z1w & Z2w \\ \hline
$m_{\rm dm}[{\rm M_\odot}]$ & $2.22$ & $1.52$ \\
$m_{\rm gas}[{\rm M_\odot}]$ & $0.48$ & $0.328$ \\
$M_{200}[{\rm M_\odot}]$ & $4.94\times 10^{5}$ & $6.9\times 10^{5}$ \\
$z_{\rm end}$ & $20.65$ & $16.75$\\ \hline
\end{tabular}
\end{center}
\caption{Parameters of simulations where the three-year {\small WMAP}
cosmological parameters are adopted, but the shape of a
standard cosmology power spectrum is retained.} 
\label{Table:t3}
\end{table}

Finally, we carried out simulations with both a higher and a lower
baryon fraction than in the Z1 simulation in order to investigate how
this parameter influences the formation of the first stars. For the
higher baryon fraction case, we chose $f_b=\Omega_b/\Omega_m=0.178 $
which is consistent with the $3$-year {\small WMAP} cosmology while,
for the lower baryon fraction case, we adopted $f_b=0.06$ which
matches the value used by ABN02 (for their assumed standard
$\Omega_m=1$ CDM cosmology). This allows a more direct comparison of
their results with ours. See Table~\ref{Table:t4} for more information
about these two simulations.

\begin{table}
\begin{center}
\begin{tabular}{lcc}
\hline
& Z1hb & Z1lb \\ \hline
$m_{\rm dm}[{\rm M_\odot}]$ & $2.94$ & $3.36$ \\
$m_{\rm gas}[{\rm M_\odot}]$ & $0.63$ & $0.21$ \\
$M_{200}[{\rm M_\odot}]$ & $3.25\times 10^{5}$ & $6.9\times 10^{5}$ \\
$f_b$ & 0.178 & 0.06 \\
$z_{\rm end}$ & $26.26$ & $24.26$ \\ \hline
\end{tabular}
\end{center}
\caption{Numerical parameters of our simulations where we explored a
  higher and lower baryon fraction version than used in our default
  simulation of the Z1 object.  The standard cosmology is assumed.}
\label{Table:t4}
\end{table}

For all our simulations we used a fixed comoving gravitational
softening length for dark matter particles equal to $1/30$ of the mean
interparticle separation, while for gas particles the gravitational
softening was set equal to the SPH smoothing length. We experimented
with a fixed gravitational softening length for gas particles, but
this caused serious numerical problems for collapse simulations such
as those presented in this paper (Bate \& Burkert 1997; Abel et al.,
in preparation). Our simulations were performed with the widely used
cosmological TreeSPH-code {\small GADGET2} (Springel 2005).

\begin{figure*}

\hspace{0.13cm}\resizebox{8cm}{!}{\includegraphics{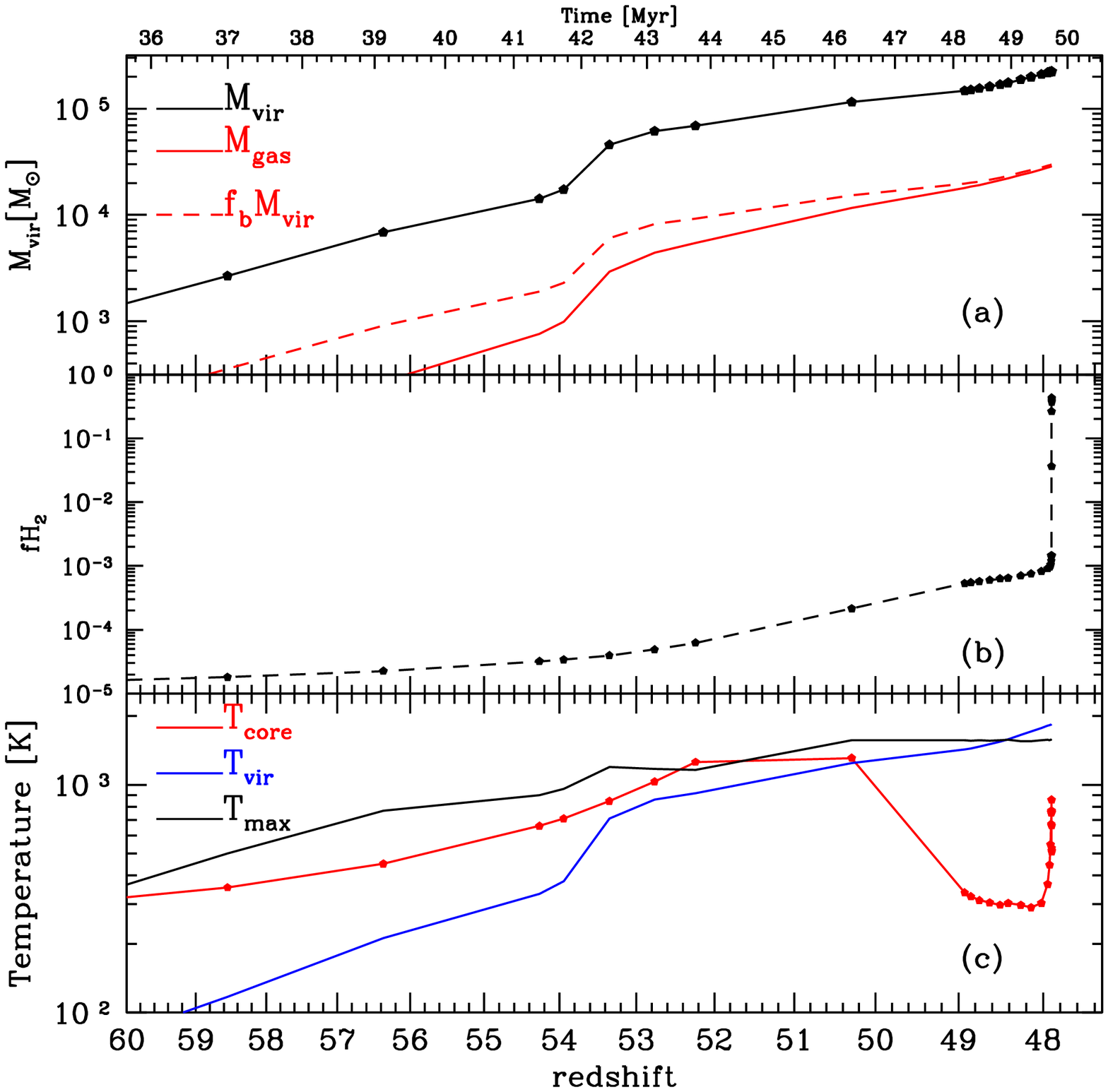}}
\hspace{0.13cm}\resizebox{8cm}{!}{\includegraphics{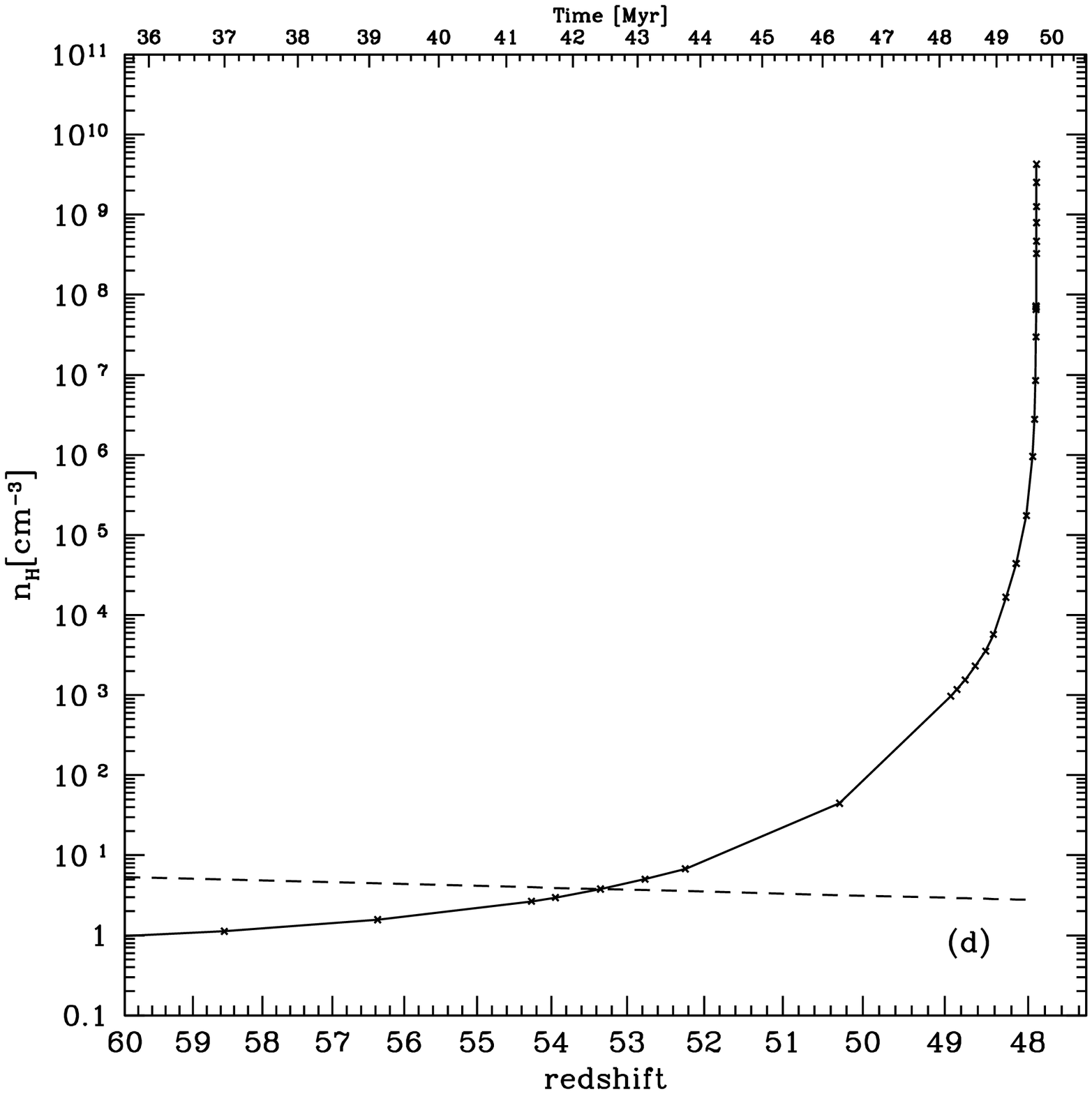}}
\caption{Evolution of different physical properties of the $100$
{\small SPH} particles ($\sim 20\, {\rm M_\odot}$) which end up
closest to the centre of the halo at the final output time. Panel (a):
the black solid lines with symbols show the virial mass of the main
halo progenitor; the lower (red) solid lines show the mass of the gas
component within the virial radius of the halo; the (red) dashed lines
are the fiducial gas mass that the halo would contain at the universal
baryon fraction, $f_b=\Omega_b/\Omega_m$. Panel (b): the evolution of
the ${\rm H_2}$-fraction for the 100 selected gas particles. Panel
(c): the gas temperature evolution. The lower (blue) solid line is an
estimate of the virial temperature of the main halo, as described in
the text; the black solid lines are the maximum temperature defined as
the maximum of the spherically mass-averaged temperature profile of
the the gas in the main halo. The (red) solid lines with symbols show
the evolution of the median value of the temperature for the 100
selected particles. Panel (d): the solid lines with symbols are the
median values of the number density of hydrogen nuclei in the 100
selected particles.  The dashed lines indicate $200$ times the
cosmological mean number density.} 
\label{fig:evor5}
\end{figure*}

\section{Formation path of the first stars}

After recombination, the temperature of the cosmic gas remains
initially coupled to that of the radiation due to the presence of a
small residual fraction of free electrons. After $z \sim 120$, this
coupling becomes weak and the intergalactic medium (IGM) temperature
starts to fall adiabatically as $T\propto (1+z)^2$ (e.g. Peebles
1993).  While the assembly of cold dark matter into halos proceeds
hierarchically starting at tiny mass scales, the accretion of gas into
halos is affected by thermal pressure (e.g. Yoshida, Sugiyama, \&
Hernquist 2003).  The density of gas only rises significantly above
the mean cosmic gas density within halos that have a mass comparable
or greater than the Jeans' mass $M_{\rm J}=4/3\pi^{5/2}c_s^3(G
\rho)^{-1/2}$, where $c_s$ is the sound speed of the gas, and $\rho$
is the density.

Once a dark matter halo becomes massive enough for its baryon content
to overcome the thermal gas pressure, it can accrete enough material
to reach a baryon fraction close to the universal value. When the
temperature is high enough ($\sim 1000$K), the processes described in
Section~2 can promote the formation of molecular hydrogen which
then acts as a coolant. A runaway cooling and compression process
begins at the halo centre, the details of which are the main topic of
this and the next section. When the central density of hydrogen
reaches values of around $n_{\rm H} \sim 10^{10} {\rm cm^{-3}}$, the
gas ceases to be optically thin to the molecular hydrogen lines. At
this point, our simulations are no longer able to follow the collapse
process any further in a physically meaningful way because the
appropriate radiative transfer processes are not modelled. For the
purposes of our analysis we therefore define the final time of our
simulations to be the point where the central density reaches a value
of $n_{\rm H} \sim 10^{10} {\rm cm^{-3}}$.

In the following subsections, we illustrate the path leading to the
first stars by discussing three of our simulations, R5, Z1, and Z2wt,
in detail. These form a comprehensive set in terms of the formation
redshift of their first star: R5 is the first to produce a
star-forming object, at a redshift of $z \sim 50$; in Z1 the halo
collapses at an intermediate redshift of $z\sim 26$, close to what is
often assumed in the literature to be the typical formation epoch of
the first stars; Z2wt forms its first star as late as $z \sim 10$.

\subsection{Overview of the formation of the first stars at the
  highest redshifts}
A useful way to describe the formation path of the first stars is to
track the evolution of the properties of gas particles that end up in
the central region at the final time. To this end, we select the
$100$ innermost {\small SPH} particles which lie closest to the centre
of the R5 halo at the end of the simulation.  We take the centre of a
halo to be the position of the minimum of the gravitational potential.
Our results are quite similar when we consider the innermost $1000$
particles instead.

The median values of various physical properties of the innermost 100
particles as a function of time are shown in
Figure~\ref{fig:evor5}. Panel (a) shows the mass accretion history of
the gas and dark matter components of the object in R5. The solid
black line shows the evolution of the virial mass of the dominant
halo. The virial mass is defined as the total mass within the
spherical region around the centre that encloses a density of $200$
times the mean cosmic value. The red dashed line is a fiducial gas
mass, given by the gas mass that the dark matter halo would contain if
it had the universal baryon fraction, $f_b=\Omega_b/\Omega_m$. The
solid red line shows the actual gas mass within the virial radius. It
is not surprising that at $z>55$ the baryon fraction of the halo is
much lower than the cosmic mean, because the gas thermal pressure
keeps the gas out of the halo. The baryon fraction is less than 10\%
of the fiducial value at redshift $z=60$ when the halo has a mass of
$1000\, {\rm M_\odot}$, but rises to $65\%$ at redshift $z=50$ by
which time the mass of the halo has grown to $10000\,{\rm
M_\odot}$. The universal baryon fraction is approximately reached when
the halo has grown to a mass of $10^5\,{\rm M_\odot}$, at $z \sim 49$.

Panel (b) shows the evolution of the molecular hydrogen fraction,
$f_{\rm H_2}$, of the 100 selected gas particles, while panel (c)
shows their temperature. For comparison, the blue solid line shows the
estimated virial temperature of the halo,
\begin{equation}
T=\mu m_p\;V_c^2/(2\,k_{{\rm B}}), \nonumber
\end{equation}
where $\mu\, m_p$ is the mean molecular weight of the gas, $V_c$ is
the circular velocity of the halo and $k_{{\rm B}}$ is Boltzmann's
constant. The black solid line refers to the maximum temperature of
the main object, determined by first binning the gas particles within
the halo into concentric logarithmically spaced shells centred on the
potential minimum, then computing the mass-weighted temperature of
each bin, and finally taking the maximum value of all the bins.  The
location of the maximum temperature roughly indicates where the
accretion shock occurs. 

The evolution of the number density of hydrogen nuclei for the
selected 100 particles is shown in panel (d) as a solid line with
symbols. The dashed line corresponds to $200$ times the cosmic mean
number density. The crossing point of the two curves roughly marks the
time when these $100$ gas particles settle into the virialised halo.

Based on these plots, we can outline a simple picture of how the gas
cloud condenses to the point of forming the first star in the R5
halo. Prior to redshift $z \sim 53$, when the mass of the dark halo is
less than $2 \times 10^4\, {\rm M_\odot}$, the virial temperature of
the halo is substantially lower than that of the ambient IGM
gas. (Note that the mean temperature of the surrounding IGM is higher
than the cosmic mean of $T \sim 60\,{\rm K}$ because the R5 object
resides in a significantly overdense region; see G05).  In this phase,
the gas accretion onto the R5 halo is limited by gas pressure. At $z
\sim 53$, a major merger takes place and the mass of the halo nearly
doubles. This significant increase in mass deepens the gravitational
potential well, allowing gravity to overwhelm the gas pressure, so
that ambient gas starts to settle into the halo. This can be observed
in panel (a) as a rapid increase in the baryon fraction at this
redshift. The 100 selected gas particles settle into the halo as their
density reaches the characteristic value of $200$ times the cosmic
mean required for virialization, as seen in panel~(d).

The amount of molecular hydrogen increases by almost one order of
magnitude from $z=53$ to $z=50$ due to the substantial increase in gas
temperature as well as density. Further production of molecular
hydrogen, however, is retarded by the depletion of electrons due to
recombination, but the gas nevertheless continues to contract.
Shortly before the gas density reaches $n_{\rm H} \sim 10^8 {\rm
cm^{-3}}$, three-body formation of molecular hydrogen becomes
important and this results in very efficient cooling which exacerbates
the runaway collapse. When the gas density reaches $\sim 10^{10} {\rm
cm^{-3}}$, the gas becomes optically thick to ${\rm H_2}$-cooling
lines, and our simulation cannot proceed further in a physically
meaningful way (see Yoshida et al. 2006 for a calculation of gas
evolution beyond this point).

\subsection{The formation of first stars at lower redshifts ($z=25$ and $z=10$)}
\begin{figure*}
\hspace{0.13cm}\resizebox{8cm}{!}{\includegraphics{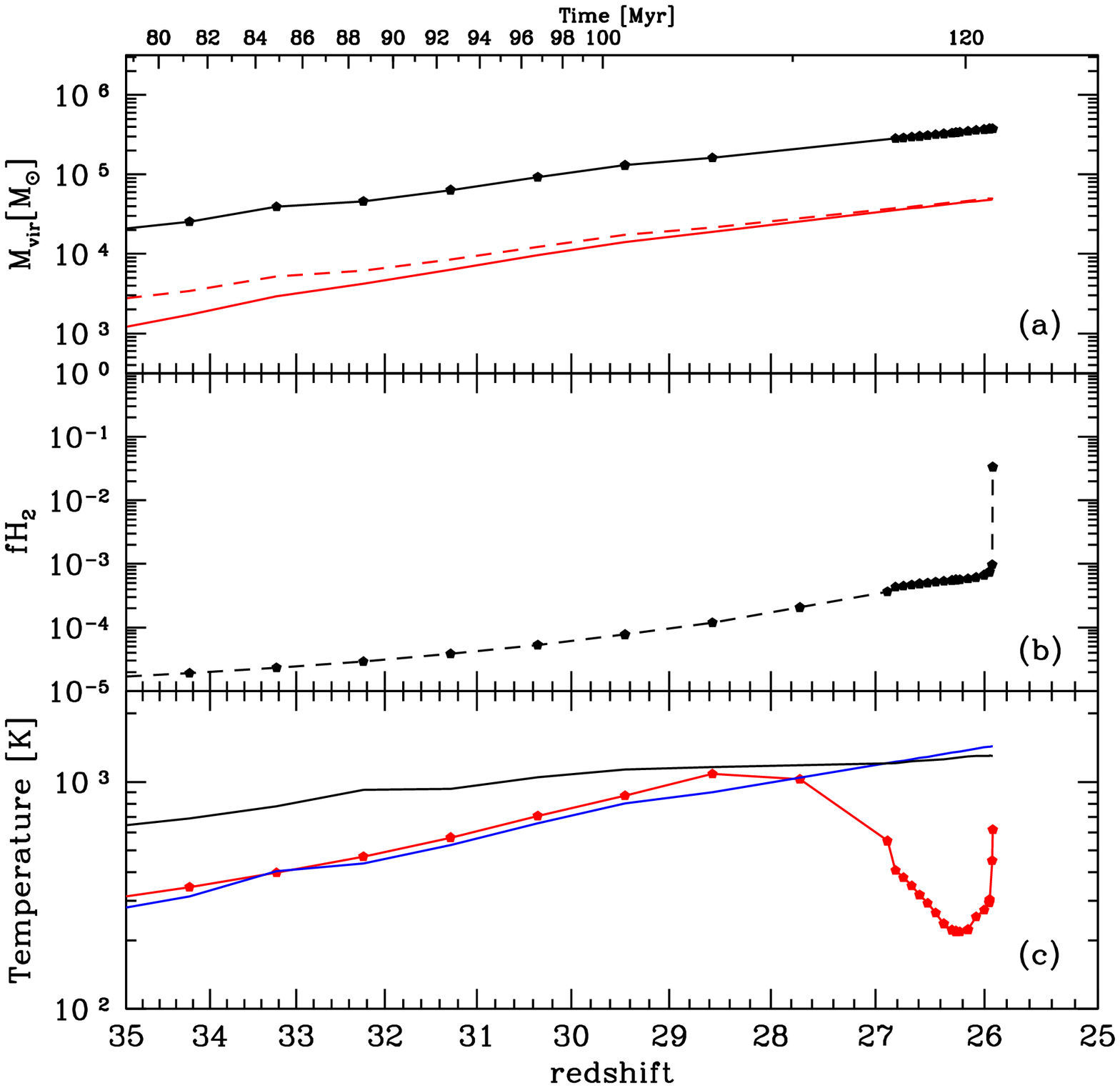}} \hspace{0.13cm}
\resizebox{8cm}{!}{\includegraphics{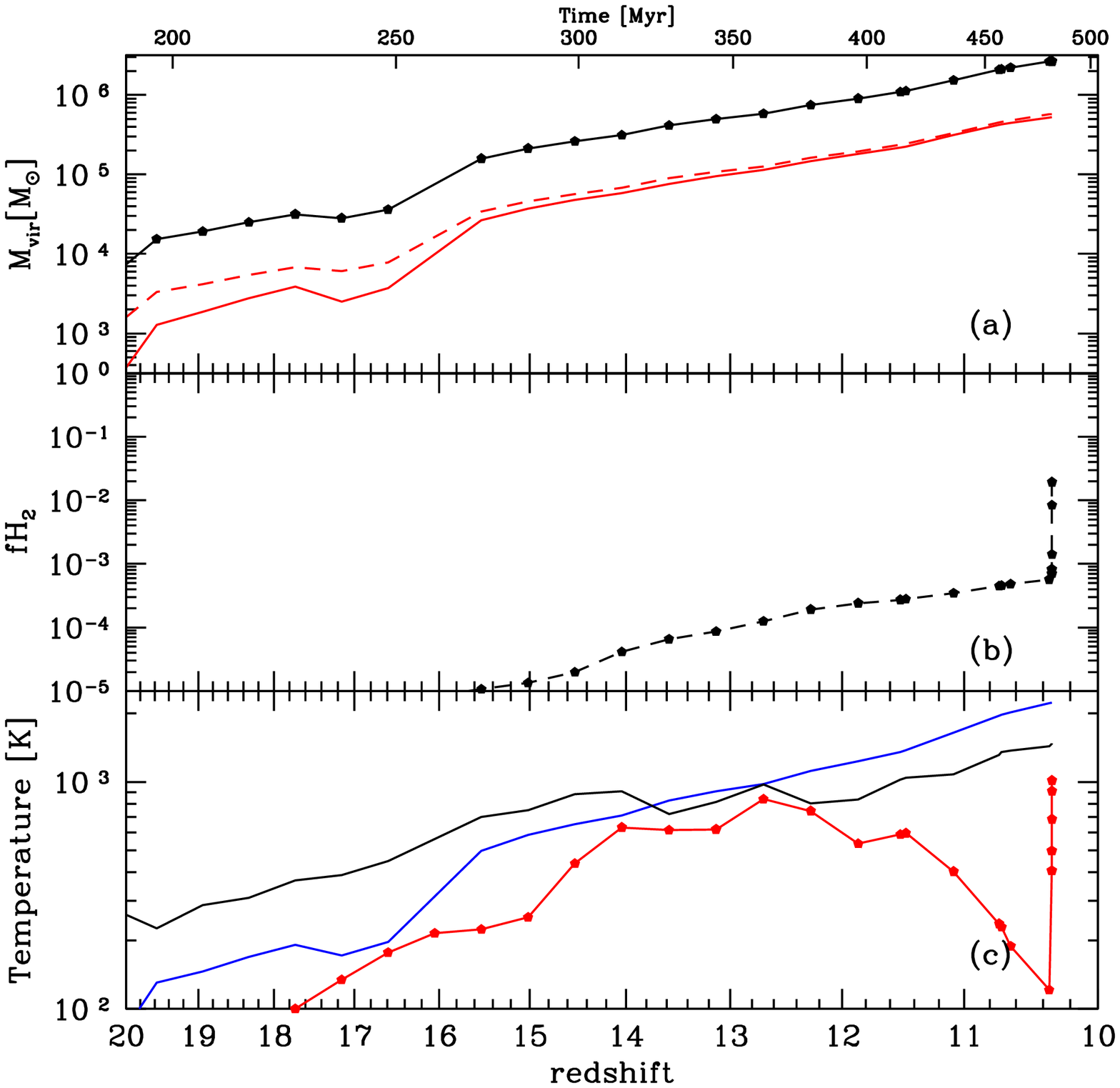}} \hspace{0.13cm}
\resizebox{8cm}{!}{\includegraphics{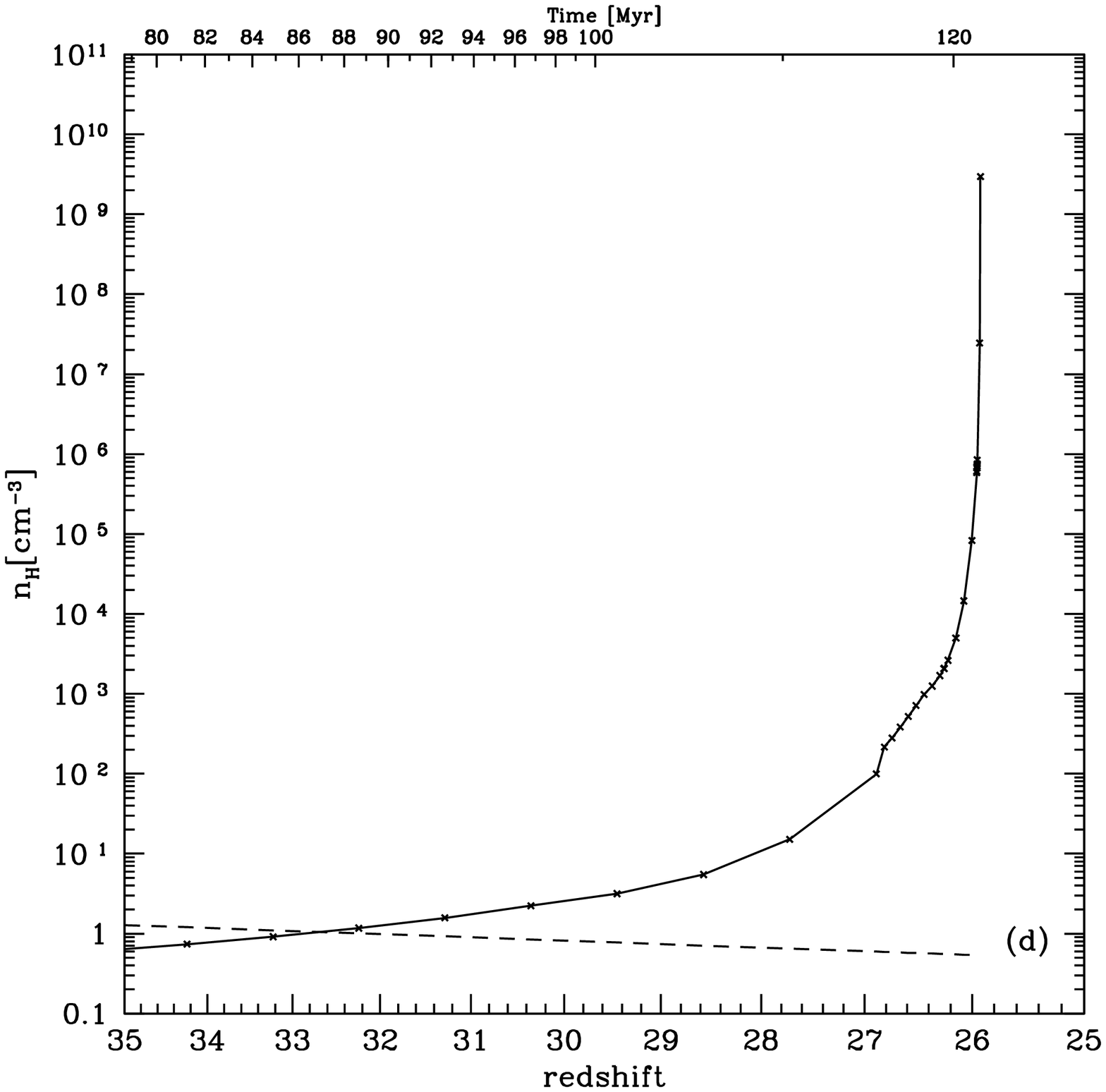}} \hspace{0.13cm}
\resizebox{8cm}{!}{\includegraphics{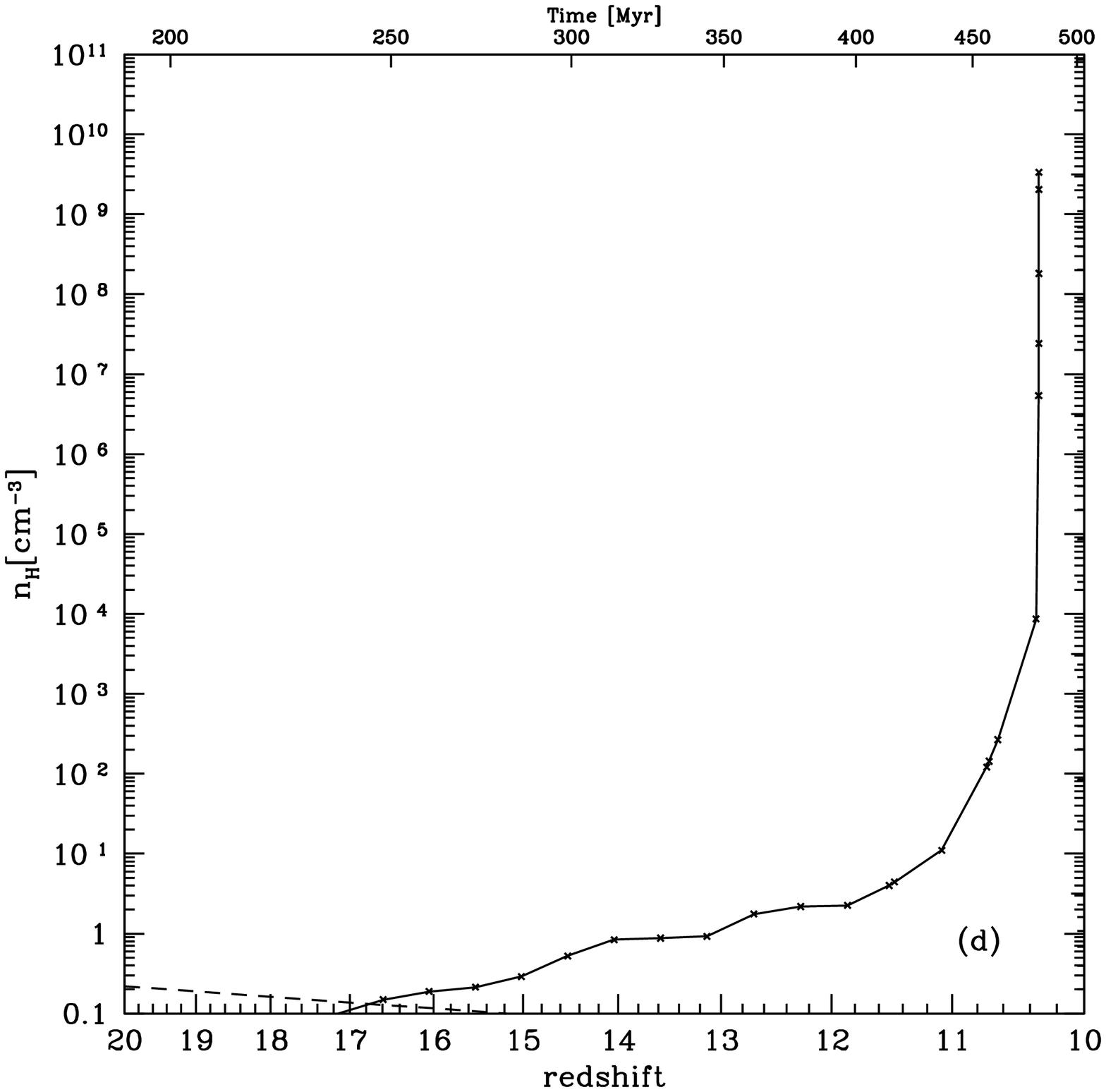}}
\caption{Evolution of various properties of the final innermost $100$
{\small SPH} particles in the Z1 simulation (left panels) and the Z2wt
simulation (right panels). Panel (a): the black solid lines with
symbols show the virial mass of the main progenitor; the (red) solid
lines show the mass of the gas component within the virial radius of
the halo; the (red) the (red) dashed lines are the fiducial gas mass
that the halo would contain at the universal baryon fraction,
$f_b=\Omega_b/\Omega_m$. Panel (b): the evolution of the ${\rm H_2}$
fraction of the final $100$ innermost gas particles. 
Panel (c): the gas temperature evolution. 
The lower (blue) solid line is an
estimate of the virial temperature of the main halo, as described in
the text; the black solid lines are the maximum temperature defined as
the maximum of the spherically mass-averaged temperature profile of
the the gas in the main halo. The (red) solid lines with symbols show
the evolution of the median value of the temperature for the 100
selected particles. Panel (d): the solid lines with symbols are the
median values of the number density of hydrogen nuclei in the 100
selected particles.  The dashed lines indicate $200$ times the
cosmological mean number density.} 
\label{fig:figz1z2wt}
\end{figure*}

The simulations labelled Z1 and Z2wt in Section~3 allow us to examine
whether or not the formation paths of protostars at lower redshifts
differ from those forming at high $z$. The first objects in these
simulations form much later than in R5. In Z1 the first protostar
forms at $z \sim 25$, which is often taken in the literature to be the
typical formation redshift of primordial stars; in Z1wt the first
protostar forms at $z \sim 10$, the lowest value in our sample. 

We carry out a similar analysis for Z1 and Z2wt to that for the R5
object presented in Figure~\ref{fig:evor5}. We identify the innermost
$100$ {\small SPH} particles at the final simulation output and then
trace these particles back to earlier times. The time evolution of the
median values of various physical quantities in the Z1 and Z2wt
simulations are shown in Figure~\ref{fig:figz1z2wt}.

On the whole, the formation paths of protostars in the Z1 and the Z2wt
simulations are similar to those in R5. The baryon fraction in the
halo is initially lower than the universal value but it then catches
up, as in R5. However, the mass scale where this happens is smaller
because the cosmological Jeans mass, $M_{\rm J} (z) \propto
T^{3/2}\rho^{-1/2} \propto (1+z)^{3/2}$, is smaller at lower
redshifts. After this stage, the gas is continuously compressed and
its temperature raises, resulting in the rapid formation of ${\rm
H_2}$. When the gas temperature reaches $T \sim 1000\,{\rm K}$ and the
${\rm H_2}$ fraction increases to a few times $10^{-4}$, the gas
begins to cool rapidly, just as in R5.

Despite the overall similarity of the formation paths of the first
objects in Z1, Z2wt and R5, the timescale for collapse is very
different.  In R5 it takes only about $3$ million years between the
time when efficient cooling starts (signalled by the decline in
temperature) to the time when the number density reaches $n_{\rm
H}\sim 10^{10} {\rm cm^{-3}}$, but this phase takes $10$ and $100$
million years in Z1 and Z2wt respectively. The initial collapse of
protostellar objects proceeds much more rapidly at very high redshift
because of the higher density which accelerates the formation of
molecular hydrogen, enhances initial cooling, and shortens the
dynamical time.

These results suggest a relatively simple picture of the formation
path of the first stars although the detailed evolution can be
complicated and may differ from object to object depending on the
particular infall pattern. As the halo virial temperature approaches a
critical temperature, $T \sim 1000\,{\rm K}$, the rapid production of
molecular hydrogen starts once the ${\rm H_2}$ fraction reaches a
critical value of a few times $10^{-4}$, allowing the gas to cool
efficiently. The timescale for the subsequent evolution to a protostar
varies systematically depending on the redshift when the collpase is
taking place, with halos at higher redshift collapsing faster. In our
simulations the timescale varies by a factor as large as $20$ between
redshifts $z \sim 50$ and $z \sim 10$.

Panel~(a) in Figures~\ref{fig:evor5} and \ref{fig:figz1z2wt} also
suggests that the virial temperature provides only a crude estimate of
the gas temperature in the halos because the gas accretion process in
the protostellar clouds is rather complicated. The maximum
temperature, which marks the site of the accretion shock, can often
differ significantly from the virial temperature. These two
temperatures only roughly match at the time when the cosmic Jeans mass
is reached, which is also the time when the baryon fraction in the
halo becomes close to the universal value. After this, the maximum
temperature is usually smaller than the estimated virial temperature
by a factor as large as $50\%$.

Our results imply that a criterion for deciding whether or not a halo
hosts a star-forming cloud based purely on the mass or virial
temperature of the host halo, as used extensively in the literature,
is unlikely to give the correct answer. While it is true that the gas
starts to collapse and cool when the halo attains a virial temperature
close to $\sim~1000\,{\rm K}$, the processes eventually leading to the
formation of a protostar can take a significant amount of time and
this must be taken into account.

\subsection{Star formation in the R5 simulation}

\subsubsection{The distributions of dark matter and gas}
\begin{figure*}
\hspace{13cm}\resizebox{5cm}{!}{\includegraphics{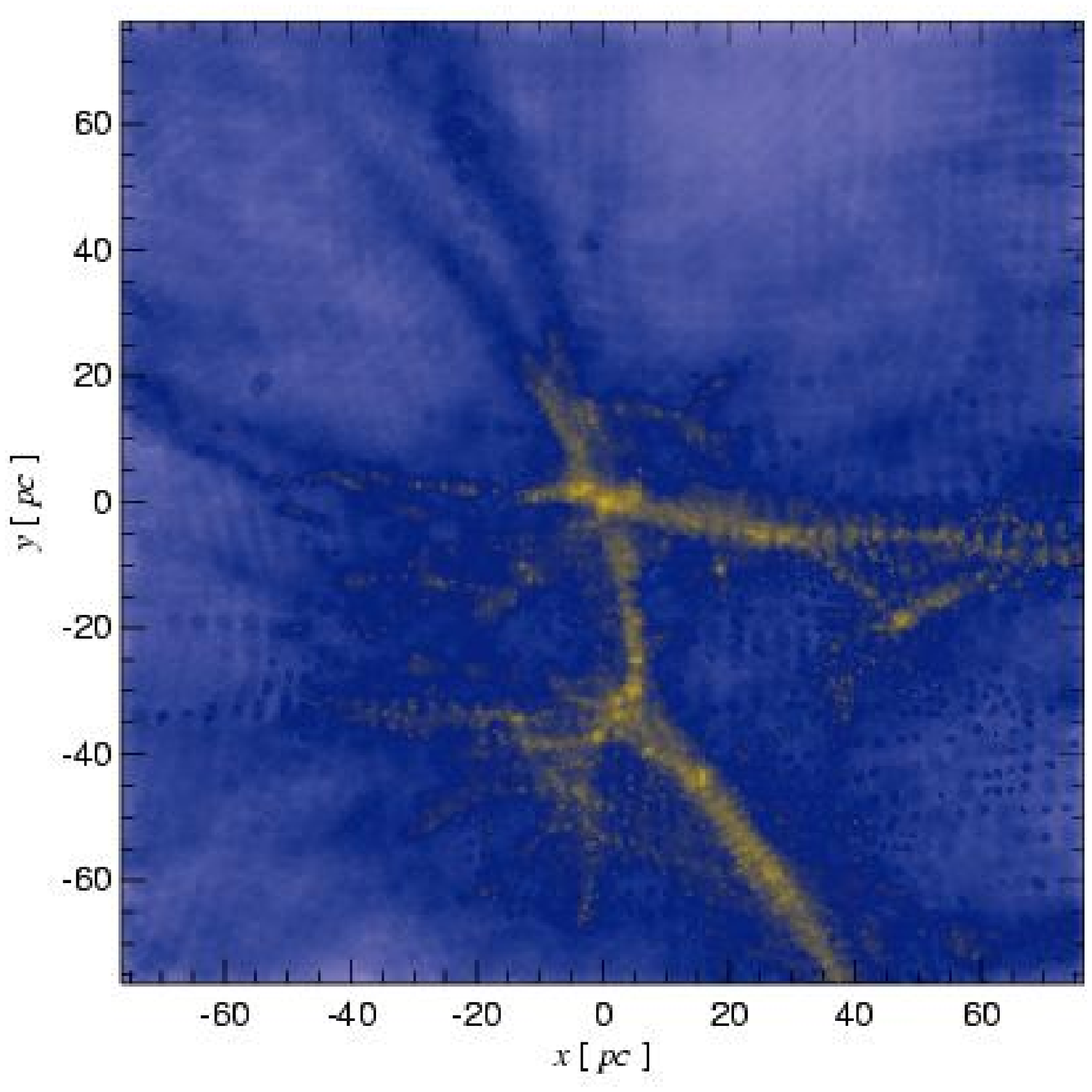}}
\resizebox{5cm}{!}{\includegraphics{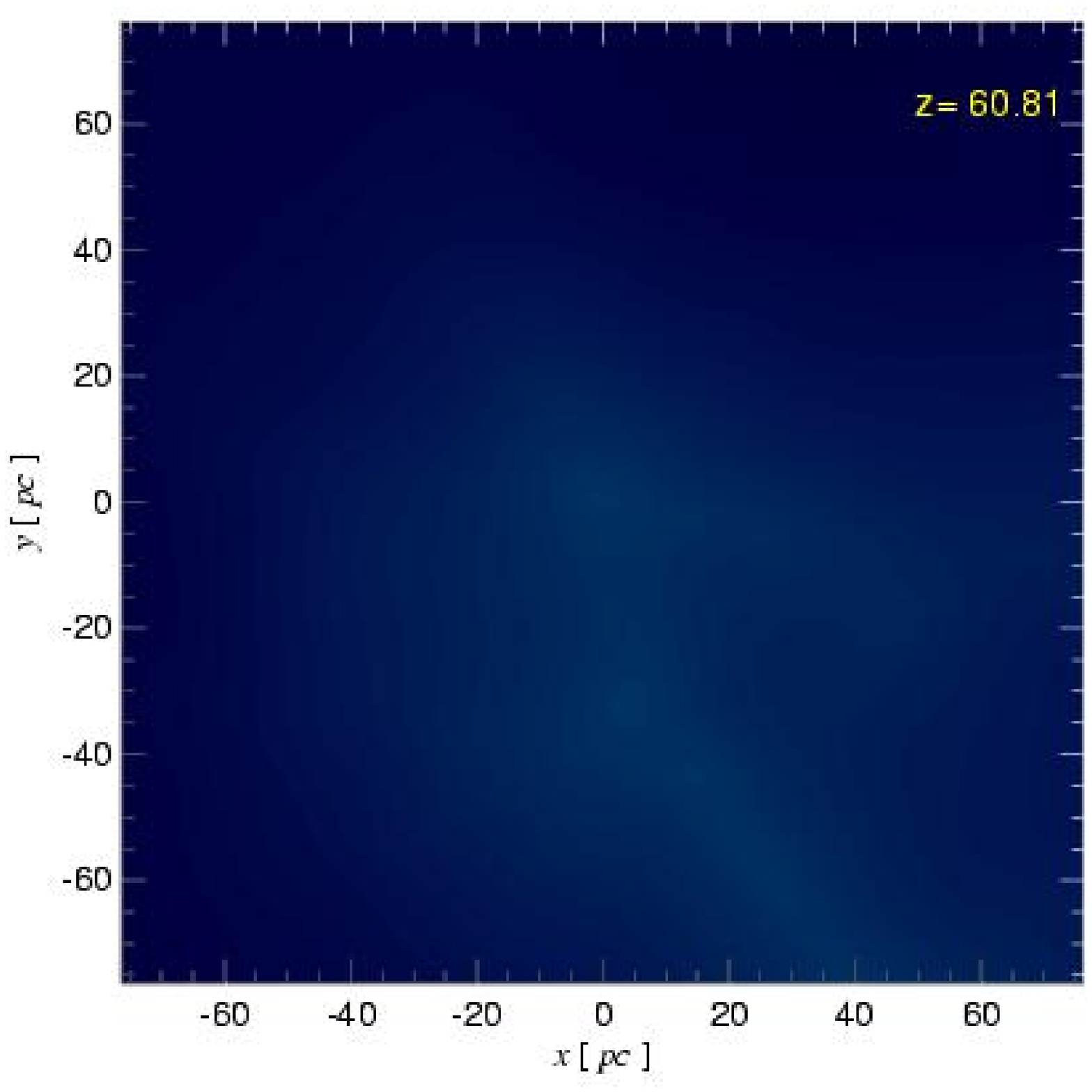}}
\resizebox{5cm}{!}{\includegraphics{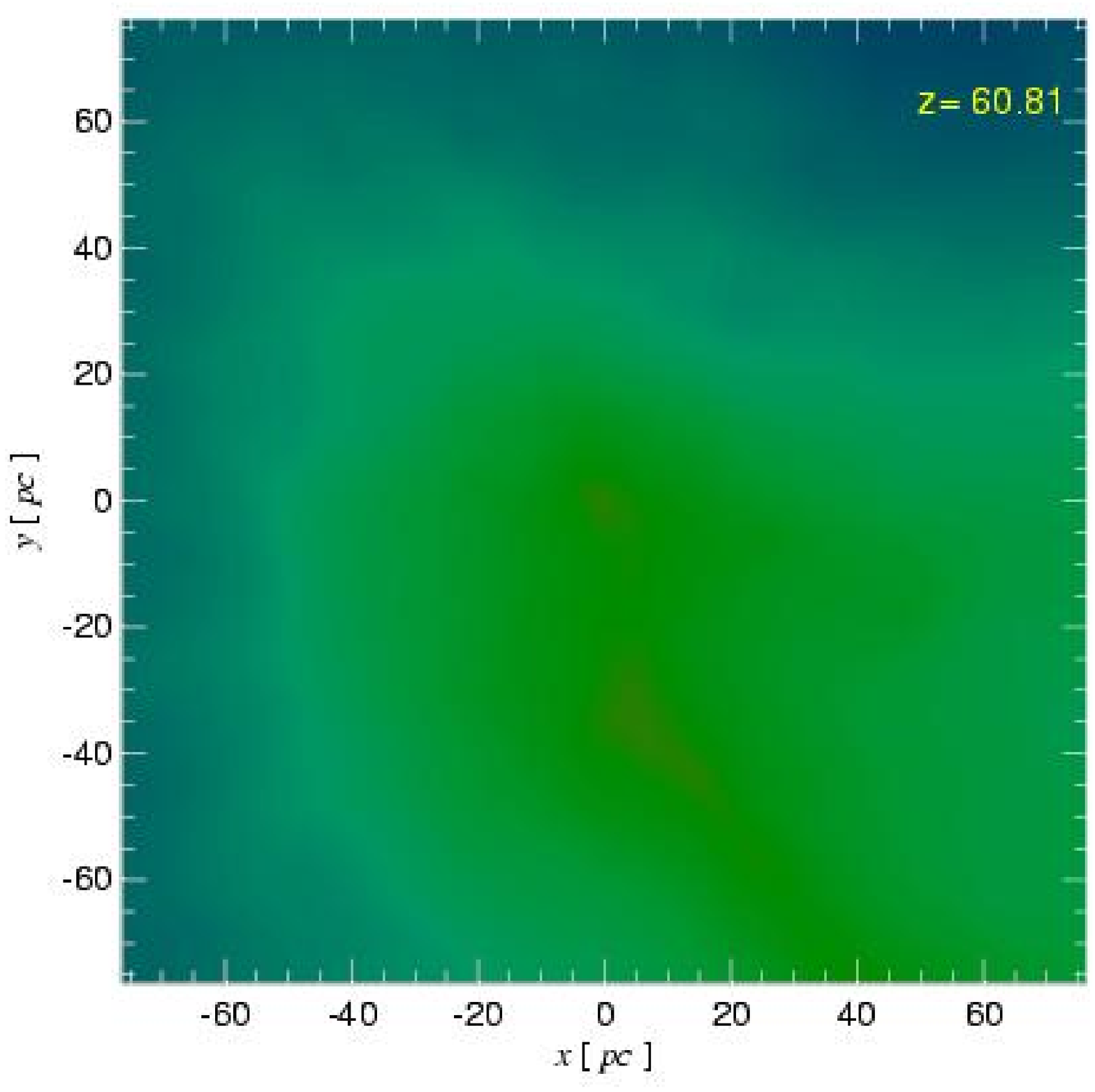}}
\hspace{13cm}\resizebox{5cm}{!}{\includegraphics{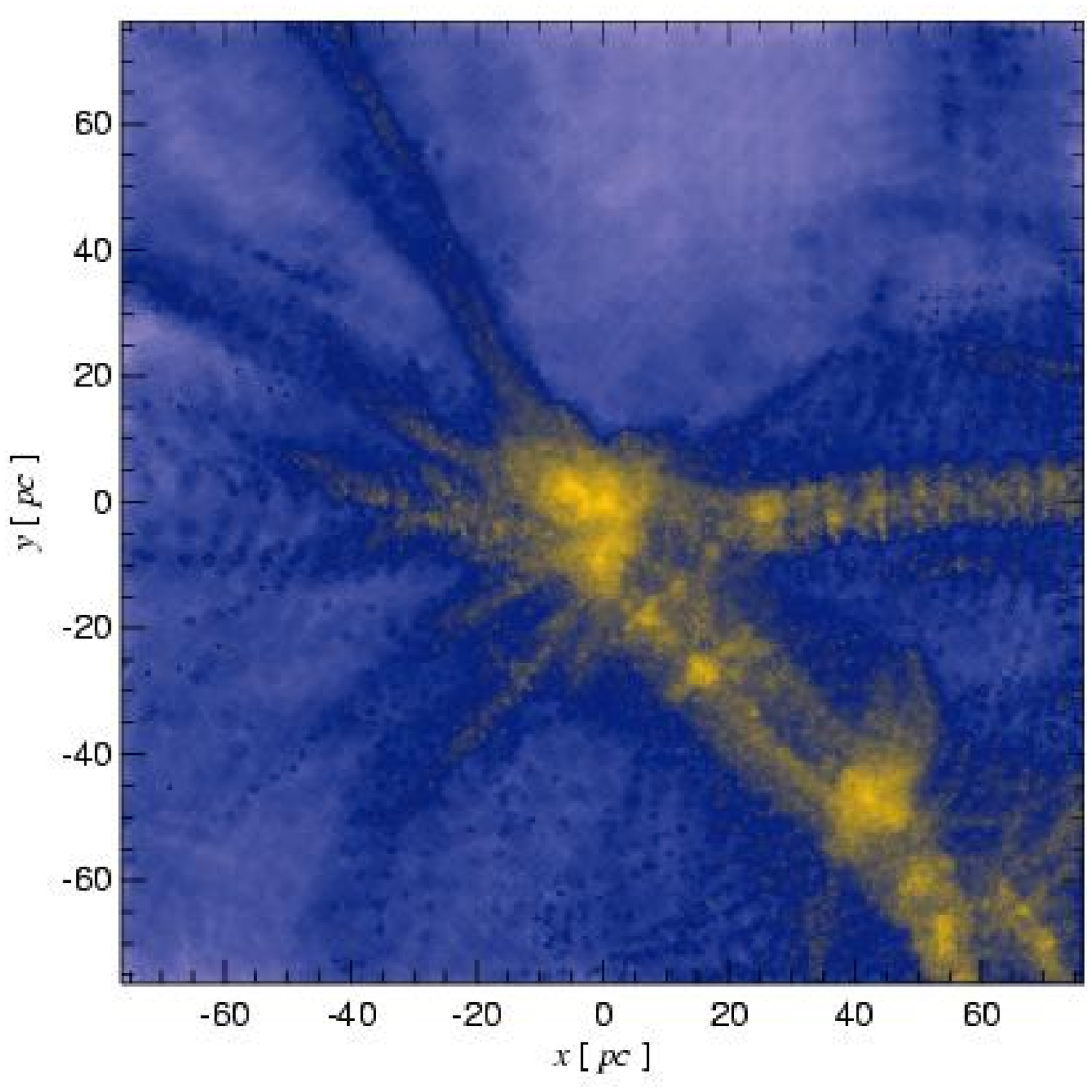}}
\resizebox{5cm}{!}{\includegraphics{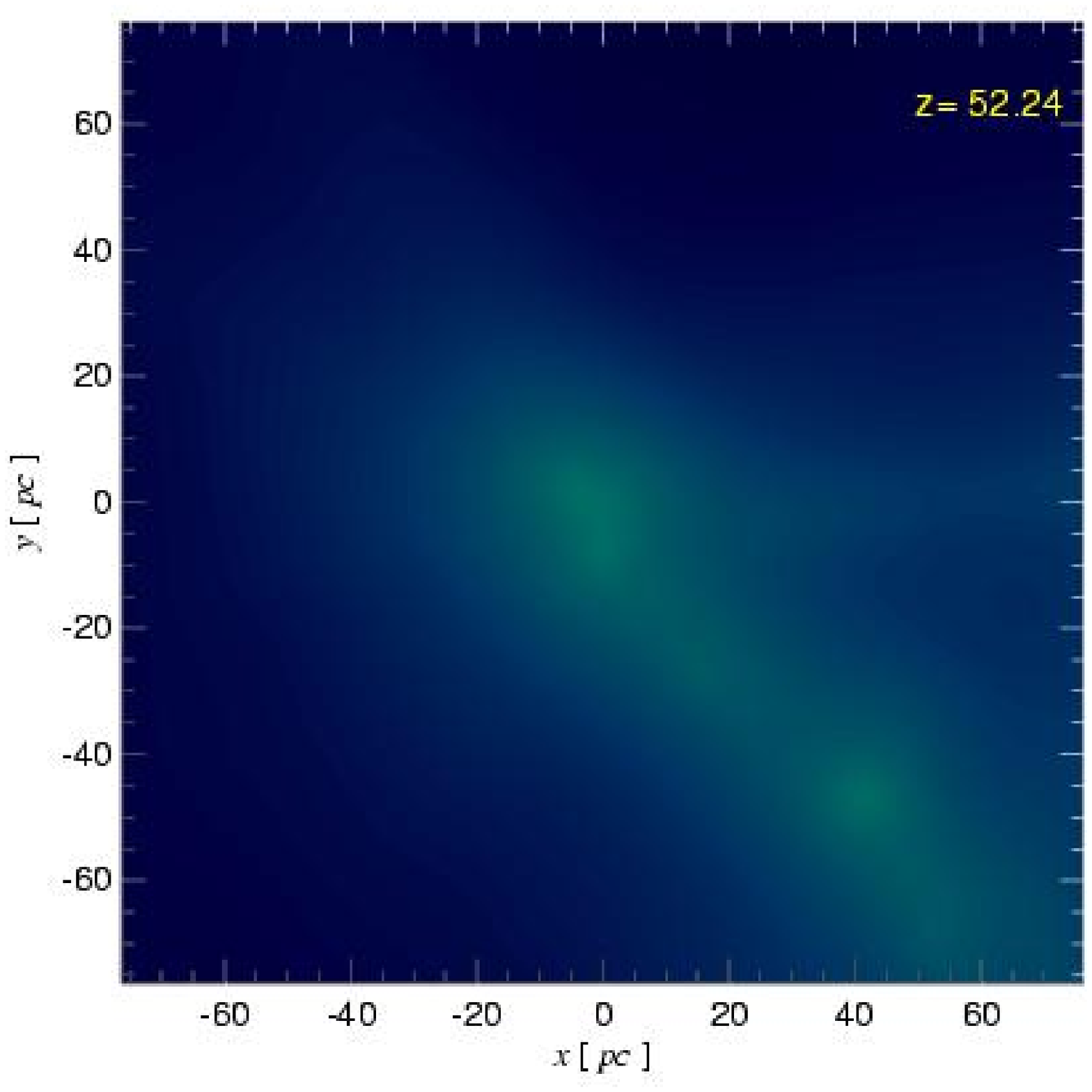}}
\resizebox{5cm}{!}{\includegraphics{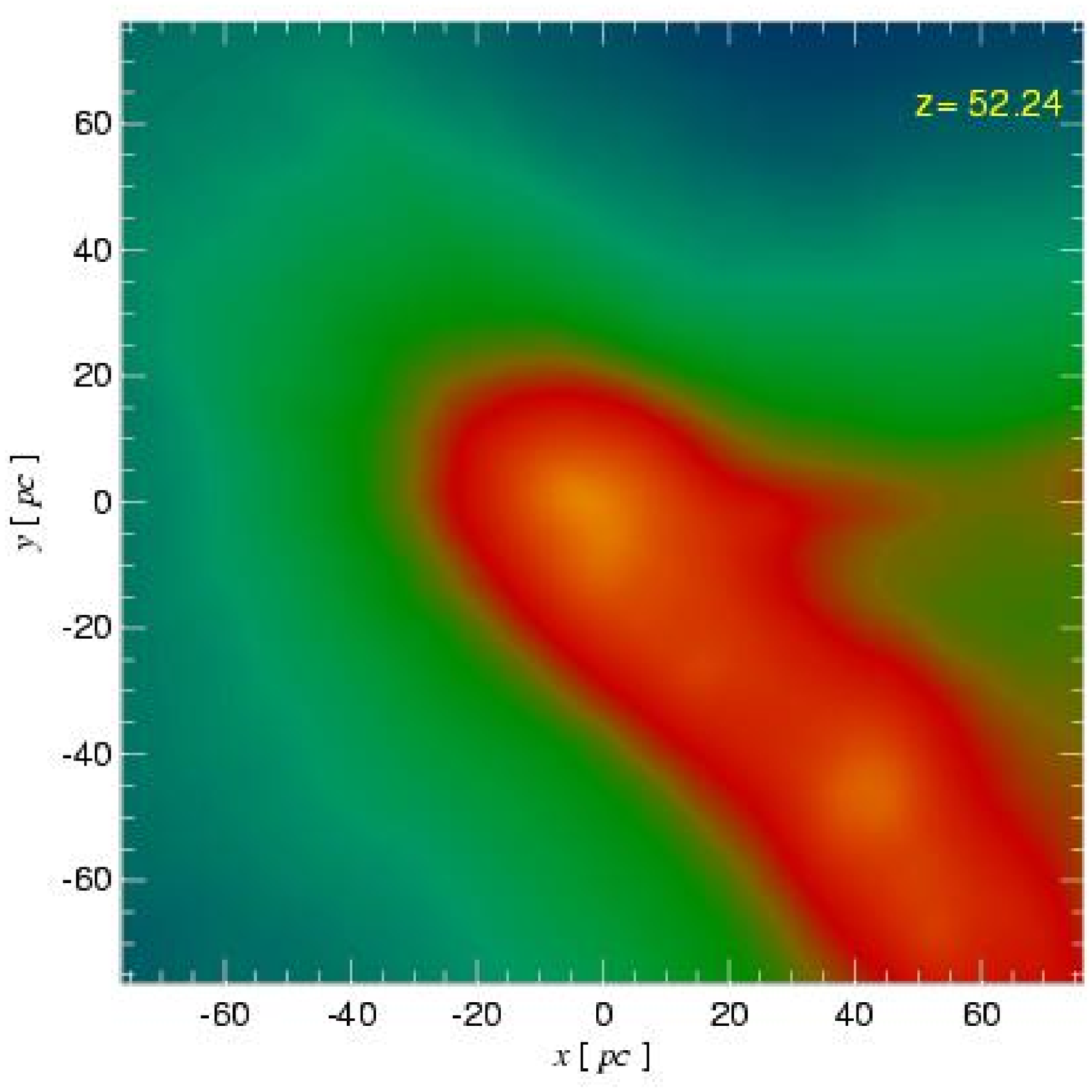}}
\hspace{13cm}\resizebox{5cm}{!}{\includegraphics{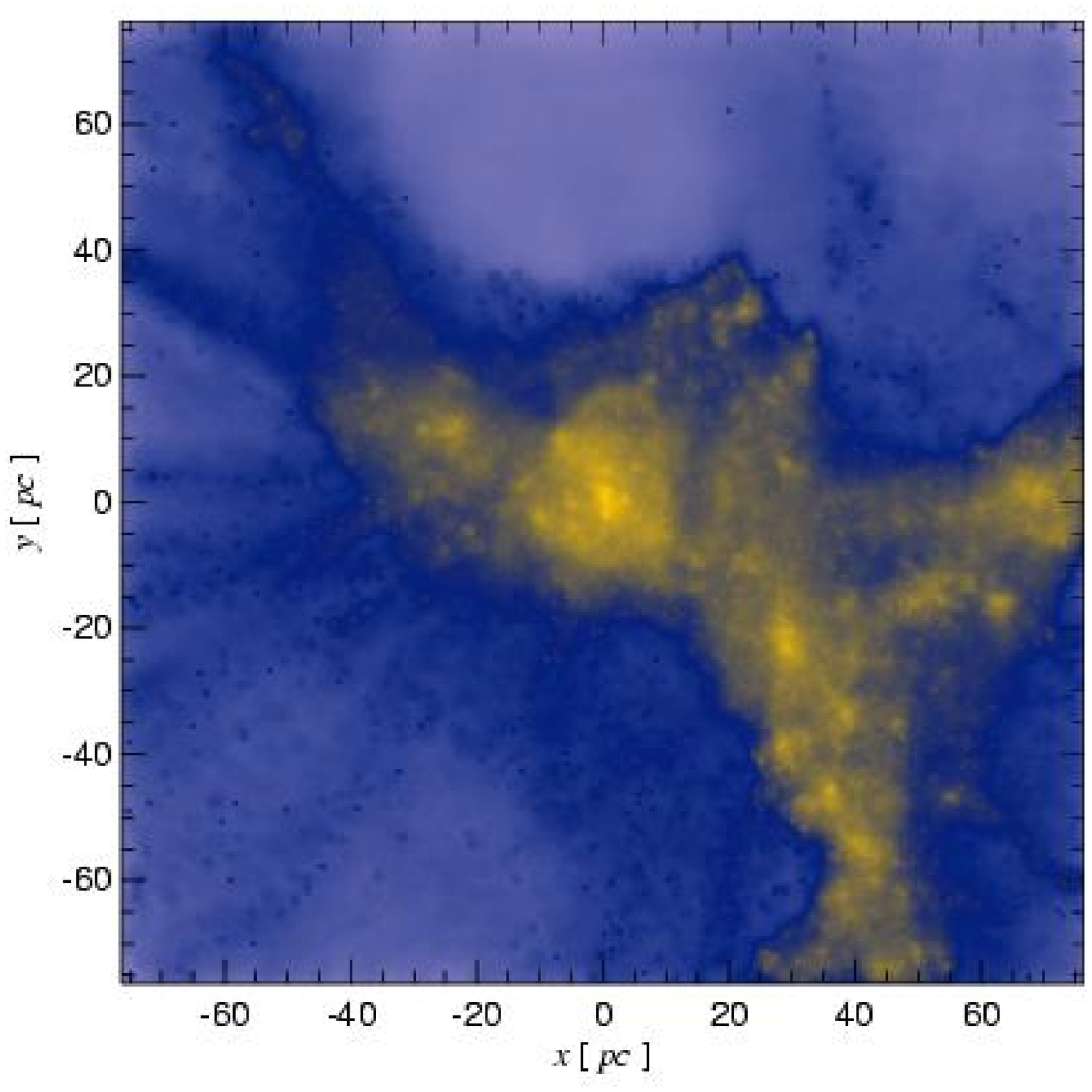}}
\resizebox{5cm}{!}{\includegraphics{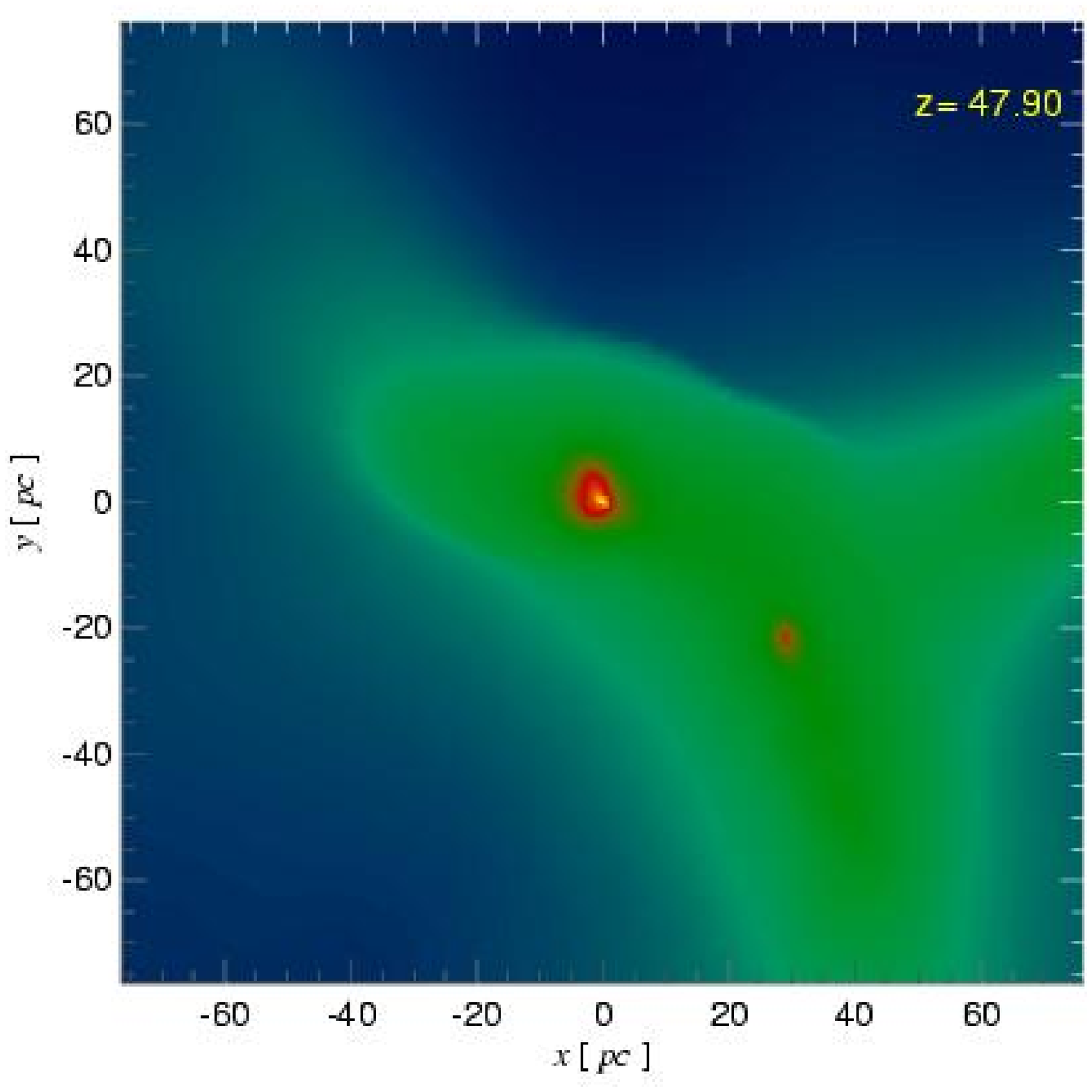}}
\resizebox{5cm}{!}{\includegraphics{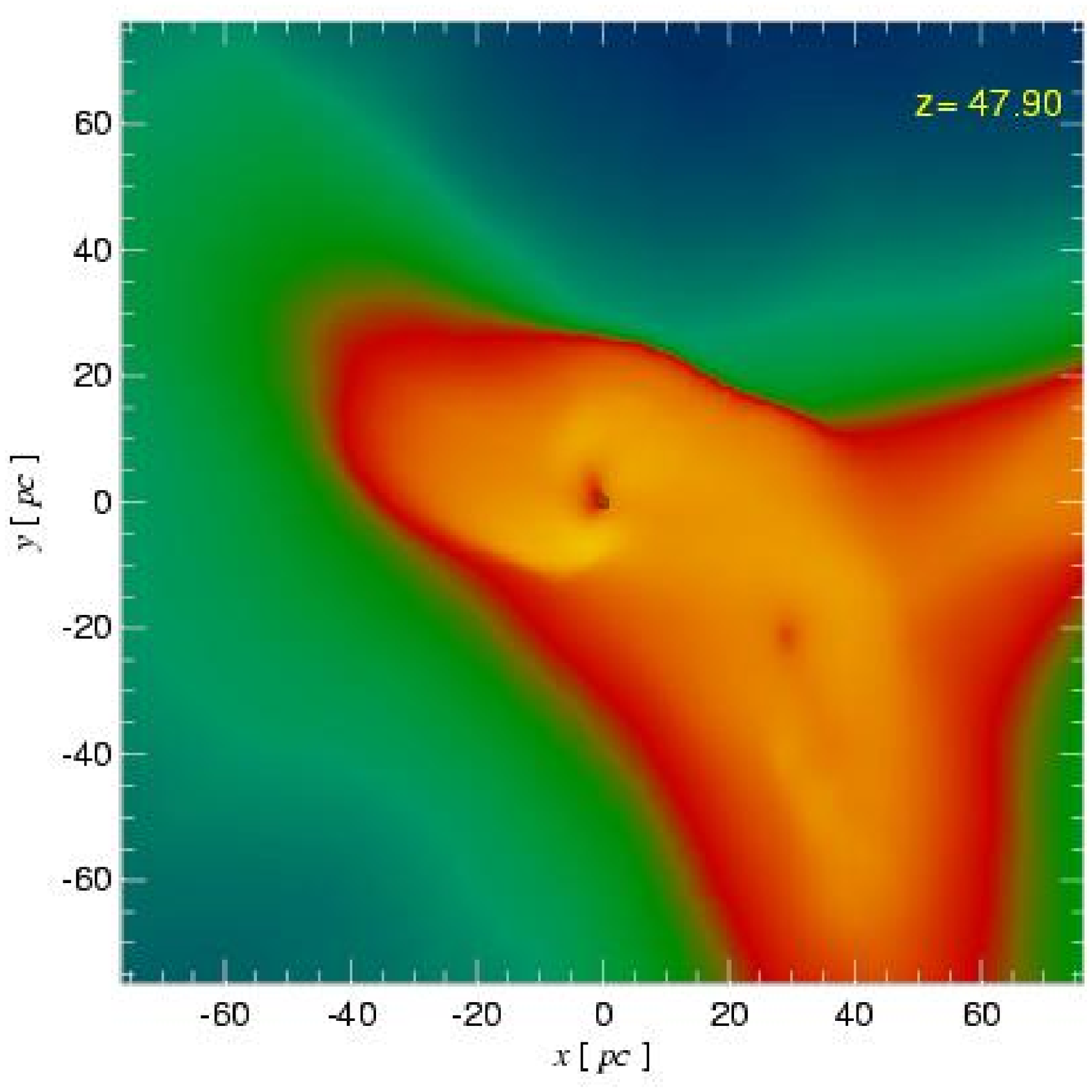}}
\hspace{13cm}\resizebox{5cm}{!}{\includegraphics{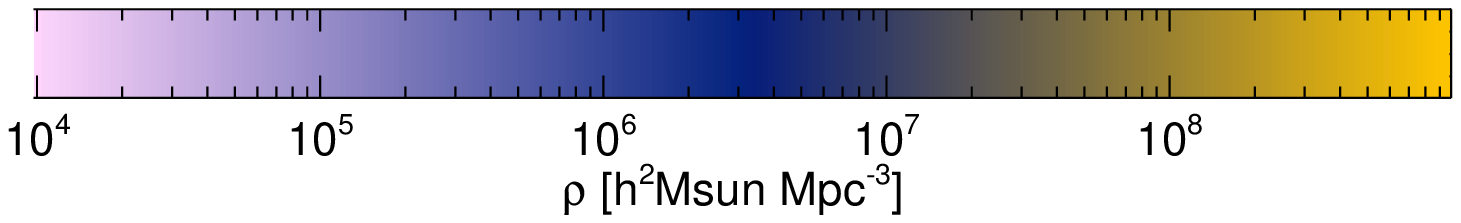}}
\resizebox{5cm}{!}{\includegraphics{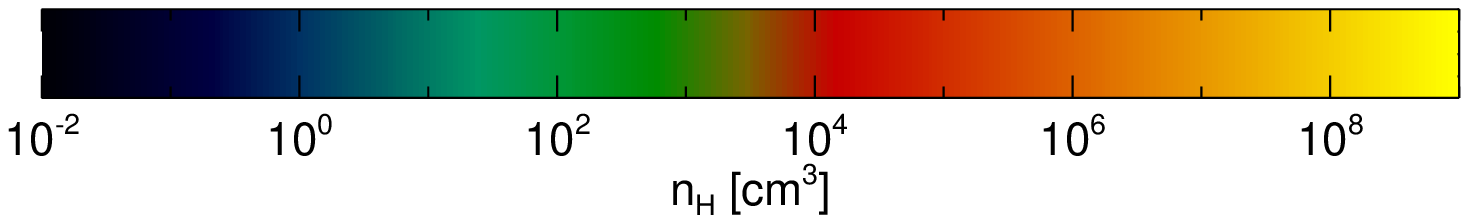}}
\resizebox{5cm}{!}{\includegraphics{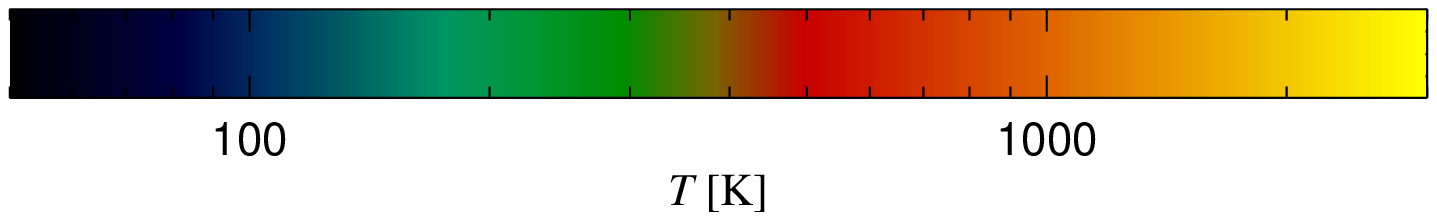}}
\caption{Projected dark matter density (left), gas density (middle) and gas
temperature (right), weighted by mass at three redshifts in the R5
simulation. In all cases, the projected regions have dimensions of $4$
times the virial radius of the R5 object in the plane of the plot and
are projected over a depth of $1$ virial radius at the final time.}
\label{fig:picevor5}
\end{figure*}

\begin{figure*}
\hspace{13cm}\resizebox{5cm}{!}{\includegraphics{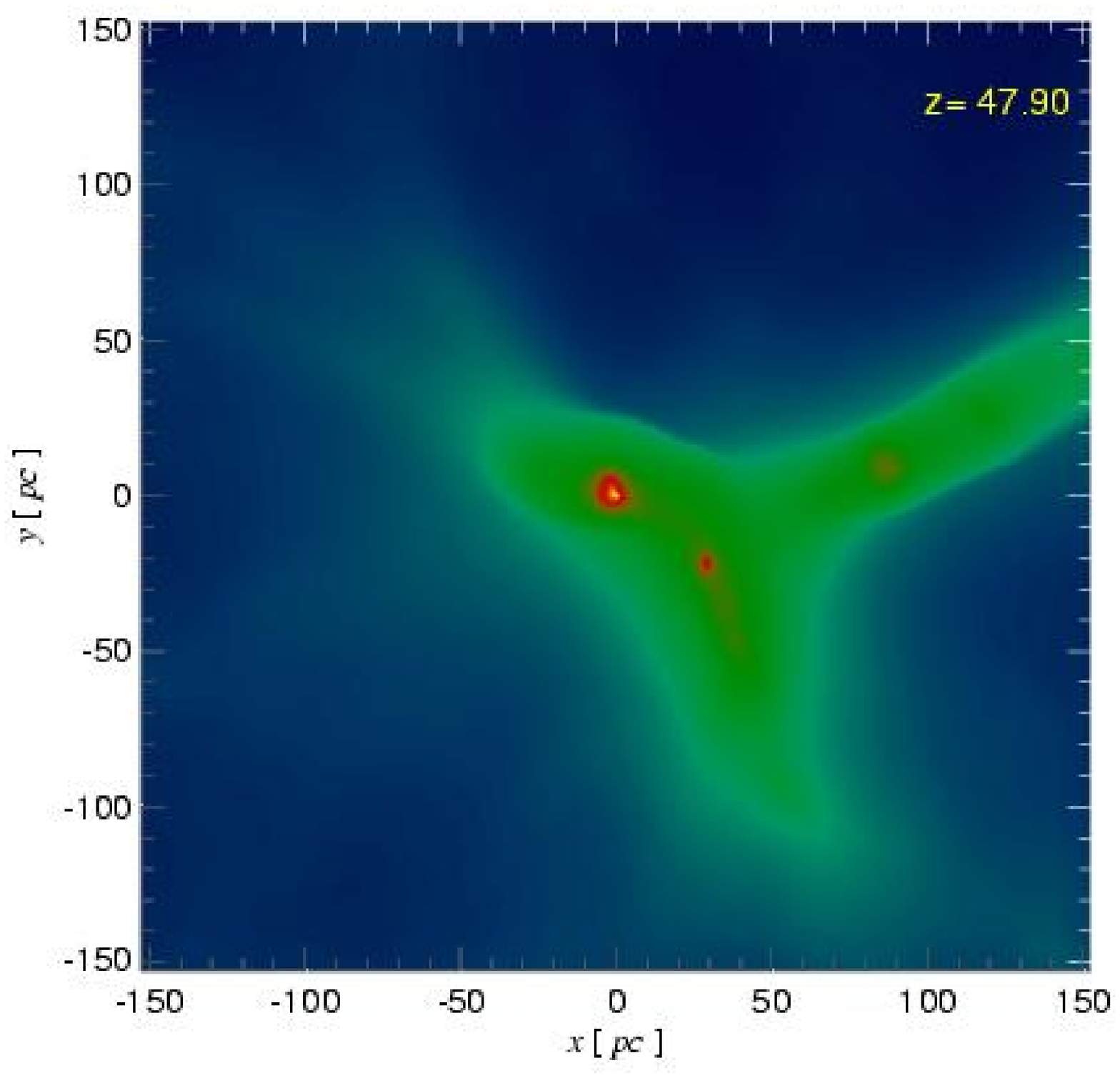}} %
\resizebox{5cm}{!}{\includegraphics{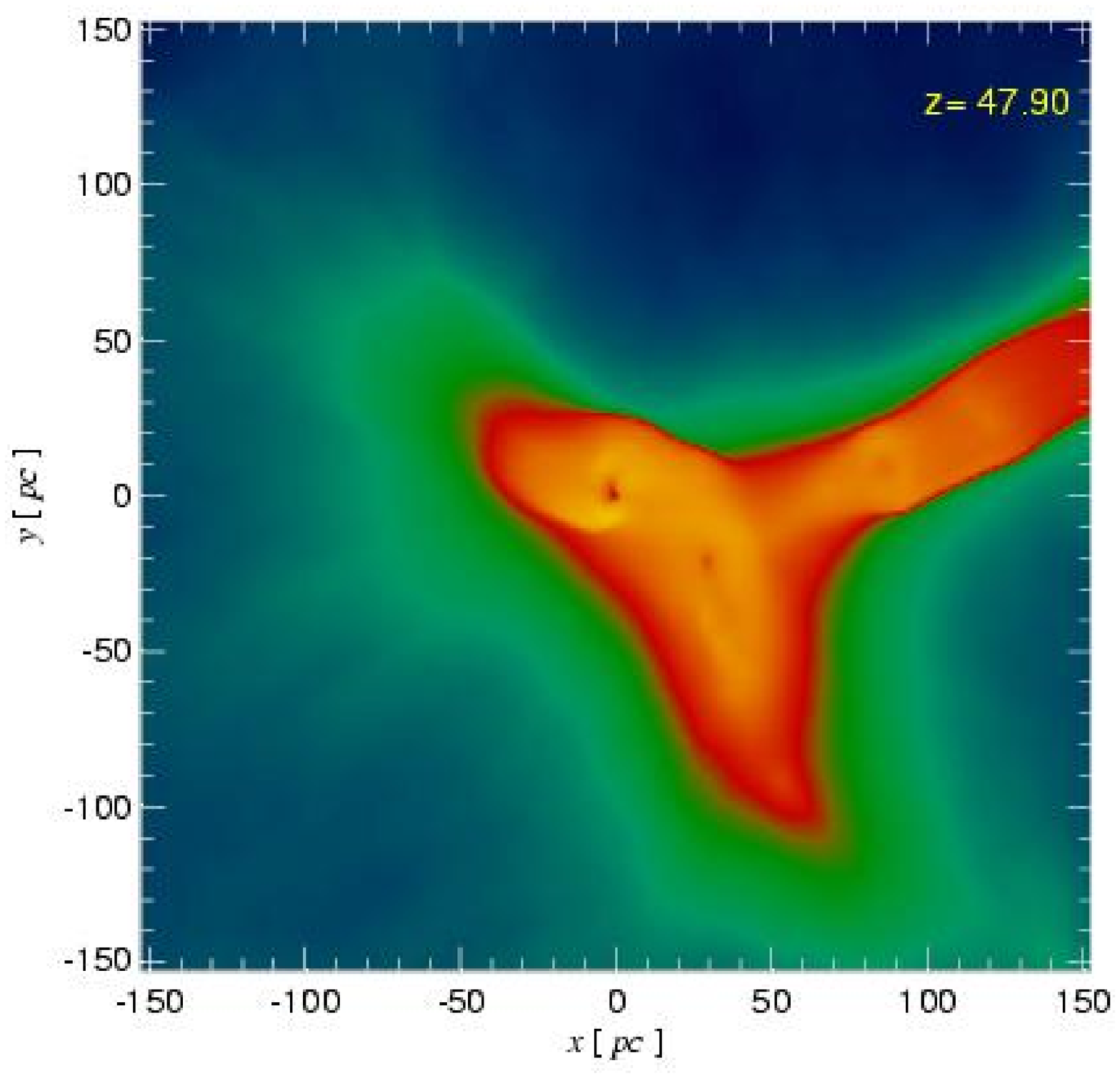}} \resizebox{5cm}{!}{%
\includegraphics{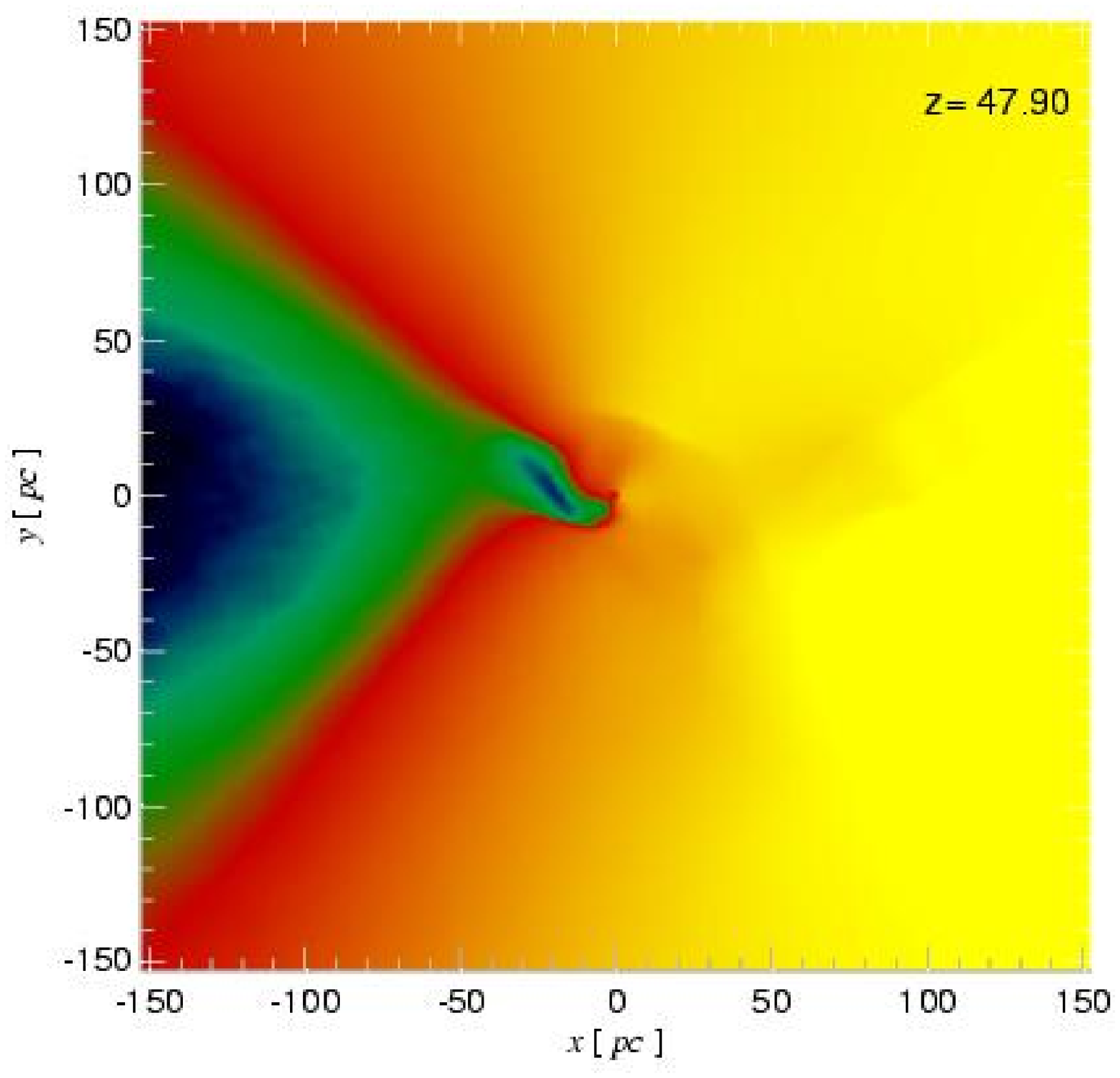}}
\hspace{13cm}\resizebox{5cm}{!}{\includegraphics{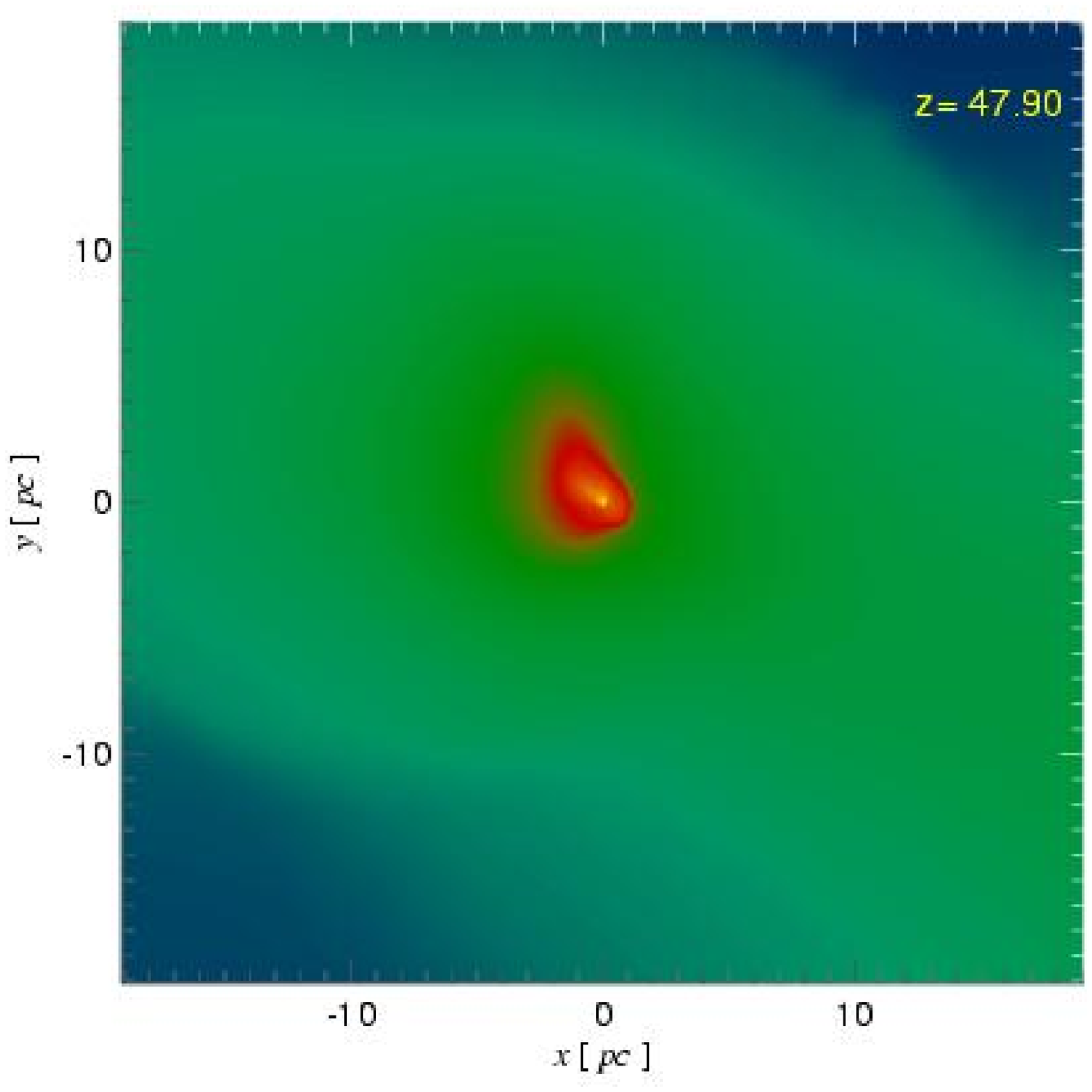}} %
\resizebox{5cm}{!}{\includegraphics{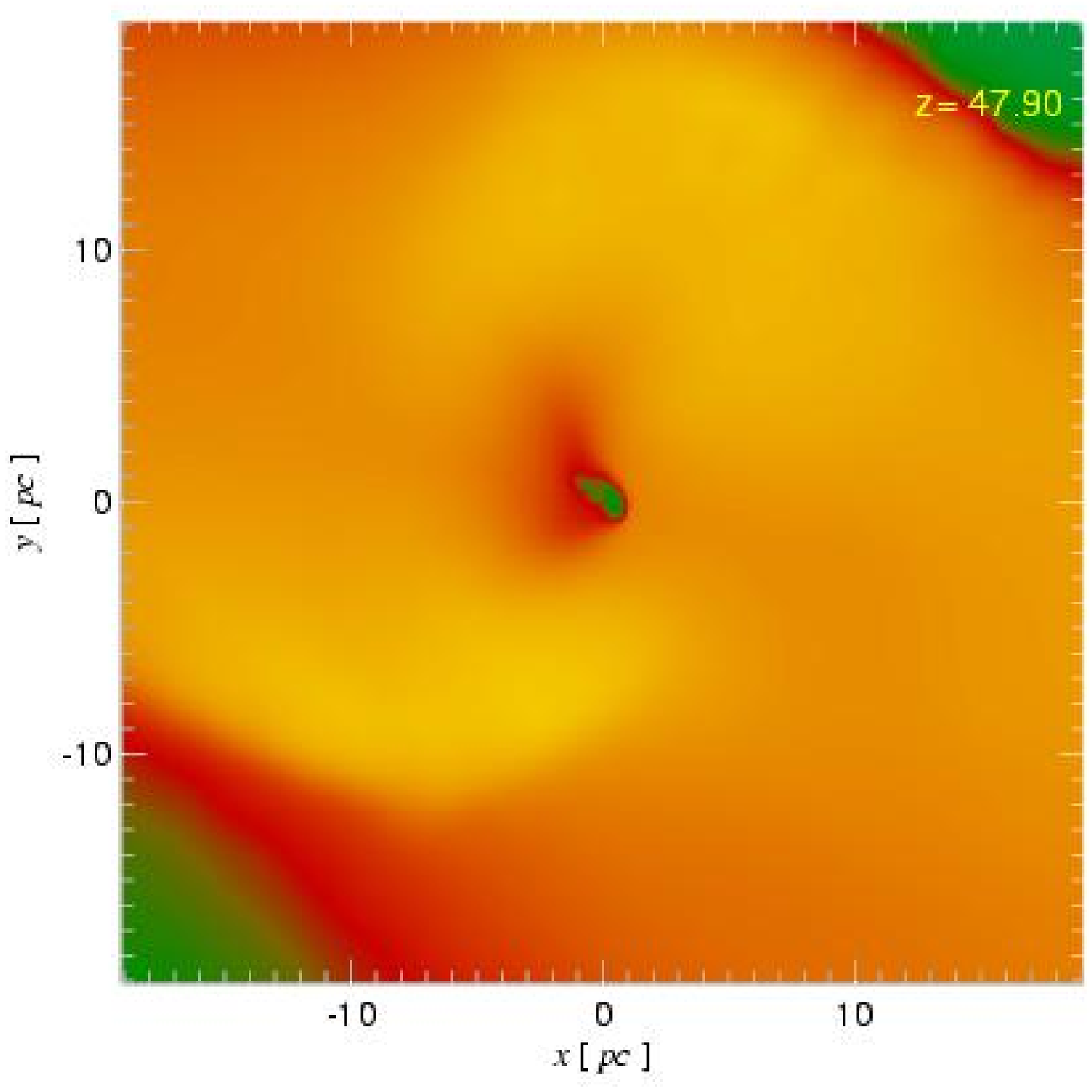}} \resizebox{5cm}{!}{%
\includegraphics{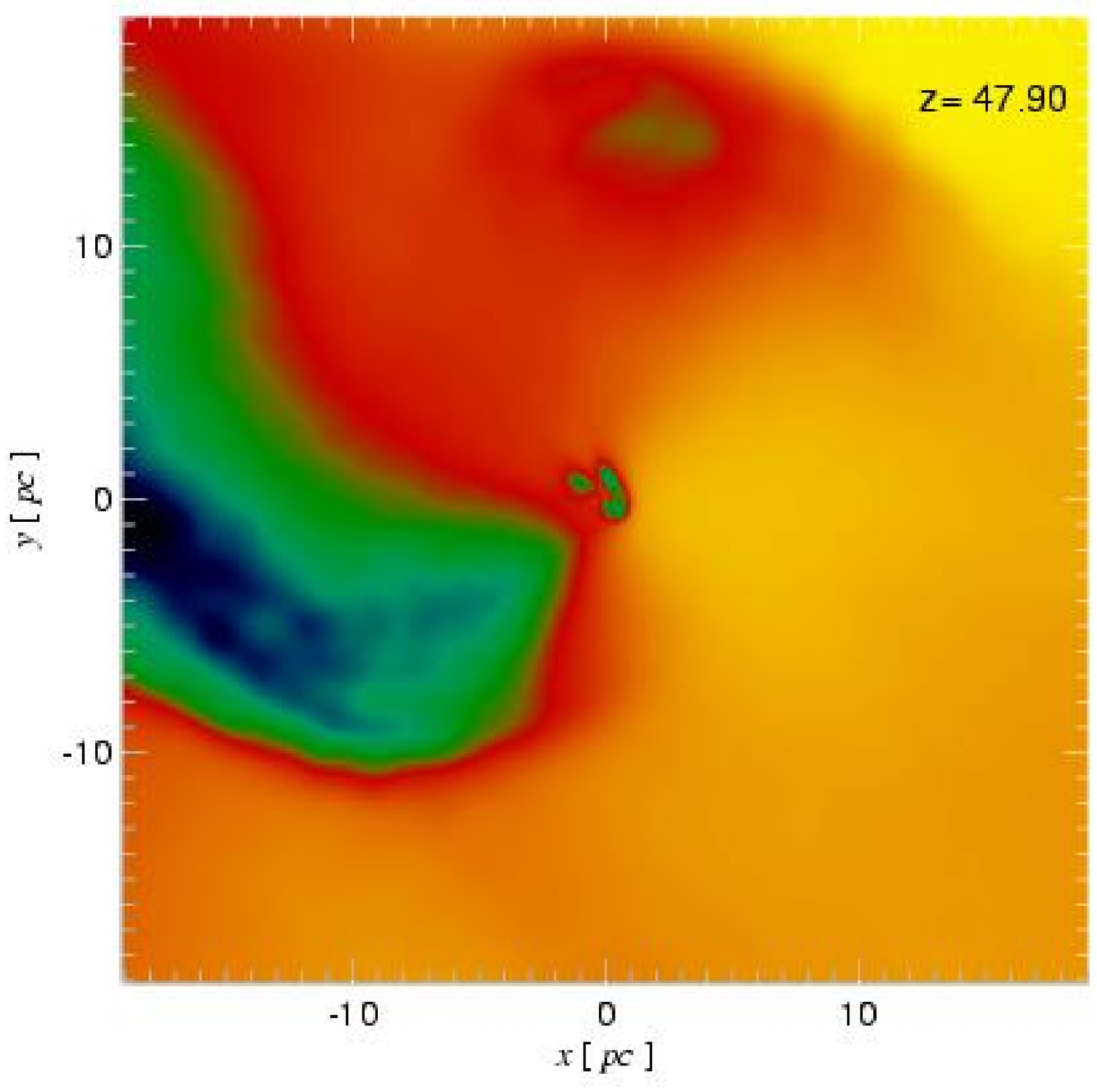}}
\hspace{13cm}\resizebox{5cm}{!}{\includegraphics{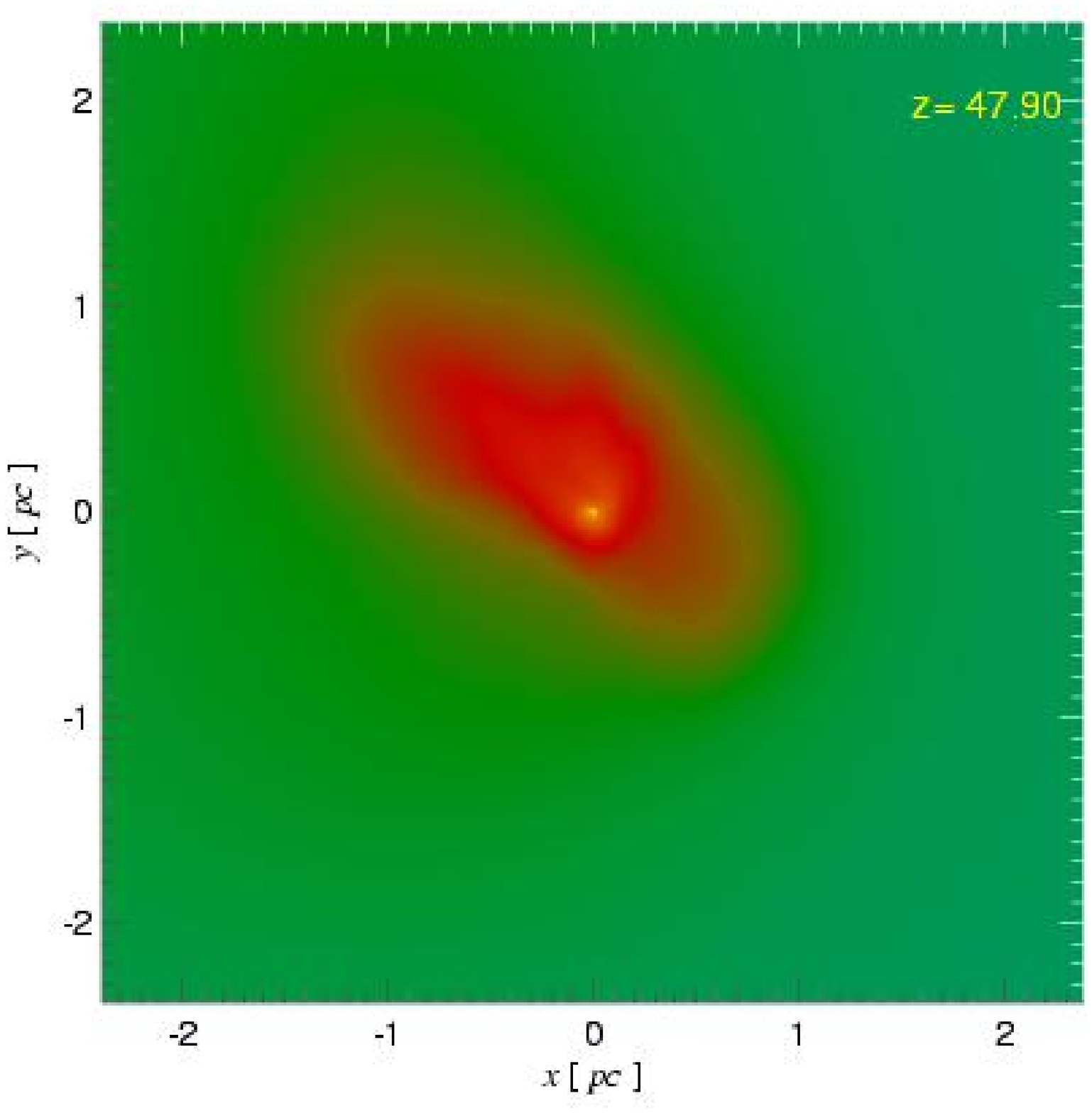}} %
\resizebox{5cm}{!}{\includegraphics{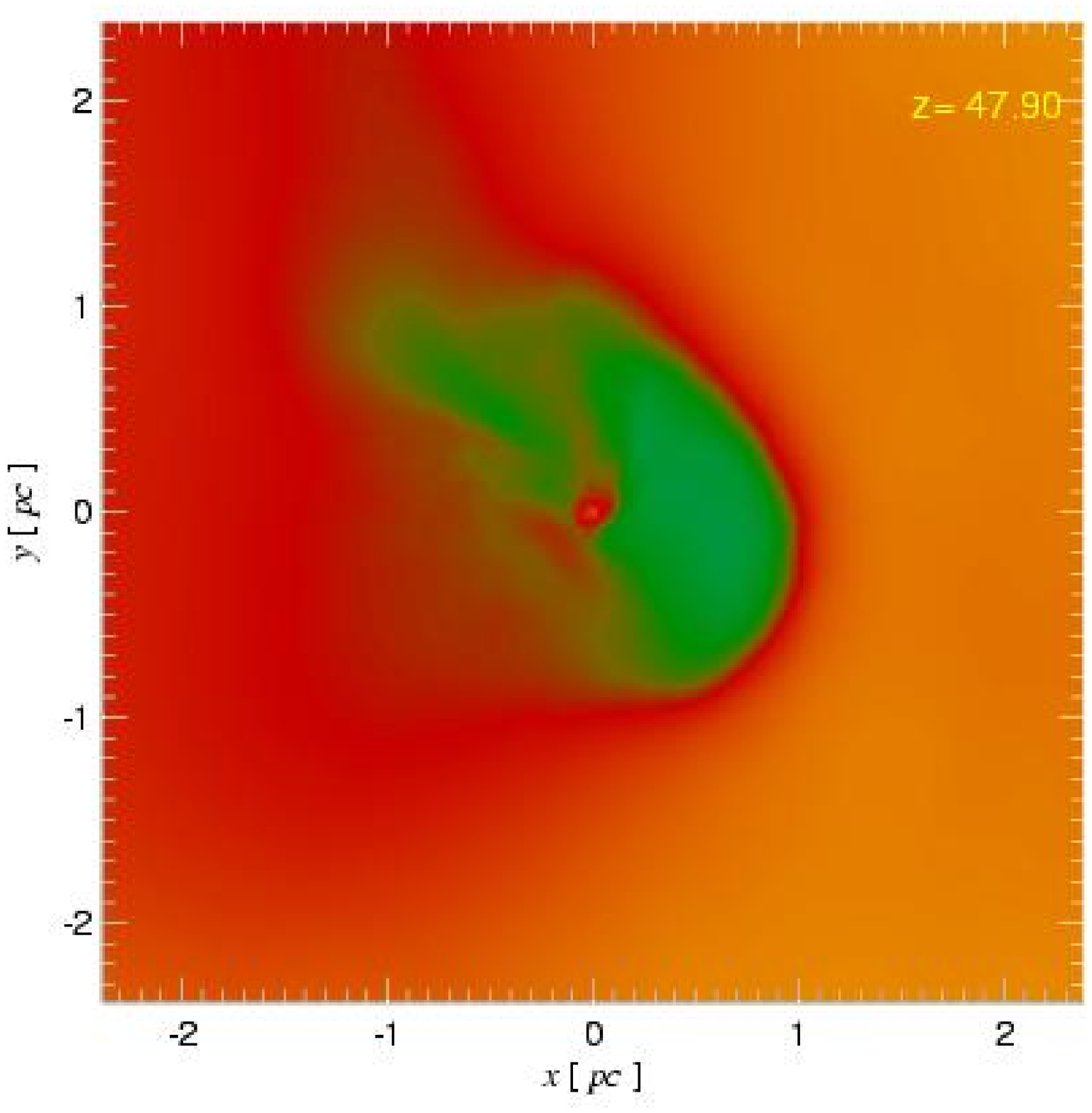}} \resizebox{5cm}{!}{%
\includegraphics{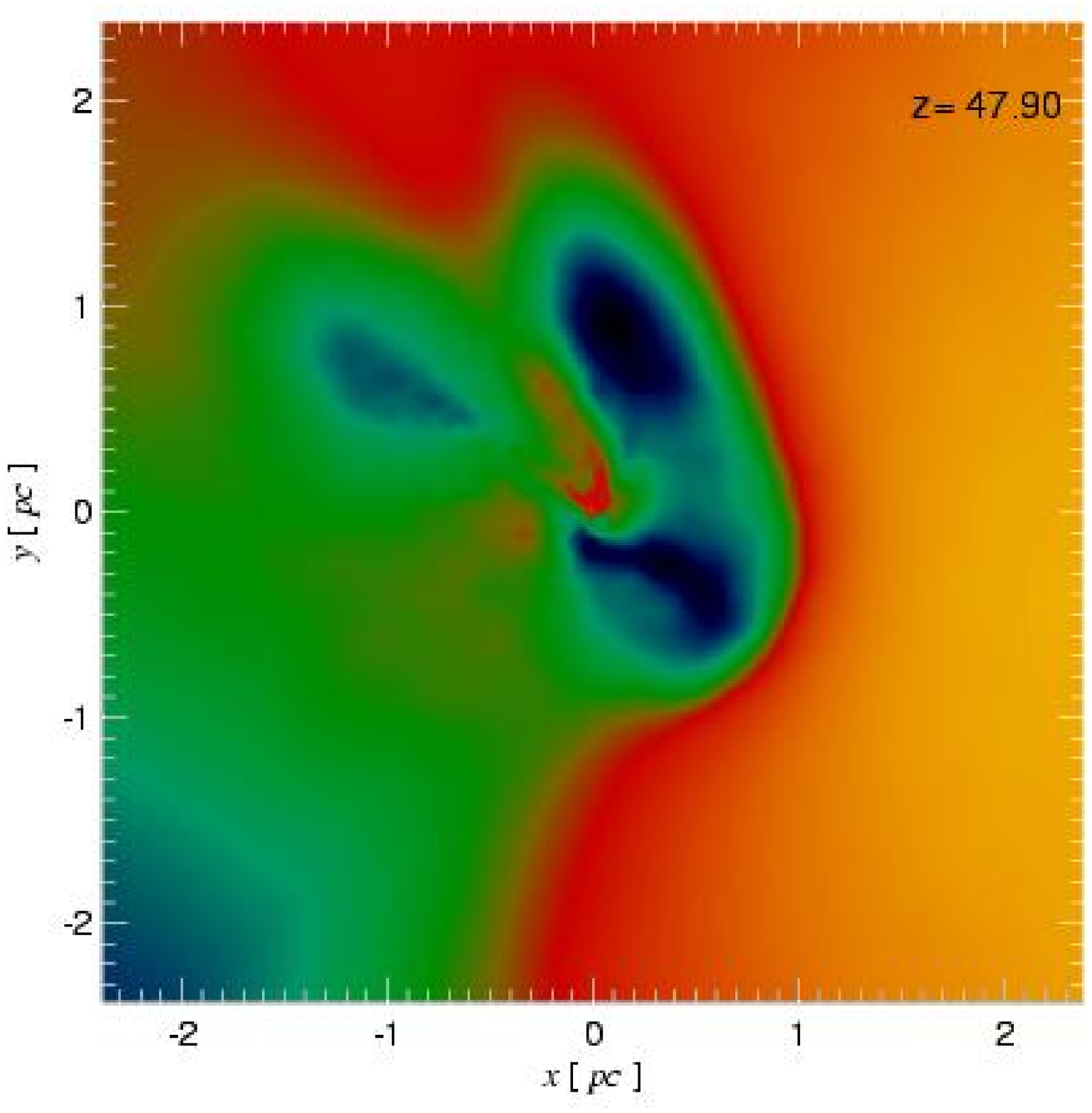}}
\hspace{13cm}\resizebox{5cm}{!}{\includegraphics{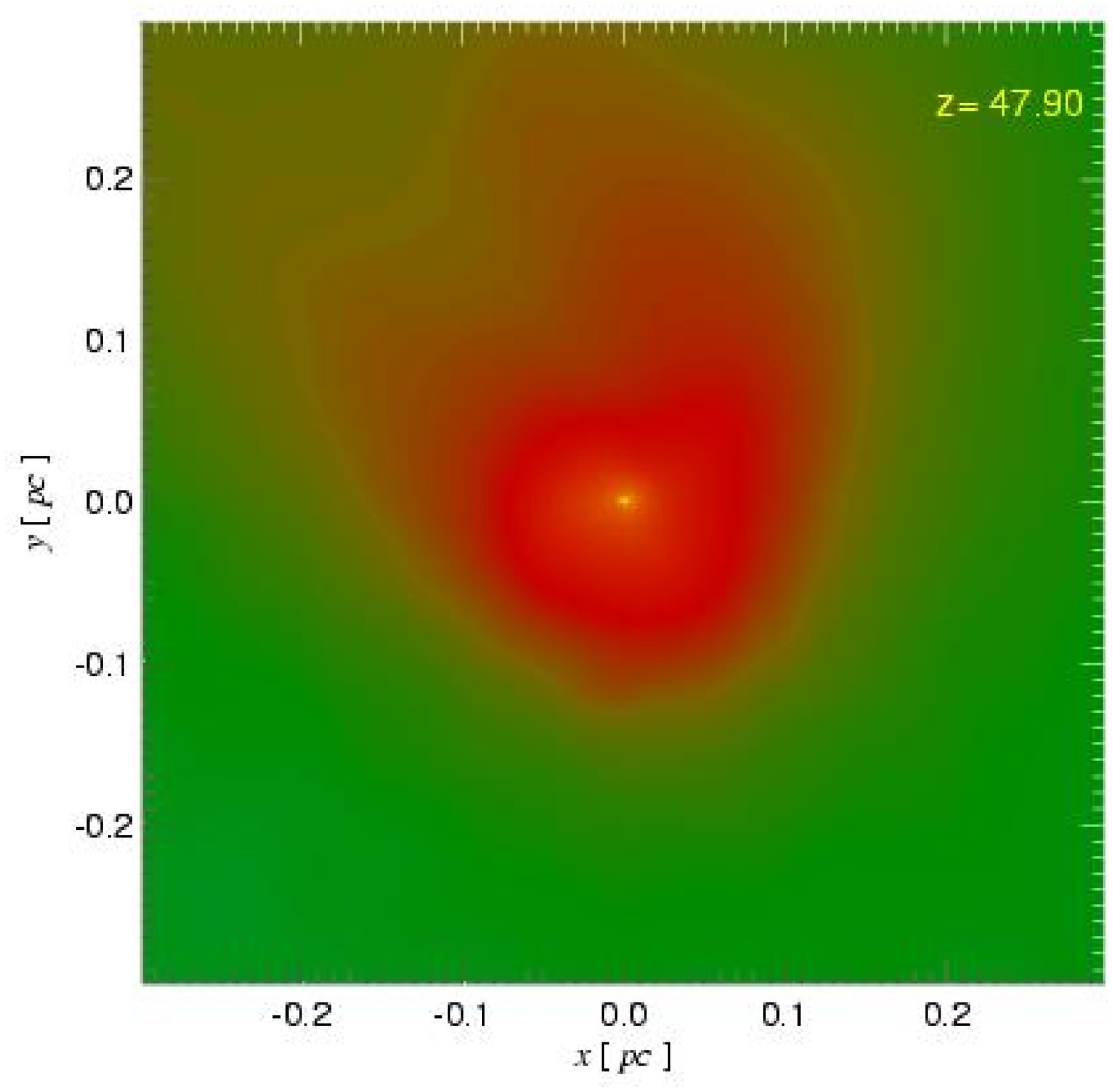}} %
\resizebox{5cm}{!}{\includegraphics{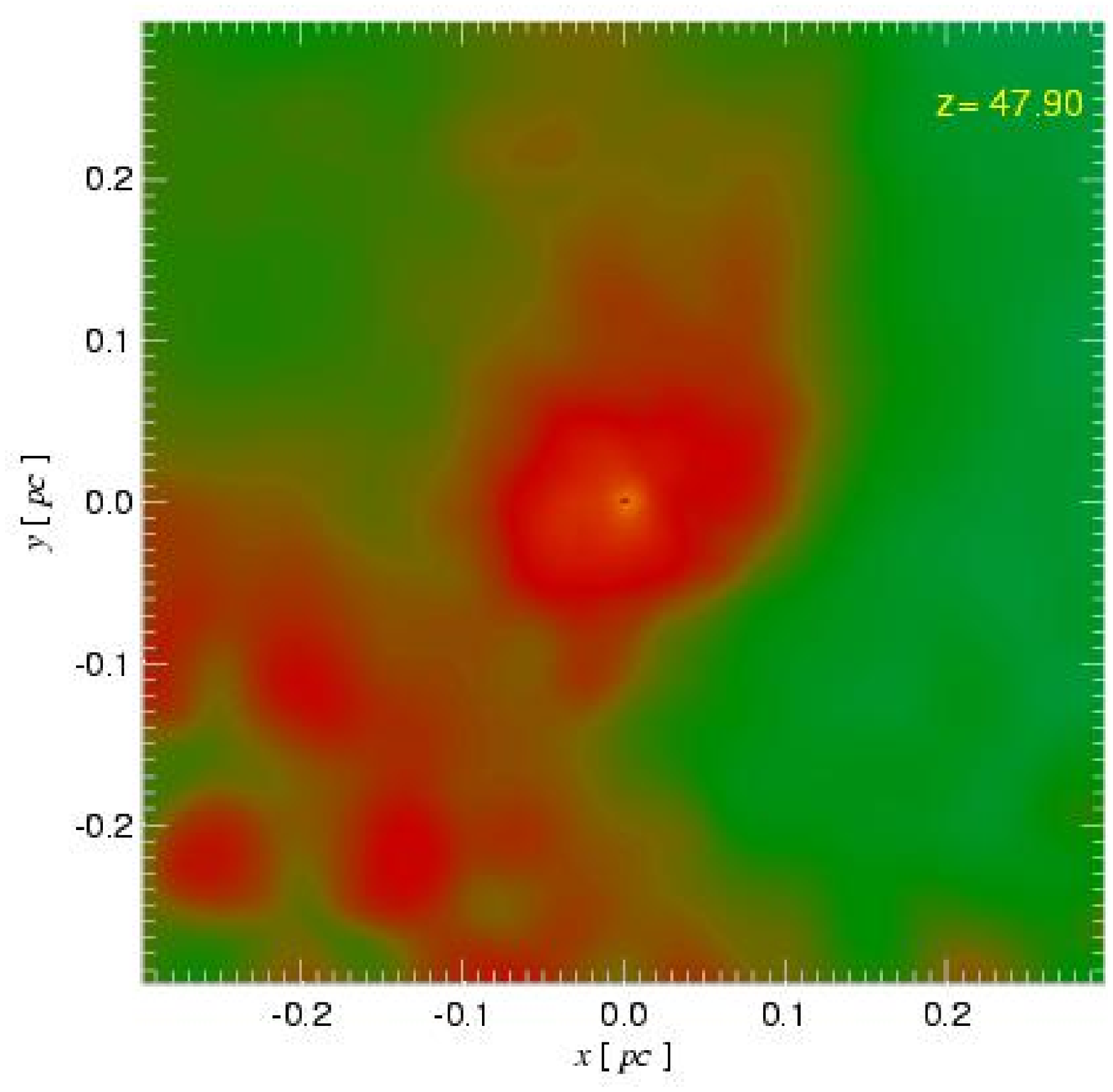}} \resizebox{5cm}{!}{%
\includegraphics{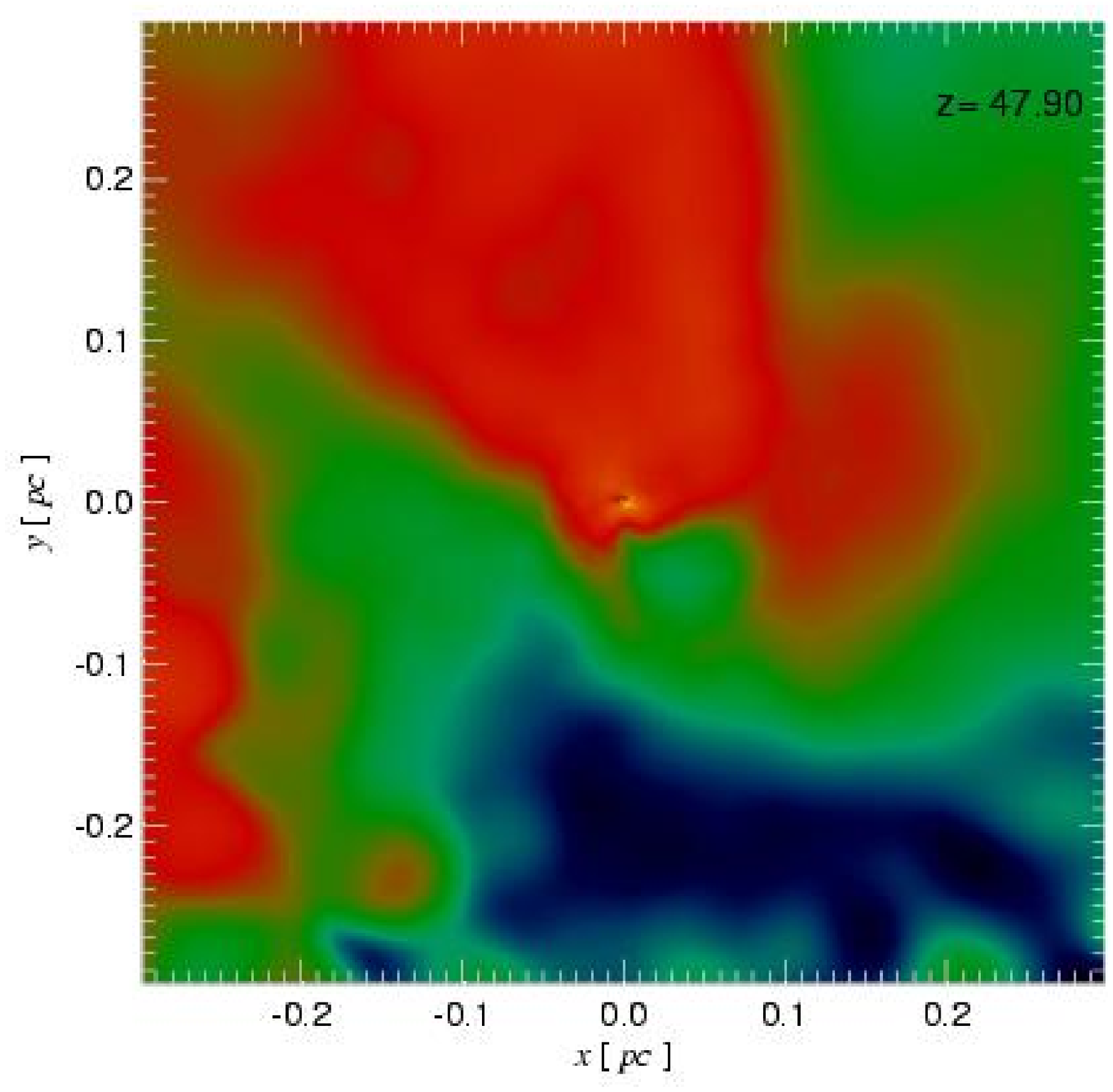}}
\hspace{13cm}\resizebox{5cm}{!}{\includegraphics{rho_legnd.eps}}
\resizebox{5cm}{!}{\includegraphics{temp_legnd.eps}}
\resizebox{5cm}{!}{\includegraphics{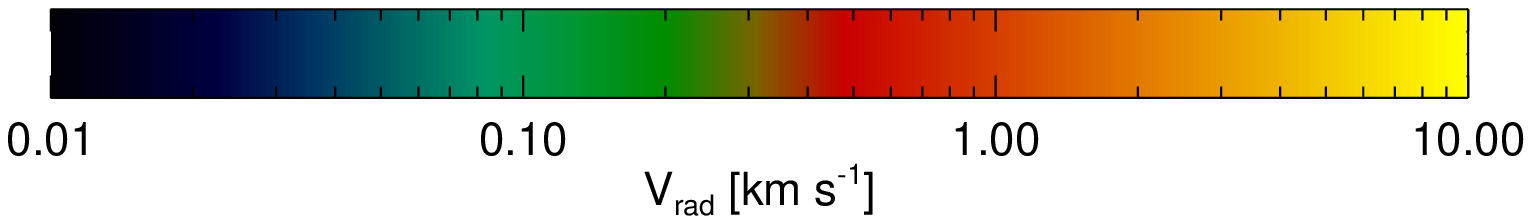}}
\caption{Projected gas density (left), mass-weighted temperarture (middle) and
radial velocity with respect to the star-forming region, at the final
time in the R5 simulation. The sequence of images in each column, from
top to bottom, shows zooms by successive factors of 8 in linear scale
onto the centre of the object.}
\label{fig:zoomr5}
\end{figure*}

The distributions of dark matter density, gas density and
mass-weighted temperature for the R5 object are shown, at three
different epochs, in Figure~\ref{fig:picevor5}. The images are of
length $4\,r_{200}$ on a side and are projected over a depth of
$r_{200}$ centred on the potential minimum. Here $r_{200}$ is defined
as the virial radius of the dominant object at the final time. The
scales on the axes give the length in physical units and a consistent
colour table has been used within each column. The three epochs shown
are representative of the system at the time when its mass is below
the Jeans mass, close to the Jeans mass, and at the end, when runaway
cooling is occurring at the centre. From left to right, the columns
give the dark matter surface density in physical units, the gas
density in physical units, and the mass-weighted gas temperature,
respectively.

At redshift $z=60$, the dark matter structure is very filamentary, as
is typical in CDM models (e.g. Abel, Bryan \& Norman 2000; Yoshida et
al. 2003; Gao et al. 2005). However, the gas is noticeably less clumpy than the
dark matter. Nonetheless, the gas does get compressed due to the
significant concentration of the dark matter.  As a result, the entire
region has a temperature of $\sim 300\,{\rm K}$, several times the
temperature of $\sim 80\,{\rm K}$ expected for the cosmic mean density
at that epoch.

At redshift $z=52$, the dark matter structures are already much more
pronounced than at $z=60$.  Following the rapid growth of the dark
matter structures, the gas begins to settle into the dark matter
halos; two dense gas clumps are clearly seen in the density map. The
increase in gas density results in substantial heating of the gas due
to compression. This can be seen in the large region around the
filaments where the temperature is $\sim 600\,{\rm K}$. In the central
region of the main dark matter halo, two small brighter regions are
visible with temperature of $\sim 1000\,{\rm K}$.

The bottom panels show the system at the final time, when star
formation is beginning in the innermost regions.  Note that a second
concentration is also evident at a distance of about 40 proper parsecs
in projection. In the bottom-right panel, two shocks develop along the
filaments. On a scale of $\sim 10\,{\rm pc}$, the shock around the
dominant object is even more pronounced. Note that the shock occurs in
the very inner regions rather than at the viral radius, $r_{\rm
vir}=35\,{\rm pc}$.  Just inside the shock front, a molecular hyrdogen
rich and rapidly cooling region is found.

A more detailed set of images of the R5 object at the final time is
displayed in Figure~\ref{fig:zoomr5}. From left to right, the columns
show surface density, mass-weighted temperature, and the radial
velocity with respect to the star-forming core.  From top to bottom,
the images show successive zooms by factors of 8 in linear scale onto
the centre of the gravitational potential at the final time. The
radial velocity is defined as the velocity relative to the average
velocity of the innermost $100$ {\small SPH} particles. On small
scales, the gas cloud reveals a complex structure, especially in the
radial velocity map. The gas accretion is clearly asymmetrical. The
infall velocity is much larger for the gas on the right hand side,
where the secondary gas concentration is closing in towards the
dominant object, accompanied by a one-sided accretion shock at a scale
of $\sim 1\,{\rm pc}$. Clearly, describing these processes as
spherically symmetric gas accretion is unrealistic.

\subsubsection{Time evolution of radial profiles around the dominant object}

\begin{figure*}
\vspace{-1cm} \resizebox{15cm}{!}{\includegraphics{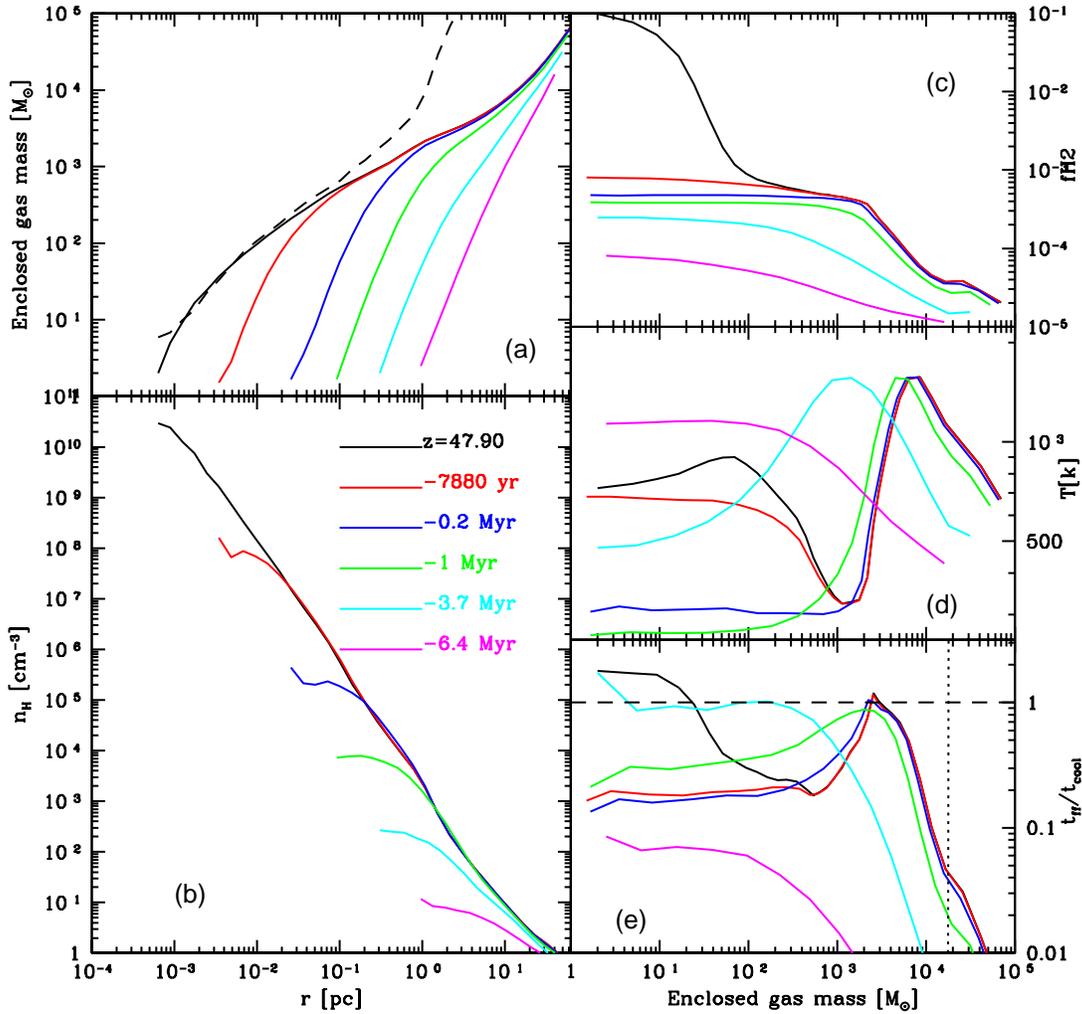}}
\caption{Radially averaged profiles of various physical quantities at 
different epochs in the R5 simulation. The colour lines refer to the
epochs indicated in panel (b), where the time is expressed as the
number of years before the final time, and the corresponding redshifts
from top to bottom are: $47.90$ (the final time), $47.91$, $z=48.02$,
$z=48.50$, $z=50.29$, and $z=52.24$. Panel~(a): enclosed gas mass as a
function of radius. Panel (b): number density profile of hydrogen
nuclei. The dashed lines are the local Bonnor-Ebert mass, which is
similar to the Jeans mass but assumes an isothermal rather than an
adiabatic distribution for the gas.  Panel (c): molecular hydrogen
fraction as a function of the enclosed gas mass. Panel (d): gas
temperature. Panel (e): the ratio of free-fall time to the cooling
timescale. The vertical doted lines in all panels indicate the virial
radius of the halo at the final time.}
\label{fig:denr5}
\end{figure*}

The time evolution of the gas in the R5 object is described in
Figure~\ref{fig:denr5} which shows the mass-weighted radial profiles
of various physical quantities of interest. We define the cooling
timescale as $t_{\rm cool} = -{\rm d}t/{\rm d}\ln T$, where $T$ is the
gas temperature, and $t$ is time. The free-fall time is defined as
$t_{\rm ff} = 2\pi\sqrt{3/32G\rho}$, where $\rho$ is the density.  We
show $6$ epochs corresponding to $z=52.24$, 50.29, 48.50, 48.02 and
47.91; the final epoch corresponds to $z=47.90$.

We start $6.4$ million years before the final time, at $z=52.24$, when
the R5 halo has only just reached the Jeans mass.  Gravity is then
strong enough to cause gas to settle into the potential well of the
halo. By then, the central part of the gas distribution is heated to
more than $1000\,{\rm K}$. However, the cooling time is still much
longer than the free-fall time because the gas density and molecular
hydrogen fraction are still relatively low. At this stage, the gas
contracts slowly.  About $2.7$ million years later, at $z=50.29$, the
central gas density has increased by a factor of $10$, and the
molecular fraction has also increased by a factor of $10$. This
results in a much shorter cooling time for the central $2000\, {\rm
M_\odot}$ of gas, which is now comparable to the free-fall time. At
this point, the gas cloud begins to collapse rapidly.

Another $2.7$ million years later, at $z=48.50$, the central $1000\,
{\rm M_\odot}$ of gas approach the critical density, $n_{\rm H}
\sim 10^{4}{\rm cm^{-3}}$, which marks the transition of ${\rm
H_2}$-cooling from NLTE rotational level populations to an LTE
state. Above this density, the cooling rate per molecule of hydrogen
becomes independent of density. The central gas temperature falls to
close to $200\,{\rm K}$, which is roughly the minimum temperature that
can be reached by molecular hydrogen cooling.  By now the production
of molecular hydrogen has entered a regime where the abundance scales
logarithmically with time due to the continuous depletion of electrons
through recombination (Tegmark et al. 1997).

Finally, a further 1 million years later, at $z=48.02$, the central
gas density approaches the critical density required for three-body
reactions, resulting in rapid $H_2$ formation and an increase of the
gas temperature due to heating from both compression and the release
of molecular binding energy. After $10^5$ years, there is a
substantial increase in the ${\rm H_2}$ fraction in the central few
hundred solar masses of gas with density exceeding $10^8{\rm
cm^{-3}}$. As a consequence, the cooling time of the central $20\,
{\rm M_\odot}$ of gas becomes shorter than the free-fall time, and the
mass eventually exceeds the locally estimated Bonnor-Elbert mass
(Elbert 1955; Bonnor 1956):
\begin{equation}
M_{\rm BE}\sim 20{\rm M_\odot}T^{3/2}n^{-1/2}\mu^{-2}\gamma^2. \nonumber
\end{equation}
Runaway collapse is triggered at this time. Here $n$ is the number
density of hydrogen nuclei, $\mu$ is the molecular weight and $\gamma$
is the adiabatic index.  Note that the Bonnor-Elbert mass is similar
to the Jeans mass but it assumes an isothermal rather than an uniform
gas distribution.

\section{Object-to-object comparisons: differencens and similarities}
\begin{figure*}
\hspace{0.13cm}\resizebox{8cm}{!}{\includegraphics{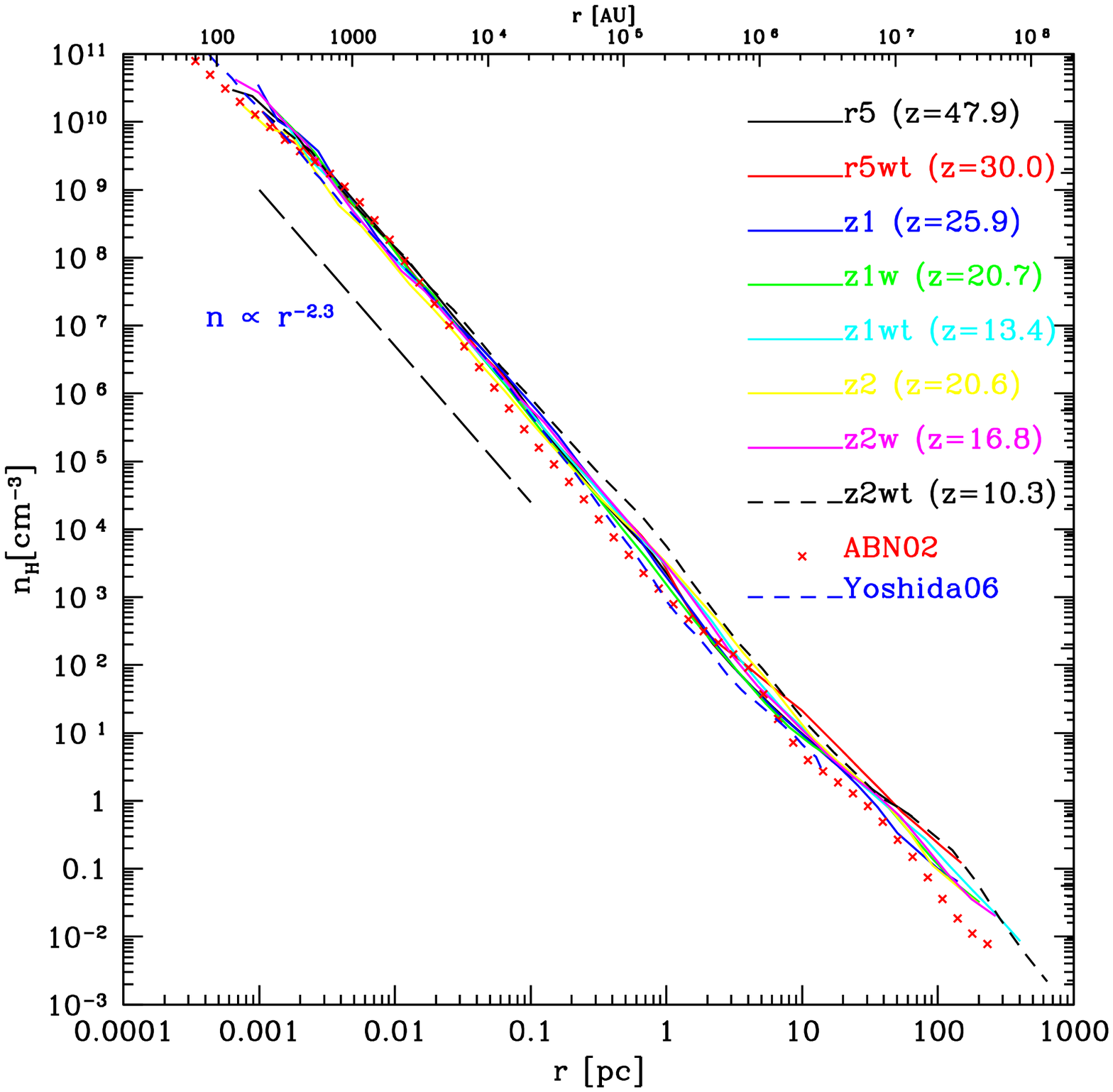}} \hspace{
0.13cm}\resizebox{8cm}{!}{\includegraphics{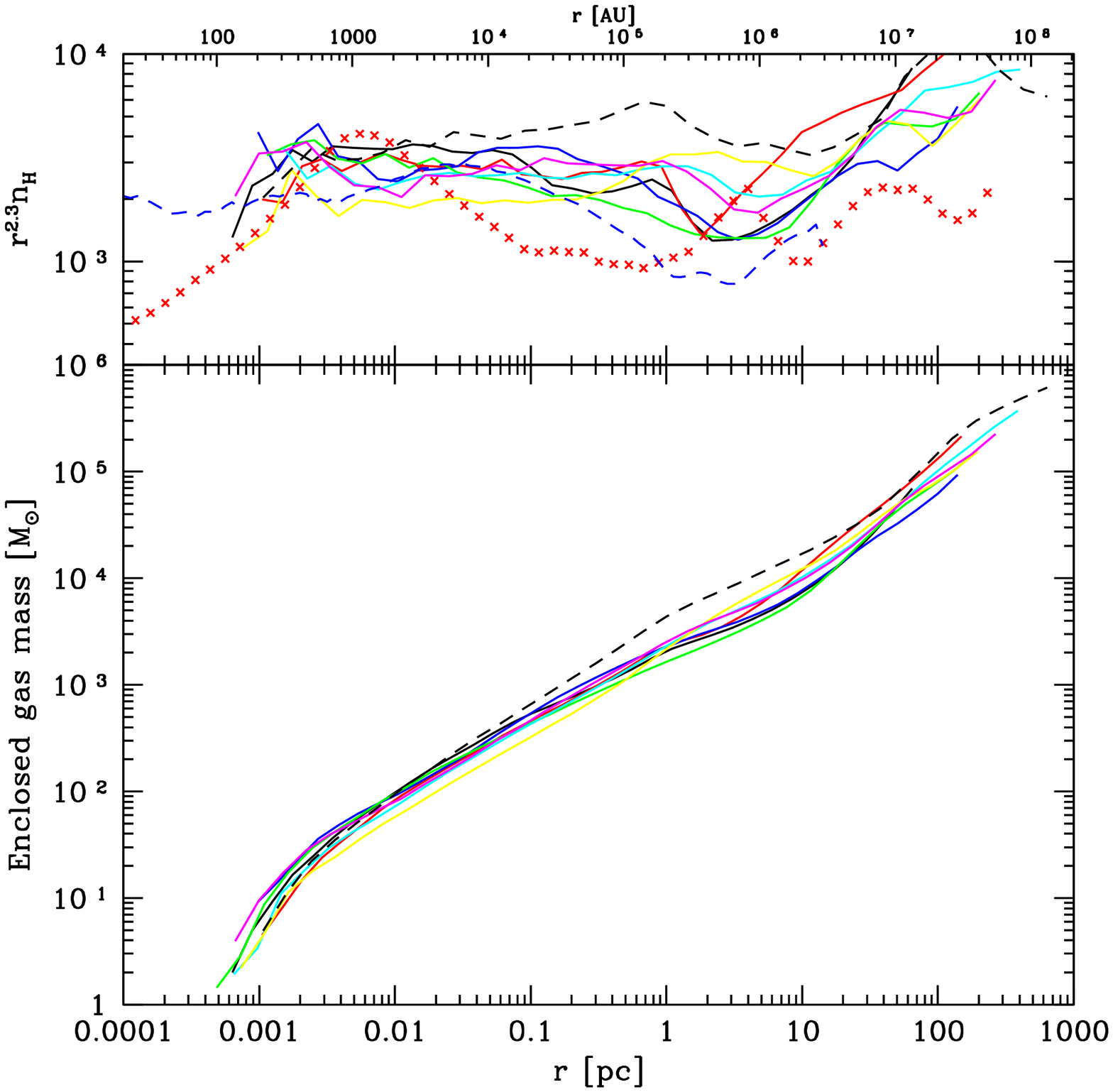}} \hspace{0.13cm}
\resizebox{8cm}{!}{\includegraphics{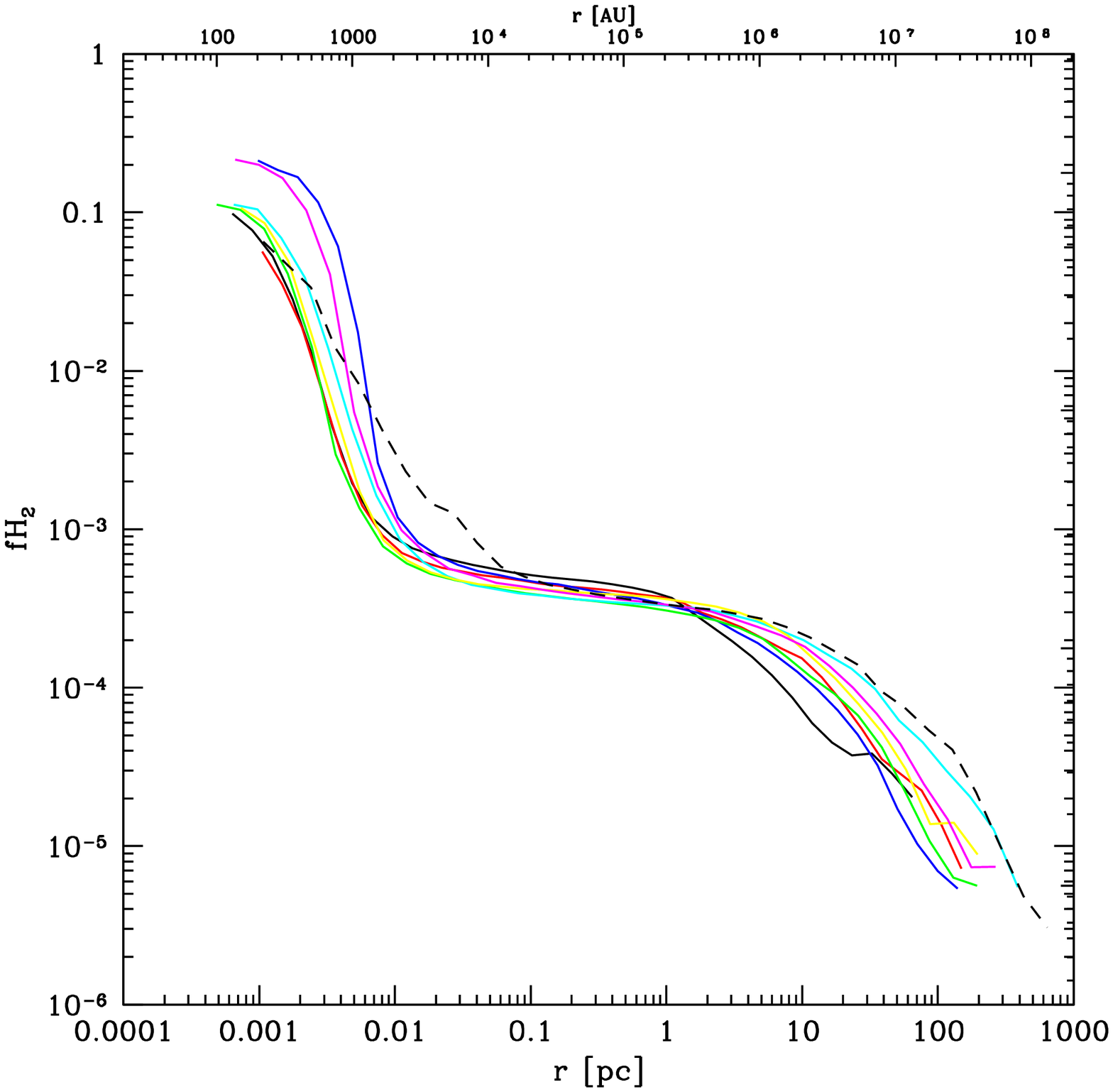}} \hspace{0.13cm}
\resizebox{8cm}{!}{\includegraphics{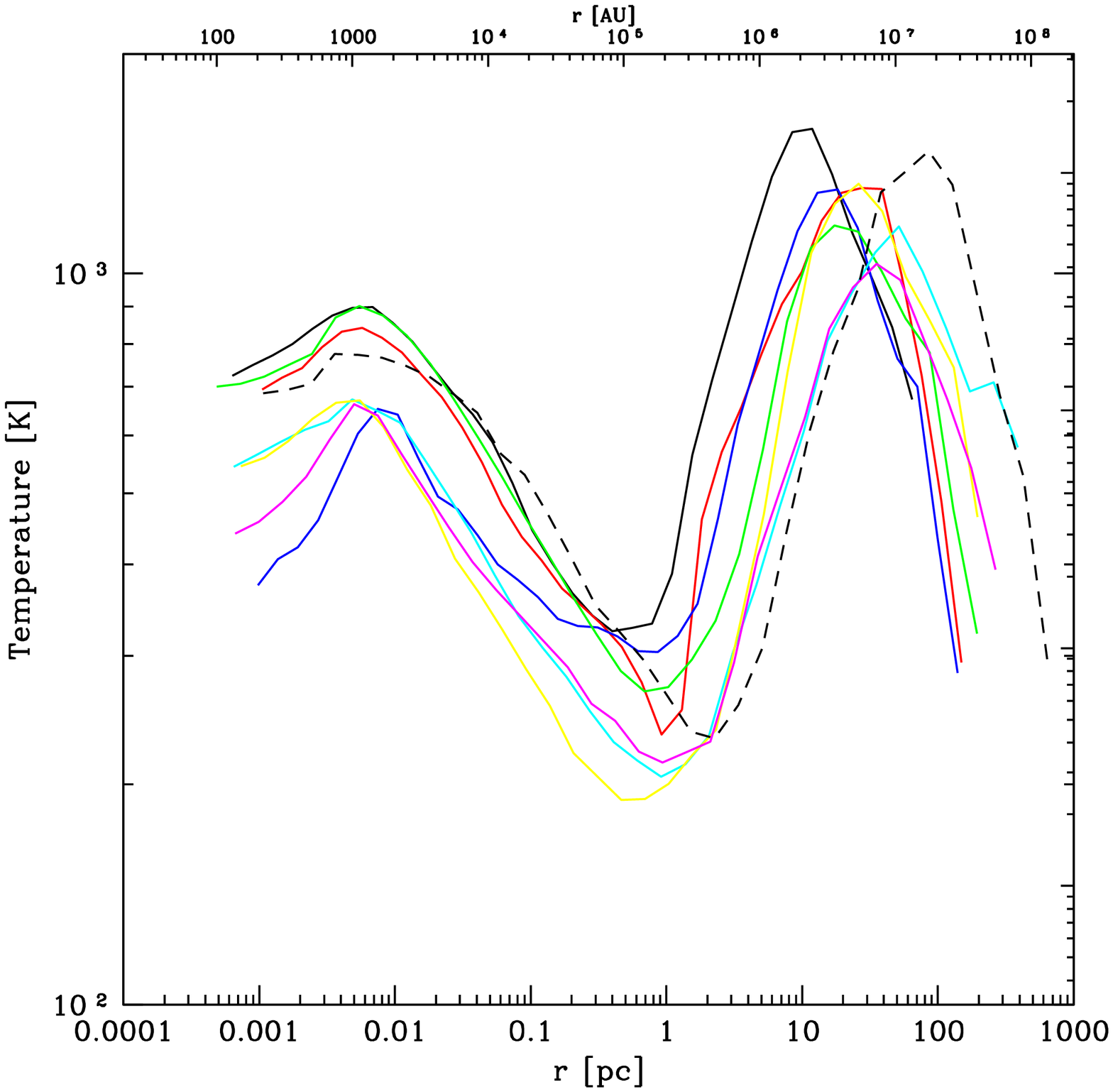}}
\caption{Radially averaged profiles of various physical quantities as
a function of radius in the $8$ simulations. The different lines refer
to the different simulations as indicated by the labels in the top
left-hand panel. Top left: number density of hydrogen nuclei. Top
right: enclosed gas mass. Bottom left: molecular hydrogen
fraction. Bottom right: gas temperature.}
\label{fig:allden}
\end{figure*}

\begin{figure*}
\resizebox{16cm}{!}{\includegraphics{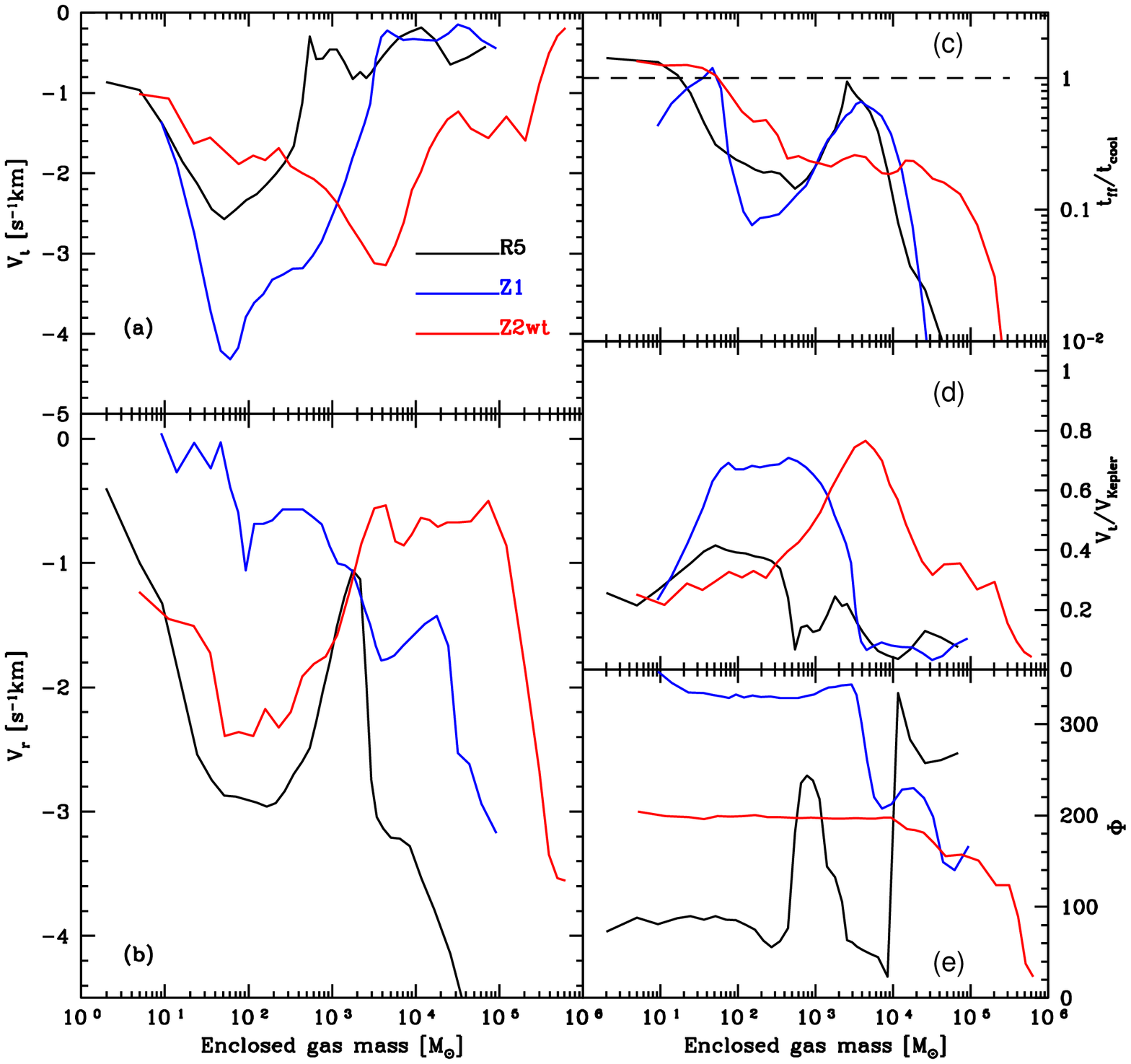}} \caption{Radial
profiles of various physical quantities as a function of the
enclosed gas mass at the final time in $3$ representative
simulations: R5, Z1 and Z2wt.  Panel (a): profiles of negative
tangential velocity. Panel (b): radial velocity profiles. Panel (c):
the ratio of the free-fall time to the cooling time. Panel (d): the 
ratio of the tangential velocity to the Keplerian velocity
$v_{Kepler}=(GM/r)^{1/2}$.  Panel (e): the orientation of the angular
momentum. Note that only the angle $\phi$ is shown.}
\label{fig:allvel}
\end{figure*}

As we have seen in the previous section, the formation epoch of the
first stars can span a broad redshift range, and there are systematic
differences in the formation process of protostellar clouds that
depend on redshift. An additional source of variation comes from the
differences in the assembly histories of these objects at a given
redshift (e.g. Lacey \& Cole 1993). It is therefore important to
analyse a number of realisations in order to assess the expected
scatter in the properties of star-forming clouds. We now address this
issue by jointly considering all of our simulations.

\subsection{Radial profiles}

Figure~\ref{fig:allden} shows the radial profiles of the density,
enclosed gas mass, molecular hydrogen fraction and temperature in all
our simulations.  The comparison is made at the final time when
the central gas has roughly reached the limiting density, $n_{\rm H}
\sim 10^{10}{\rm cm^{-3}}$, in all the simulations. At this time,
runaway collapse has set in and the minimum simulation timestep
becomes extremely small.  The top left panel shows the number density
profile of hydrogen nuclei. Each object is represented by a different
colour line as indicated in the top-left panel. A striking feature in
this plot is that the gas density profiles of the objects in all our
simulations coincide with one another on scales less than $\sim
10\,{\rm pc}$, in spite of their very different collapse redshifts
and environments.

The scatter is remarkably small, only a factor of $\sim 2$ about the
mean, which is smaller than in the simulations of O'Shea \&Norman
(2006b). For comparison, the red crosses and the blue dashed line show
the fiducial density profiles of ABN02 and Yoshida et al. (2006),
respectively. Our simulations agree well with ABN02 on scales smaller
than $0.02\,{\rm pc}$, but not very well at larger radii. As we will
show in Section~6, the different baryon fractions assumed in the two
studies largely account for this discrepancy. Yoshida et al. (2006),
on the other hand, adopt the same cosmology as us. Our results agree
well within 0.2 pc, but not as well on larger scales. This suggests
that the actual variation in density profile from object to object may
be bigger than indicated by the limited samples we have simulated
here. The top right panel shows the enclosed gas mass as a function of
radius, which also confirms the rather small variation in mass profile
from object to object in our simulations.

The molecular hydrogen fractions in all our simulations also agree
well with each other at radii less than $10\,{\rm pc}$. The
temperature structures are similar as well. The latter can be roughly
divided into $4$ physically distinct zones. Zone $1$ is the outermost
region and corresponds to cosmic infall, where the gas is continually
heating up, culminating at the accretion shock front where the maximum
temperature occurs. In the region immediately behind the shock front,
which makes up zone 2, the gas temperature decreases inwards up to a
radius, $\sim 1\,{\rm pc}$, where the temperature reaches a minimum of
$\sim 200\,{\rm K}$, below which the molecular hydrogen cooling rate
drops exponentially.  Within this radius, in zone $3$, there is a
second infall region in which the temperature rises back up again and
approaches $1000\,{\rm K}$ at a radius of $\sim 0.01\,{\rm pc}$. Here,
the gas density reaches a value of $n_{\rm H} \sim 10^8{\rm cm^{-3}}$.
At this characteristic density, three-body reactions which rapidly
form molecular hydrogen start to operate and the enhanced cooling
associated with the rising molecular hydrogen fraction causes the
temperature to decrease towards smaller radii temporarily. The
temperature profiles are all qualitatively similar but they can be
grouped into two classes: R5, R5wt, Z1w and Z2wt in one, and Z1, Z1wt,
Z2, Z2w in the other. As we will see in the next subsection, the
star-forming clouds in the second group have a disklike structure. As
a result, the collapse is slower and the competition between
compressional heating and molecular hydrogen cooling results in a
lower inner temperature.

Figure~\ref{fig:allvel} compares the radial profiles of various
dynamical quantities. Here, for clarity, we show results only for 3
typical simulations, R5, Z1 and Z2wt, which we analysed in
Section~3. A complete version of the figure including all the other
simulations may be found in the Appendix. Figure~\ref{fig:allvel}
reveals that despite the similarities in the gas properties seen in
Figure~\ref{fig:allden}, the internal gas motions differ dramatically
from object to object. They are only similar in the cosmic infall
region which terminates at the first accretion shock where the radial
velocity (defined to be negative for infalling gas) reaches a local
minimum and the temperature a local maximum. The mass inside this
radius is about $10^4\,{\rm M_\odot}$ for the R5 and Z1 objects, and
$10^5\,{\rm M_\odot}$ for the Z2wt object.  For the R5 object, the gas
infall velocity is relatively constant but declines rapidly at a
radius corresponding to a mass scale of $10^3\,{\rm M_\odot}$.  Just
inside this scale, the gas infall speeds up again.  For the Z1 object,
the infall speed increases just behind the accretion shock and reaches
a local maximum at a mass scale of $3
\times 10^3\, {\rm M_\odot}$, where the ratio of the free-fall time to
the cooling time is maximal (panel c). Exactly at the same scale, the
tangential velocity starts to become larger than the infall
velocity. For the Z2wt object, the radial velocity stays constant at
large radii. At a mass scale of about $5 \times 10^{3}\, {\rm
M_\odot}$, the radial infall velocity increases until a mass scale of
a few tens of solar masses. Note that a jump in the ratio of the
free-fall time to the cooling time often marks a transition in the
velocity and orientation of the angular momentum.

Panel (d) gives the ratio of the actual tangential velocity of the gas
to the velocity required for Keplerian motion, defined as
$|\vec{V_t}|/V_{{\rm Keplerian}}$. The curves for the three
simulations differ substantially, implying that the degree of
rotational support is rather different.  Finally, panel (e) shows the
azimuthal angle between the z-axis in the simulation and the angular
momentum vector of the gas interior to a given radius, or equivalently
an enclosed mass.  It is clear that the different parts of the halos
often spin in different directions.  For the R5 object, the gas only
rotates in the same direction inside an enclosed mass of $300\, {\rm
M_\odot}$, while the same is true for the Z1 and Z2wt halos inside
$3000\, {\rm M_\odot}$ and $10000\, {\rm M_\odot}$, respectively.
Comparing with the other panels in the Figure, we see that a jump in
tangential velocity typically accompanies a jump in the orientation of
the angular momentum.

\subsection{Morphology of the star-forming clouds}

In spite of the similarities in radial profiles seen in the previous
subsection, the final star-forming clouds exhibit a variety of
morphologies. We define the edge of the star-forming cloud as the
largest radius inside which the orientation of the angular momentum of
the enclosed material (determined by the azimuthal angle introduced in
the previous subsection) does not change significantly. We then
calculate the principal axratios, $q=b/a$ and $s=c/a$ (taking
$a>b>c$), of the inertia tensor of all gas particles in the
star-forming cloud.  (These axial ratios may be viewed as the
equivalent axial ratios of a homogeneous ellipsoidal distribution with
the same inertia tensor as the gas distribution.) The axial ratios for
all our simulated objects are given in Table~\ref{Table:shape}. We
also list here the ratio of the tangential velocity to the Keplerian
velocity at the edge of the cloud. It is clear that the star-forming
clouds are rarely close to spherical. Only three objects in our
sample, R5, R5wt and Z1, have a relatively round shape, with a
tangential velocity that is only $\sim 30\%$ of that required for
Keplerian motion. These clouds are similar in shape to the objects
simulated by ABN02 and Yoshida et al. (2006). However, half of our
objects, Z1, Z1wt, Z2 and Z2w, have a disk-like structure of varying
size and thickness which is nearly rotational supported
($V_t/V_{Keplerian} >0.6$), while one, Z2Wt, has a bar-like structure.

\begin{figure*}
\hspace{0.13cm}\resizebox{8cm}{!}{\includegraphics{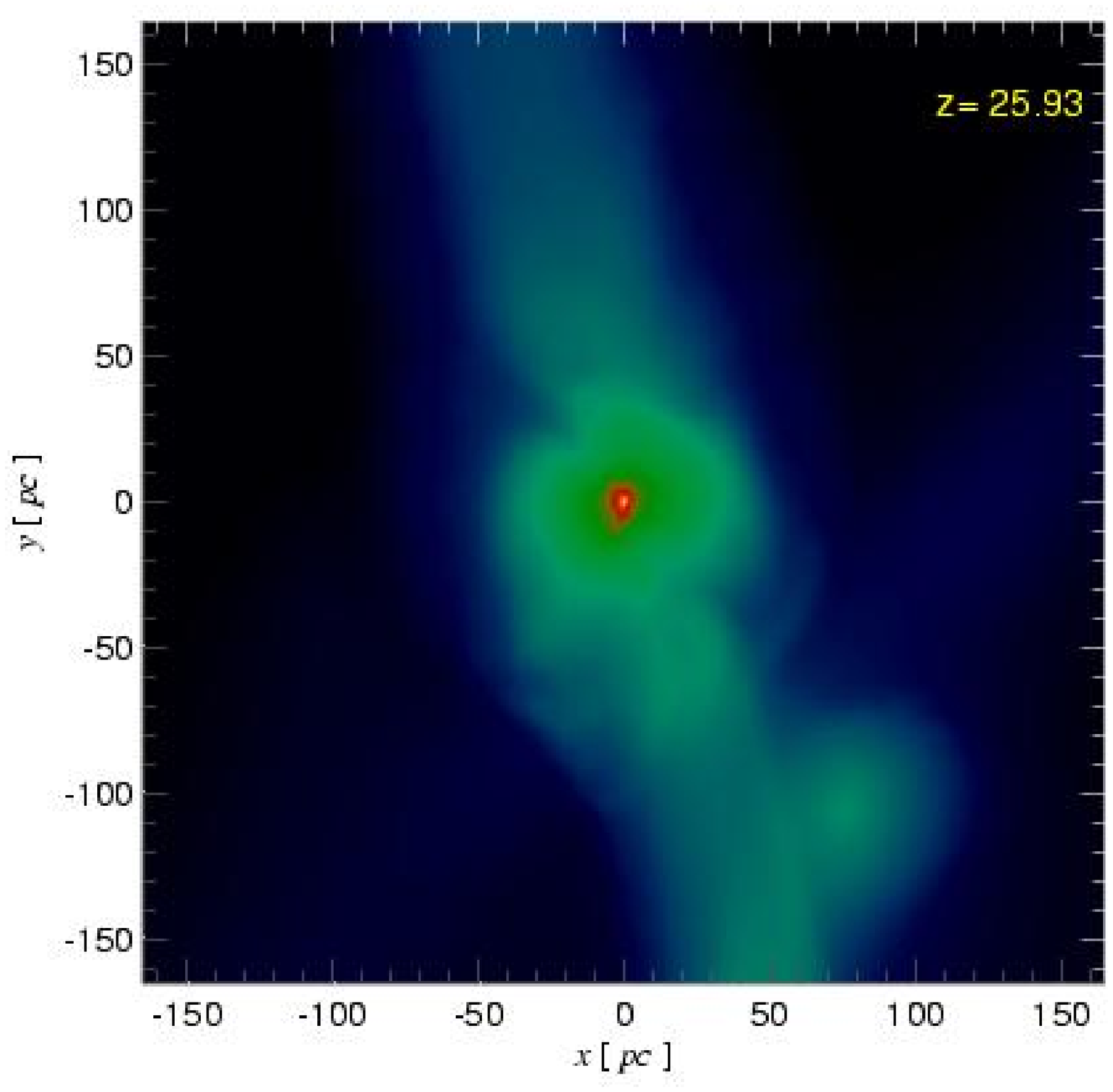}}
\hspace{0.13cm}\resizebox{8cm}{!}{\includegraphics{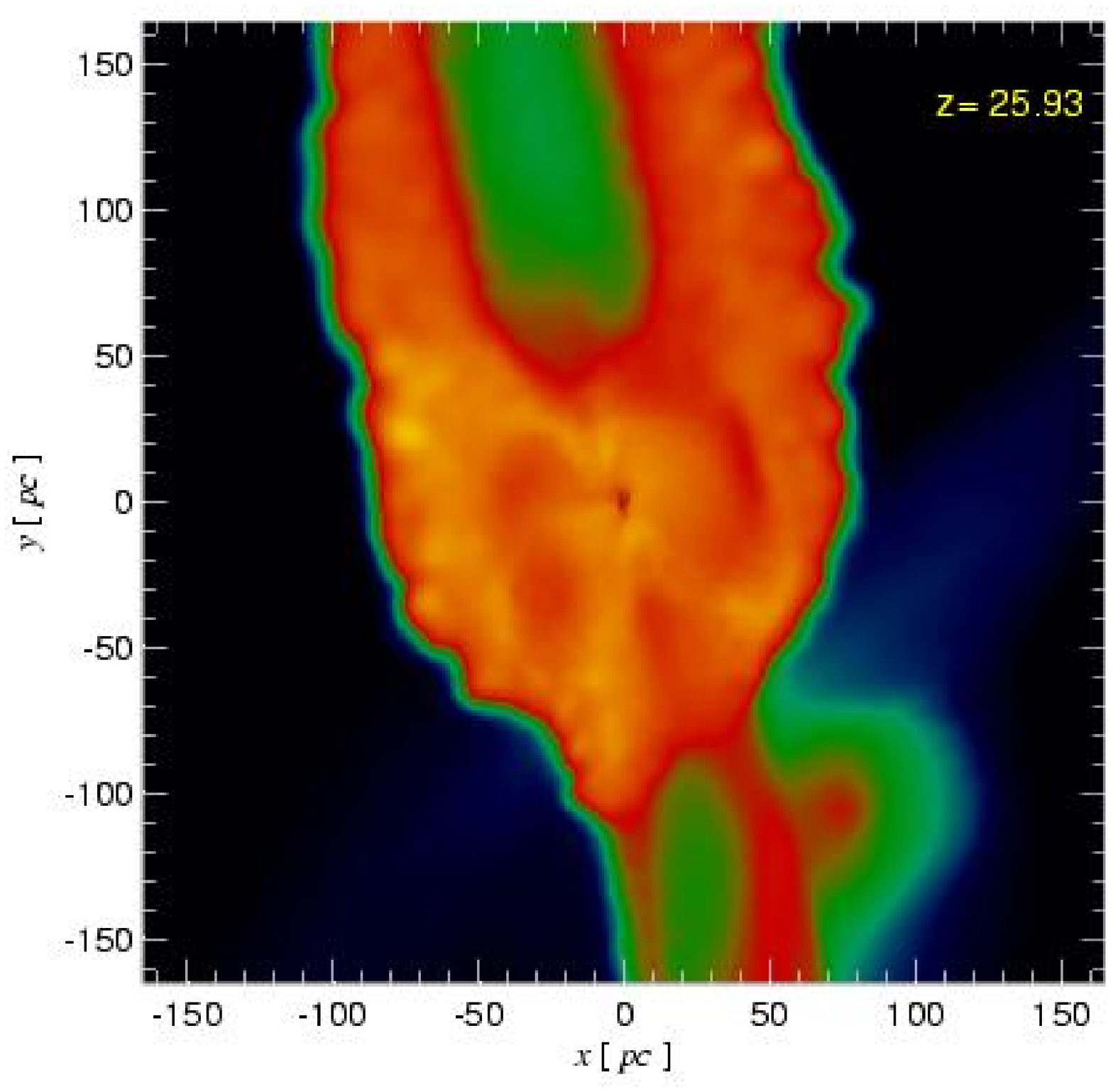}}
\hspace{0.13cm}
\resizebox{8cm}{!}{\includegraphics{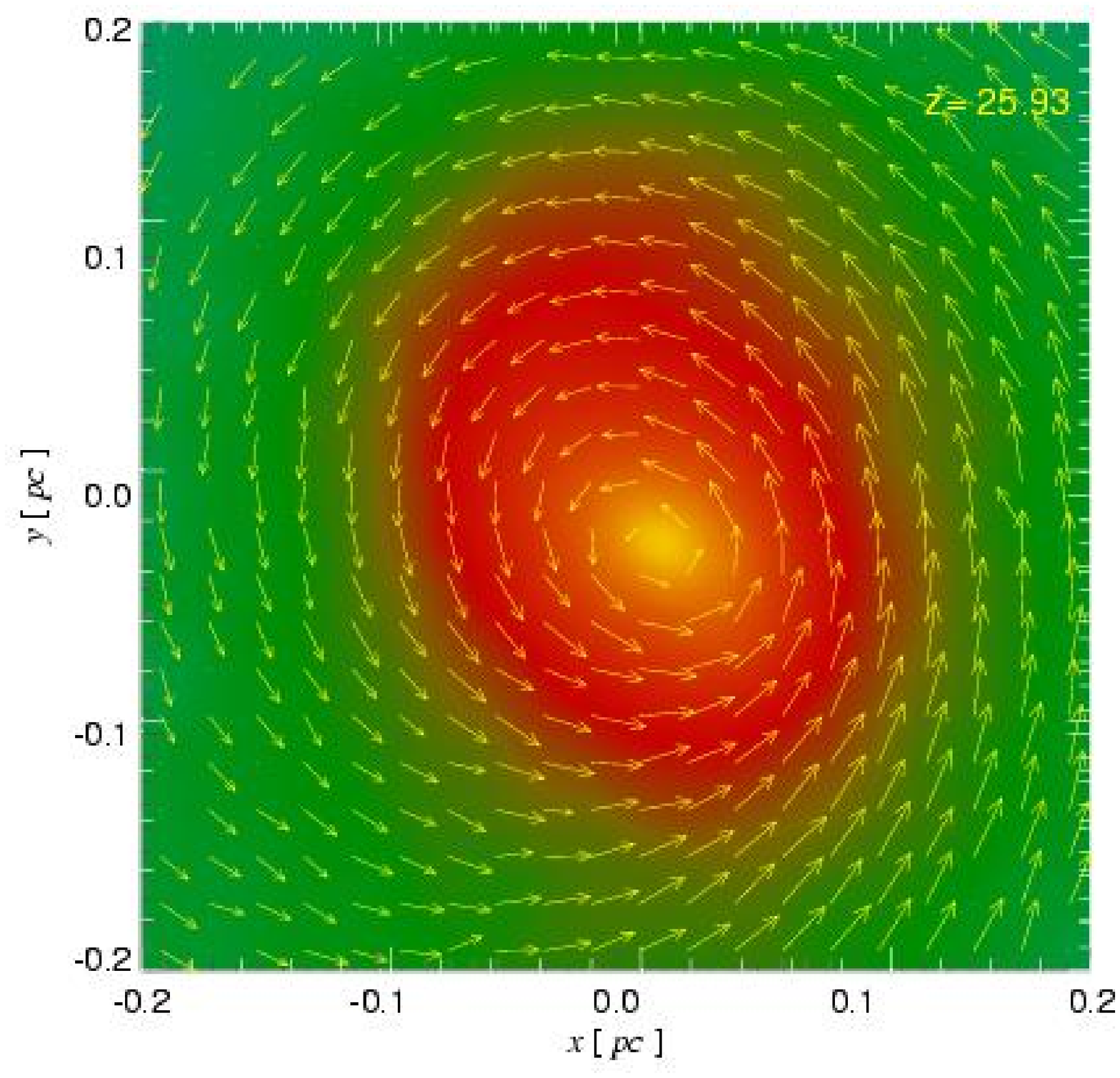}}
\hspace{0.13cm}
\resizebox{8cm}{!}{\includegraphics{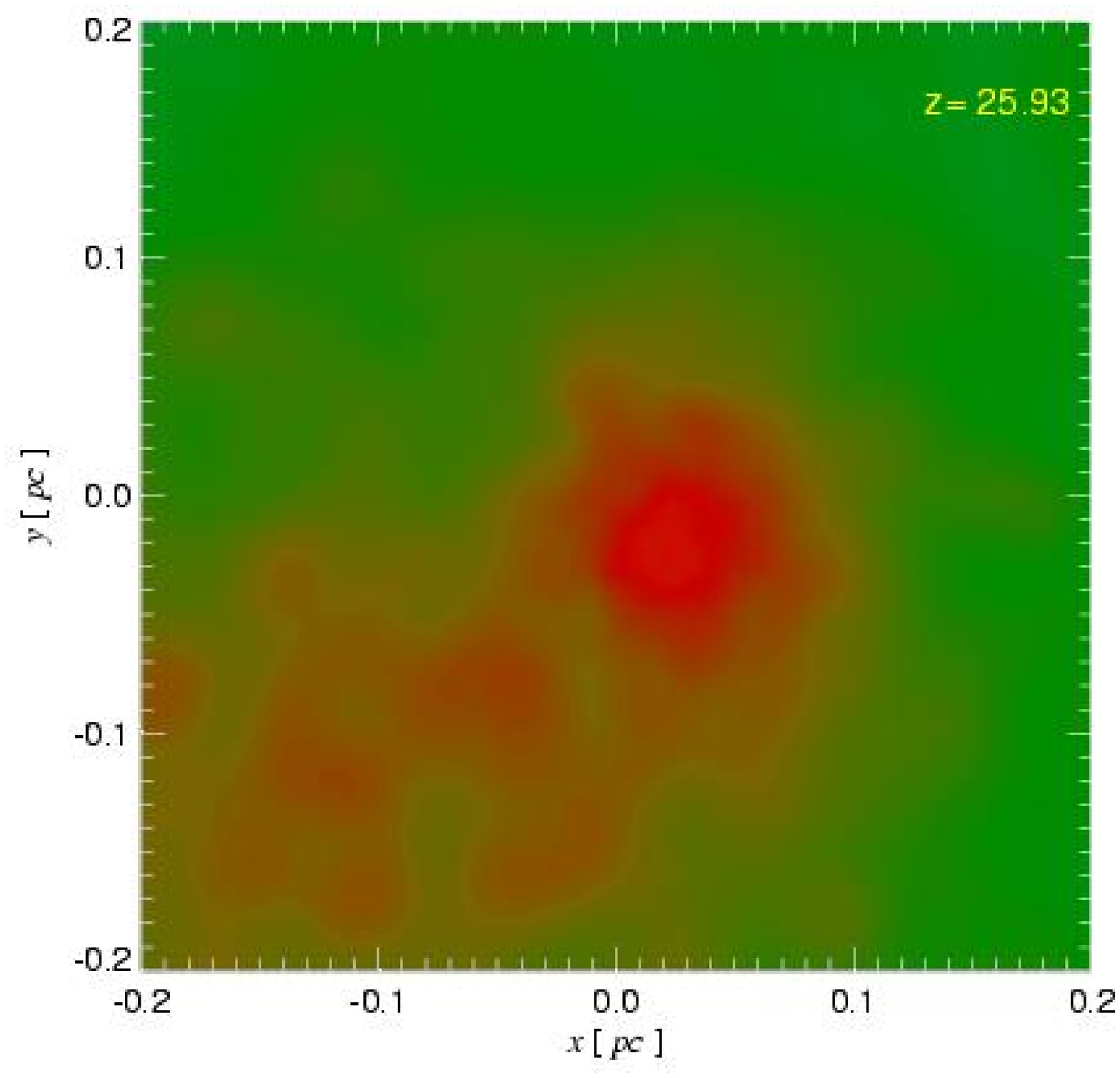}}
\caption{Density (left) and temperature (right) maps for the z1
simulation which has a disk-like structure at the final time. The
upper panel shows a slice of side $4r_{vir}$ and depth $r_{vir}$. The
bottom panel shows a cubic region of side $0.4\,{\rm pc}$ chosen to
include the disk-like structure which is shown face-on. The projected
velocity field is superposed onto the density map.}
\label{fig:slicez1}
\end{figure*}

\begin{figure*}
\hspace{0.13cm}\resizebox{8cm}{!}{\includegraphics{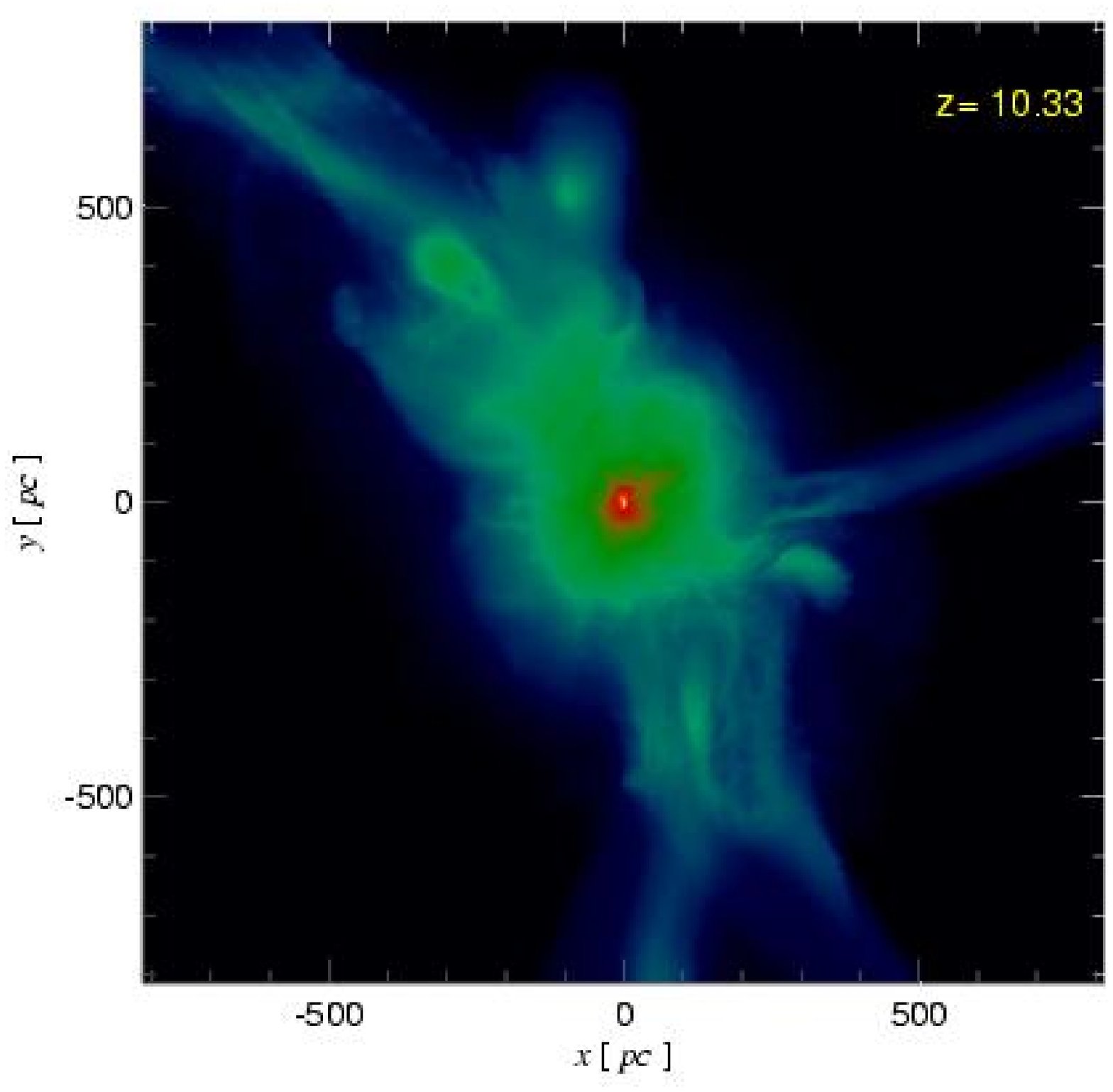}}
\hspace{0.13cm}\resizebox{8cm}{!}{\includegraphics{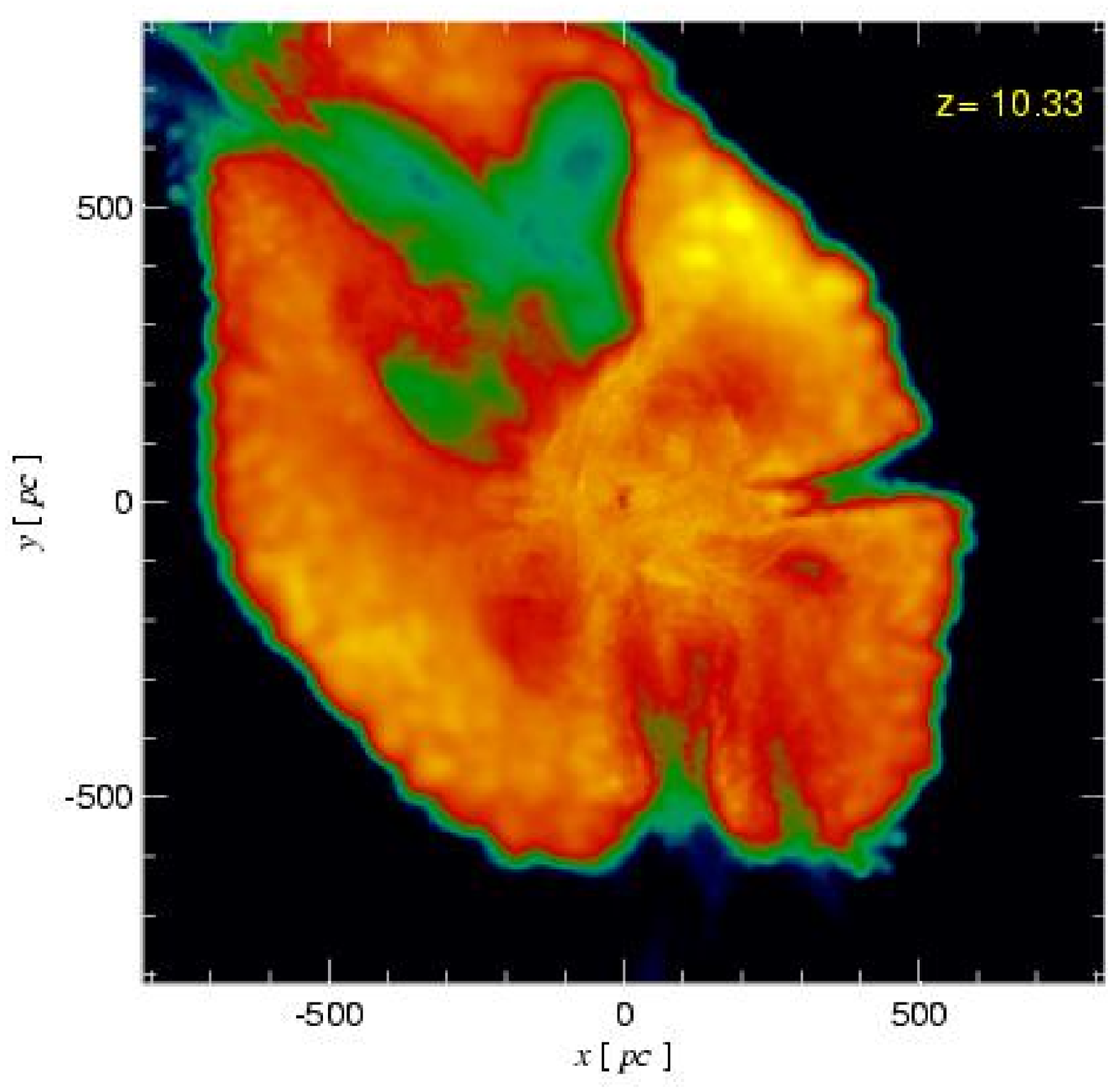}}
\hspace{0.13cm}\resizebox{8cm}{!}{\includegraphics{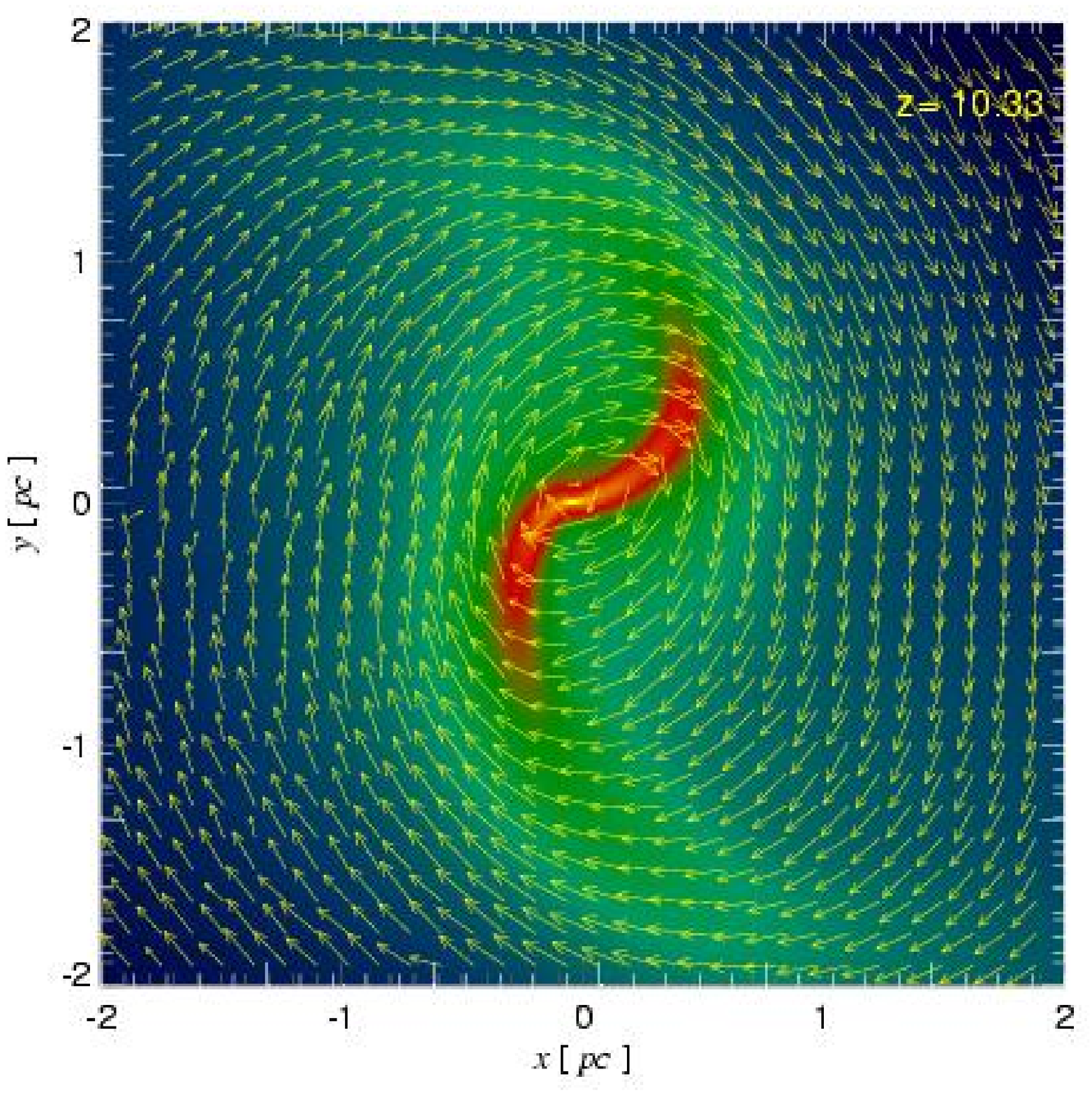}}
\hspace{0.13cm}\resizebox{8cm}{!}{\includegraphics{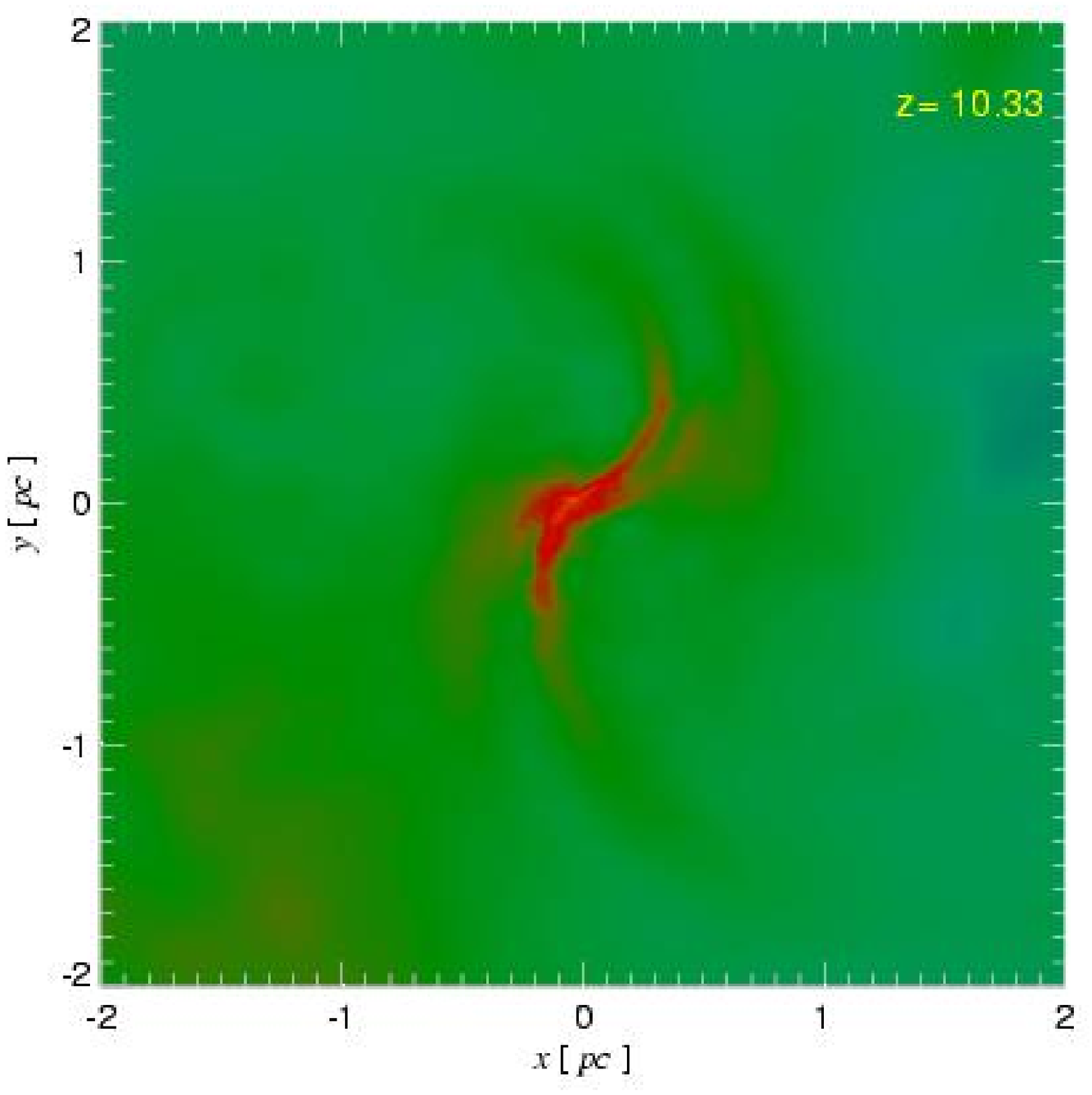}}
\caption{Density (left) and
temperature (right) maps for the Z2wt simulation which has a bar-like
structure at the final time. The upper panel shows a slice of side 
$4r_{vir}$ and depth $r_{vir}$. The bottom panel shows a
cubic region chosen to include the bar-like structure which is shown
face-on. The projected velocity field is superposed onto the density
map.}
\label{fig:slicez2wt}
\end{figure*}

Density and temperature maps at the final time in the Z1 and the Z2wt
objects are shown in Figures~\ref{fig:slicez1} and~\ref{fig:slicez2wt}
respectively. The left and right panels give the density and
temperature maps respectively. The top panels depict square regions of
length $4 \times r_{200}$ and depth $r_{200}$ in projection. In the
lower panels, the scale has been adjusted to illustrate the disk in Z1
and the bar in Z2wt, with the projected velocity field superposed onto
the density map. Both structures are shown face-on.

There are systematic differences between the internal structure of the
main object and its surroundings in R5, Z1 and Z2wt. For the R5
object, the surrounding dark matter distribution is quite
complex. There are $3$ prominent objects connected to each other by
filaments (see Fig~3) and the gas at the intersection of these
filaments is shock-heated by infall. For the Z1 and Z2wt objects, the
arrangement of the dark matter is simpler. These objects form at the
intersection of only two filaments. For the gas distribution, the
situation is the reverse: the gas around R5 is relatively featureless,
but in Z1 and Z2wt, the inflowing gas has a complex structure. This is
reflected also in the temperature maps of the Z1 and Z2wt objects
which show a distinctly chaotic pattern.

The star-forming cloud in the R5 object is relatively spherical, as
may be seen in Table~\ref{Table:shape}, while the Z1 object has a
disk-like structure extending out to a radius of $\sim 0.1\,{\rm
pc}$. The Z2wt object has a huge bar extending out to a radius of
$\sim 1\,{\rm pc}$. In both these objects the gas velocity fields are
well organized, revealing a nearly circular pattern in Z1 and a vortical
pattern around the bar in Z2wt. However, in the very inner regions of
Z1, the gas has less rotational support (see Fig.~\ref{fig:allvel})
and is undergoing an inside-out collapse. It will be very interesting
to study the fate of this gas in future work.

Disk-like structures have been seen in previous simulations of first
stars by Bromm et al.~(1999, 2002). These authors used constrained
initial conditions and arbitrarily assigned a high value of the spin
parameter to the simulated region. The disk in our simulation is
substantially different from theirs. In Bromm et al.~(1999, 2000), the
disk-like structure formed shortly after virialization, then broke up
rapidly into several disjointed gas clumps ranging in mass from $200$
to $10^4\,{\rm M_\odot}$. In our simulation which started from proper
cosmological initial conditions, the gas acquires angular momentum
self-consistently by tidal torques from surrounding structures, rather
than being put in by hand as in Bromm et al. The disk forms at a later
stage when the central gas density approaches $10^9{\rm cm^{-3}}$, and
there is no sign that the disk-like structure is unstable to
fragmentation up to the time when we stop our simulation. This is
because the disk is hot and thick, and so is pressure supported. We
have checked that the Toomre $Q$ parameter, 
\begin{equation} 
Q=\frac{\Omega c_s}{\pi G\Sigma}, 
\nonumber
\end{equation}
of the Z1 disk lies in the range $2-3$ which also suggests that it is
stable against perturbations. Here, $\Omega$ is the local angular
frequency, $c_s$ is the sound speed, and $\Sigma$ is the surface
density.

\begin{figure*}
\hspace{0.13cm}\resizebox{8cm}{!}{\includegraphics{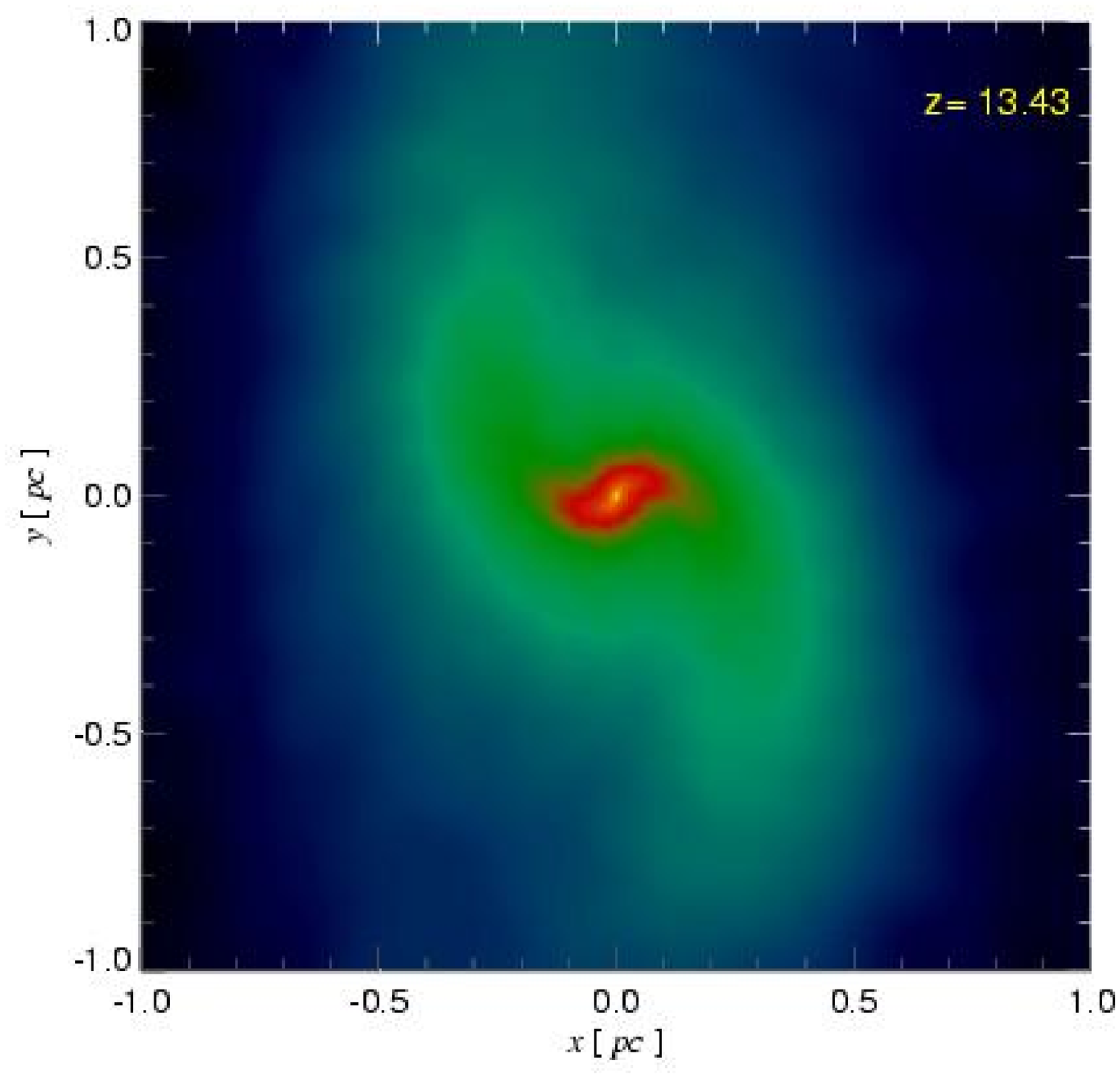}}
\hspace{0.13cm}\resizebox{8cm}{!}{\includegraphics{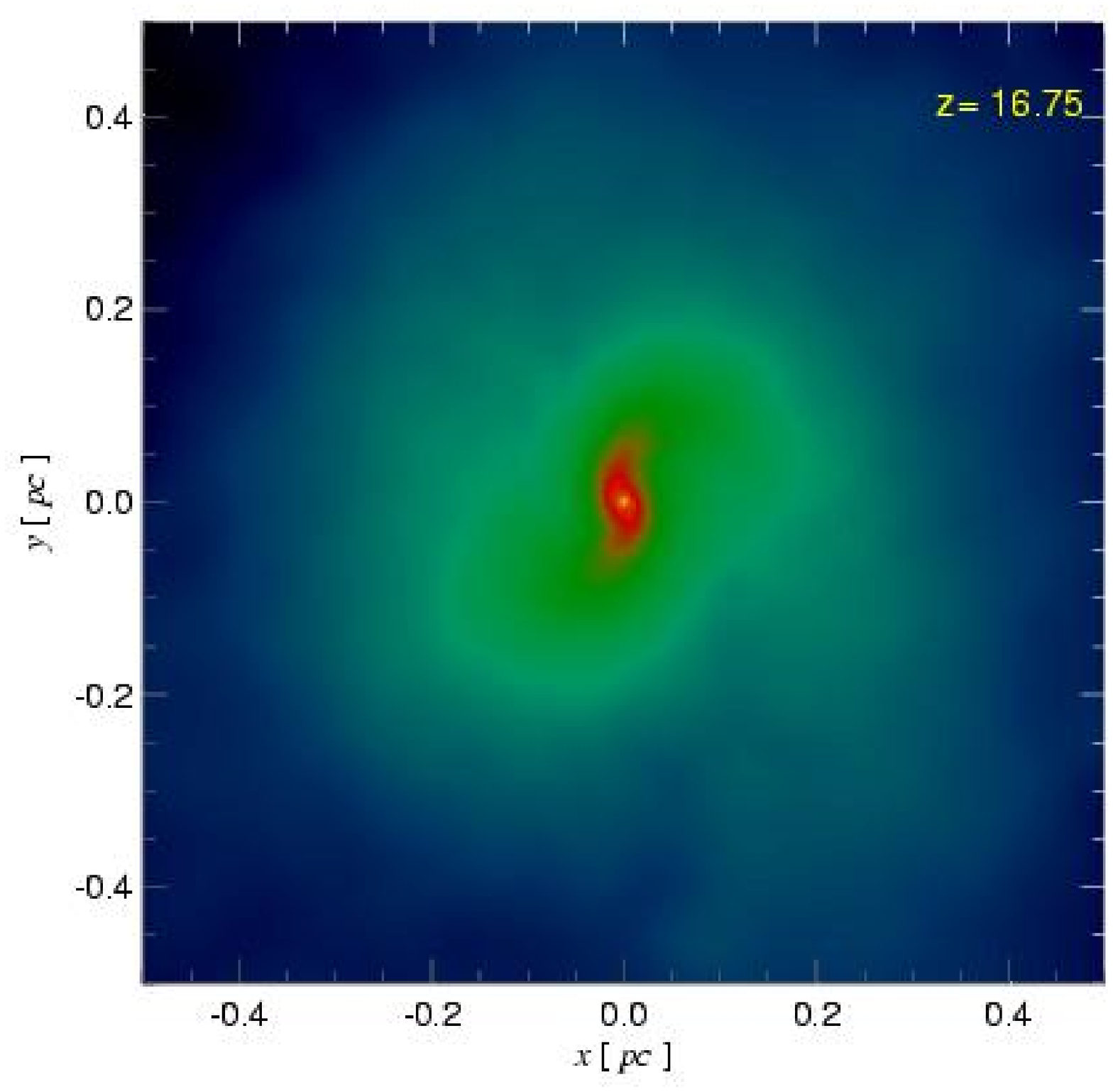}}
\caption{Density maps for the Z1wt and the Z2w
simulations which have a disk-like structure at the final time. These
images clearly reveal the presence of spiral arms contained in the
disk-like structure. The projection is in an arbitrary direction.}
\label{fig:spiral}
\end{figure*}

The Z1wt and Z2w simulations also formed thick disks and careful
inspection reveals the presence of spiral arms, as may be seen in
Fig.~\ref{fig:spiral}. The associated asymmetric density fluctuations
will produce gravitational torques that will help transport angular
momentum outwards efficiently.

The processes that drive the different morphologies exhibited by the
star-forming clouds in our simulations remain unclear. We have been
unable to find any correlation between morphology and the various
global properties that we have examined and we have found no clear
relationship to the merging history of the host dark halo. We defer a
more detailed study of the origin of the morphologies of star-forming
clouds to a future study.

\begin{table*}
\begin{tabular}{l c c c c c c c c}
\hline
& R5 & R5wt & Z1 & Z1w & Z1wt & Z2 & Z2w & Z2wt \\ \hline
$M_{{\rm core}}[{\rm M_\odot}]$ & $200$ & $250$ & $1000$ & $1000$ & $2000$ &
$3000$ & $2000$ & $4000$ \\
$R_{{\rm core}} [{\rm pc}]$ & $0.03$ & $0.05$ & $0.28$ & $0.48$ & $0.99$ & $%
1.78$ & $0.88$ & $1.1$ \\
$q=b/a$ & $0.83$ & $0.83$ & $0.87$ & $0.96$ & $0.83$ & $0.87$ &
$0.97$ & $0.40$ \\
$s=c/a$ & $0.50$ & $0.53$ & $0.28$ & $0.70$ & $0.32$ & $0.30$ &
$0.33$ & $0.11
$ \\
$V_t/V_{K}$ & $0.3$ & $0.3$ & $0.7$ & $0.3$ & $0.6$ & $0.6$ & $0.7$ & - \\
\hline
\end{tabular}
\caption{The principal axes ratios, $q=b/a$ and $s=c/a$, for our $8$
simulations, where $a>b>c$. Here, the edge of the star-forming cloud
is defined as the outer radius of the region in which the orientation
of the specific angular momentum is roughly constant. This radius is
denoted by $R_{\rm core}$; $M_{{\rm core}}$ is the enclosed gas
mass within it.}
\label{Table:shape}
\end{table*}

\subsection{Accretion rate}
The rate of mass accretion within the pre-stellar cloud is an
important quantity which governs the evolution of the protostar. The
growth of the protostar can be estimated by the mass accretion rate,
$\dot{M}$, via
\begin{equation}
M_*(t)=M_{{\rm proto}} + \int_0^t\dot{M}(t)\,{\rm d}t, \nonumber
\end{equation}
where $M_{{\rm proto}}$ is an initial protostellar mass. In our
simulations, we have neither enough resolution nor a sufficiently
complete physical model to be able to resolve the initial
protostar. One-dimensional hydrodynamic simulations including all the
relevant physics suggest that the protostar mass is of order a few
Jupiter masses (Omukai\& Nishi 1998; Ripamonti et al. 2002).

Ideally, we need to follow the protostellar evolution along with the
accretion in order to obtain a reliable final mass for the
star. However, the coupled evolution of the protostar and the
accreting envelope gas is a very complex physical problem which has
not been solved so far, even in one-dimensional radiative
hydro--dynamical simulations. At present, to estimate plausible
stellar masses one needs to resort to either analytical collapse
solutions (e.g. Larsen 1968; Penston 1967, Shu 1977), to detailed
protostellar calculations with constant accretion rates (Omukai \&
Palla 2001;2003), or to estimates based on the instantaneous accretion
rate from a particular numerical simulation (Tan \& McKee 2003; Omukai
\& Palla 2003; Yoshida et al. 2006; but see Bromm \& Loeb 2004 for a
slightly different approach). Here we present the instantaneous
accretion rates found in our own simulations.

\begin{figure}
\resizebox{8cm}{!}{\includegraphics{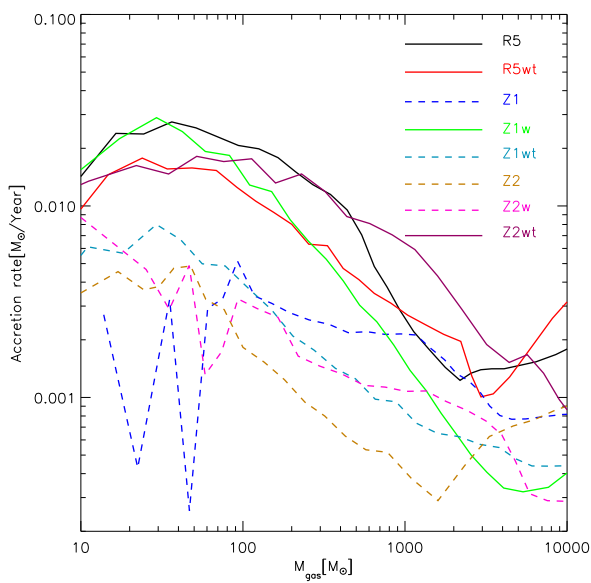}} \caption{Accretion
rate, $\dot{M(r)}=4\pi r^2\rho(r)v_{\rm infall}$, as a function of
enclosed gas mass for our 8 simulations. Here, the accretion rate is
computed at the time when the central gas density has reached a value
of $n_{\rm H} \sim 10^{10}{\rm cm^{-3}}$. The accretion rates for
objects with a central disk-like structure are ploted as dahsed lines,
while objects with other morphologies are plotted as solid lines.}
\label{fig:accretion}
\end{figure}

In Figure~\ref{fig:accretion}, we show the instantaneous accretion
rate, as a function of enclosed gas mass, for our $8$
simulations. Here, the accretion rate is defined as $\dot{M}(r)=4\pi
r^2\rho (r) |v_{\rm rad}|$ and is calculated at the time when the
central gas density has reached a value of $n_{\rm H} \sim 10^{10}
{\rm cm^{-3}}$. Since the accretion rate depends on the local gas
density and on the gas infall velocity, we expect a considerable
scatter from one object to another due to very different gas motions
and variations in the local density.


The R5, R5wt, Z1w and Z2wt objects have the largest instantaneous
accretion rates amongst all the simulations, and, at the same time,
they also have the least rotational support. The Z1, Z1wt, Z2, and Z2W
objects have the smallest accretion rates because their disk-like
structures slow down the gas infall rate. Thus, the
accretion time allows our simulated objects to be roughly divided into
two populations, depending on the morphology of the star-forming
clouds. The accretion time for rotationally supported disks is much
longer than for the other morphologies.

Omukai \& Palla (2003) identified a critical accretion rate of $\sim 4
\times 10^{-3}\, {\rm M_\odot yr^{-1}}$ (the thick solid line in
Figure~\ref{fig:accretion}) for a star to form. This rate is
physically motivated by the fact that the protostar luminosity cannot
exceed the Eddington limit; otherwise a radiation-driven expansion of
the protostar would halt or even reverse the accretion when the star
reaches a mass of $\sim 100\, {\rm M_\odot}$. Note that with this high
accretion rate, the star may in fact start burning hydrogen well
before the accretion slows down, so that $M/\dot{M}$ becomes much
longer than the lifetime of the star. Such an object would not be a
zero-age main sequence mass. If the accretion rate is smaller than the
critical rate, however, the star could continue to accrete gas after
it settles onto the zero-age main sequence ({\small ZAMS}), allowing
it to reach a larger final mass. In our simulations, objects without
disks all have instantaneous accretion rates on scales corresponding
to an enclosed mass of less than a few hundred solar masses that are a
few times higher than the critical rate, while those with disk-like
star-forming clouds have lower or comparable rates to the critical
one.

\begin{figure}
\resizebox{8cm}{!}{\includegraphics{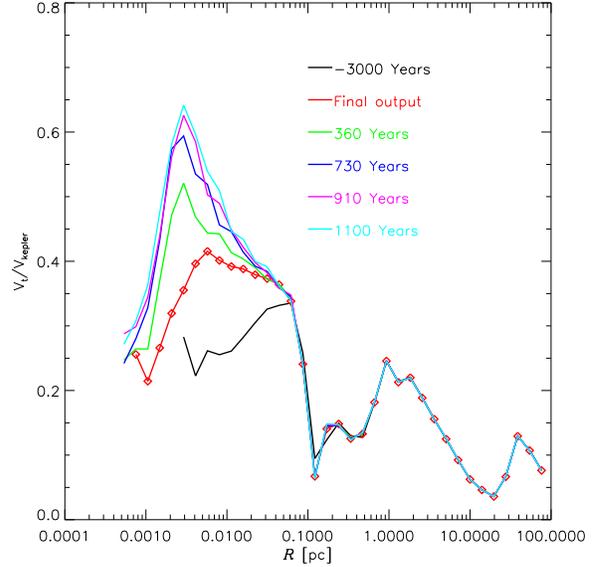}}
\caption{The ratio of the tangential velocity to the Keplerian
circular velocity, as a function of radius, at several epochs for the
R5 simulation. The red curve with open squares refers to our last
trusted epoch when the optically thin assumption is still valid. The
time intervals between the other outputs and this final time are
indicated in the labels.}
\label{fig:future}
\end{figure}

Does this mean that our simulated objects with accretion rates
larger than the critical value would be limited to a final mass of
about $\sim 100\, {\rm M_\odot}$ by radiative feedback from their own
protostars? To address this question, we carry out an experimental
calculation to examine whether the instantaneous accretion rate is
robust enough to serve as a reliable estimator of the theoretically
expected final mass of the star. To this end, we examine whether the
velocity structure of the gas in the protostellar environment changes
rapidly, particularly whether the gas would spin up after further
collapse which may slow down (or even halt) later accretion. For the
R5 simulation, which has a star-forming cloud with little rotational
support at our fiducial end state, we continued to integrate forward
in time for another $1000$ years (unlike ABN02 and Yoshida et
al.~(2006) who halted their simulation at this time). The ratio of the
tangential velocity to the Keplerian rotational velocity at several
times during this period is shown in Figure~\ref{fig:future}.

The red solid line with open diamonds refers to the final time when
our physical model is still adequate. One can see that as the
evolution is allowed to proceed, the gas cloud becomes increasingly
rotationally supported and forms a ring-like structure at a length
scale of $\sim 0.02$pc within one thousand years, which is a short
time in the lifetime of a protostar. While this evolution is not
fully realistic because neither is the adopted inner boundary
condition correct nor is the feedback generated by the protostar
considered, it nevertheless highlights the fact that the velocity
field in the protostellar environment can change rather rapidly. This
suggests that it is hard to draw definitive conclusions regarding the likely
mass of the star-forming cloud in the R5 halo based solely on the
instantaneous accretion rate.

Continuing the integration of the other systems which have similar
morphologies to R5 yields similar results: the accretion rate
typically tends to become smaller as gas becomes more and more
rotationally supported. By contrast, in the Z2wt object, whose
star-forming cloud has a bar-like structure, the accretion rate rises
at slightly later times when the asymmetrical mode of the bar
transfers angular momentum outwards very rapidly and so boosts gas
infall. As a result, gas accretion onto the protostar of the Z2w halo
might be limited by the feedback effects discussed in Omukai \& Palla
(2003) and Tan \& McKee (2003). Integrating the systems with a
disk-like structure further in time, we find that the accretion rate
falls even below that measured at the final reliable output time. We
conclude that the instantaneous accretion is not a robust indicator of
the mass of the final PoP-III star.

The instantaneous accretion rates within a radius containing $100$
solar masses found in the simulations by ABN02 and Yoshida et al. (2006)
are compatible with our results.  Their rates lie within the scatter we
find from our set of eight simulations.  Recently, O'Shea \& Norman
(2006b) carried out a set of simulations similar to ours although using an
AMR code.  While they found, as we do, that there is a considerable
scatter in the instantaneous accretion rate from object to object, their
scatter is an order of magnitude larger than ours.  Another significant
difference between their work and ours is that we find no correlation
between the redshift of formation and the instantaneous accretion rate for our
objects. We do not know the reasons for these differences and it will be
interesting to run the simulations with identical initial conditions
with both our code and Enzo.

\section{Influence of the baryon fraction}
\begin{figure*}
\resizebox{16cm}{!}{\includegraphics{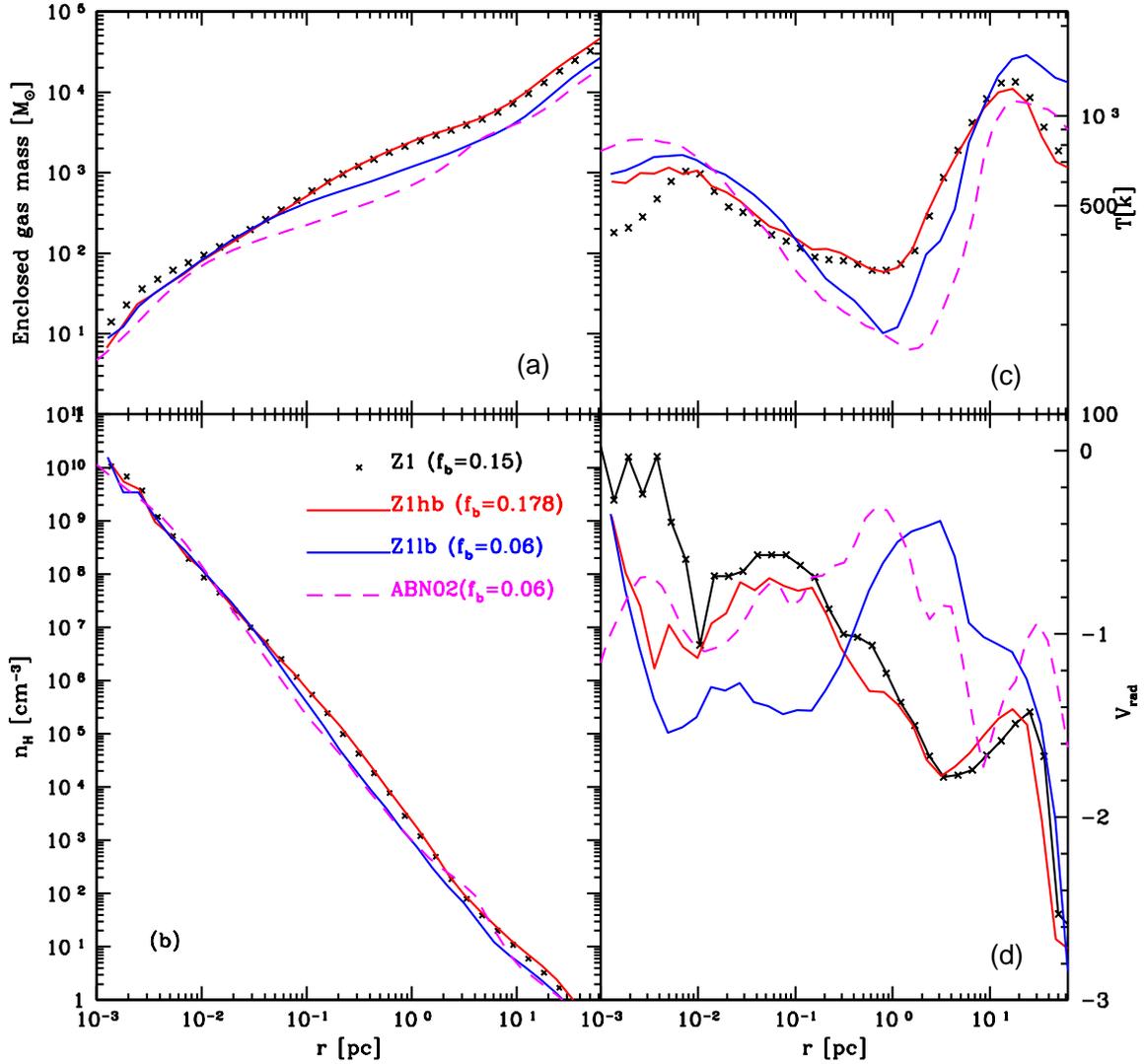}}
\caption{Comparison of various spherically averaged profiles in the Z1
simulation for different assumed baryon fractions. Panel (a): enclosed
gas mass; panel (b): number density of hydrogen nuclei; panel (c):
temperature; panel (d): radial velocity. The lines refer to different
simulations as indicated by the label. The redshifts are $z=25.93$ for
the Z1 run, $z=26.26$ for the Z1hb simulation, and $z=24.26$ for the
Z1lb run. Relative to the fiducial, $f_b=2/15$, case, star formation
in the higher baryon fraction simulation occurs $\sim 2$ million years
earlier, while in the lower baryon fraction simulation, it is delayed
by $\sim 12$ million years.}
\label{fig:bary}
\end{figure*}

It is interesting to examine how our choice of cosmological parameters
affects the properties of primordial gas clouds. In particular, the
baryonic matter density, or baryon fraction, may be one parameter
which can signficantly affect the gas evolution. Naively, one would
expect that the baryon fraction influences the gas collapse process,
since both the production of molecular hydrogen and the cooling
function at low densities depend strongly on density. Furthermore,
initially the gas distribution in dark matter halos follows closely
that of the dark matter component, with an enclosed baryonic fraction
that should be close to the mean cosmic value (e.g. Lin et
al. 2006). Thus, varying the baryon fraction may significantly change
the density distribution of the gas within a halo.

Given the uncertainties that remain regarding the precise values of
the basic cosmological parameters of our Universe (see the discussion
in Spergel et al. 2006), in this section, we investigate the effects
of varying the baryon fraction, $f_b=\Omega_b/\Omega_m$, on the
formation of the first pro-stellar objects. The precise values of the
matter-density, $\Omega_m$, of the dark energy component,
$\Omega_\Lambda$, and the shape and normalization of the initial power
spectrum are all important for determining the abundance and the
formation epoch of the first stars, but are less important than the
baryon fraction in determining the properties of individual star
forming clouds. We consider some of these other properties in the next
section. A final reason for varying the baryon fraction is that it
facilitates a direct comparison with the results of ABN02 who used a
smaller value than we have assumed so far.

We have resimulated the Z1 object with three different choices for the
baryon fraction. Our fiducial choice is $f_b=2/15$, until recently,
the value assumed in the standard (or ``concordance'') $\Lambda$CDM
model. We compare this with the lower baryon fraction, $f_b=0.06$, 
which was often used in the context of the now abandoned $\Omega_{\rm
m}=1$ standard CDM model assumed by ABN02. Finally, we also consider
$f_b=0.178$, as suggested by the latest {\small WMAP} and 2dFGRS
results (Sanchez et al. 2005, Spergel et al. 2006). We refer to the
simulations with lower and higher baryon fractions as as Z1lb and Z1hv
respectively. 

In Figure~\ref{fig:bary}, we compare the radial profiles of the gas
density, enclosed gas mass, temperature and radial velocity in the
three simulations. The magenta dashed lines show the results of
ABN02. The radially averaged physical properties of the Z1hb and Z1
simulations are generally very similar, although some small
differences are apparent. For example, the density profile of Z1hb is
$\sim 20\%$ higher than that of Z1 over much of the range shown. The
differences are much larger for the lower baryon fraction
simulation. On scales $0.1 - 1 \,{\rm pc}$, the gas density in Z1lb is
roughly a factor of $2$ lower than in the other simulations. The
radial gas velocity of Z1lb is also very different from that of Z1 and
Z1lb. In fact, surprisingly, the disk-like structure has disappeared
in the Z1lb simulation, although it is still present in the higher
baryon fraction simulation, Z1hb. Finally, we note that matching the
lower baryon fraction assumed by ABN02, as in Z1lb, produces much
better agreement between our results and theirs.

We find that the baryon fraction influences the collapse time of the
star-forming cloud quite significantly. Relative to Z1, the collapse
proceeds slightly faster in Z1hb and much more slowly in Z1lb. In
Z1hb, star formation occurs as early as $z=26.26$ (that is $2$ million
years earlier than in Z1), while in Z1lb, it is delayed by $12$
million years, to a redshift of $z=24.26$. We conclude that a slight
change in the baryon fraction does not affect the properties of
star-forming clouds much, but changes by a factor $\simeq 2$ have a
significant effect.

\begin{figure*}
\hspace{0.13cm}\resizebox{8cm}{!}{\includegraphics{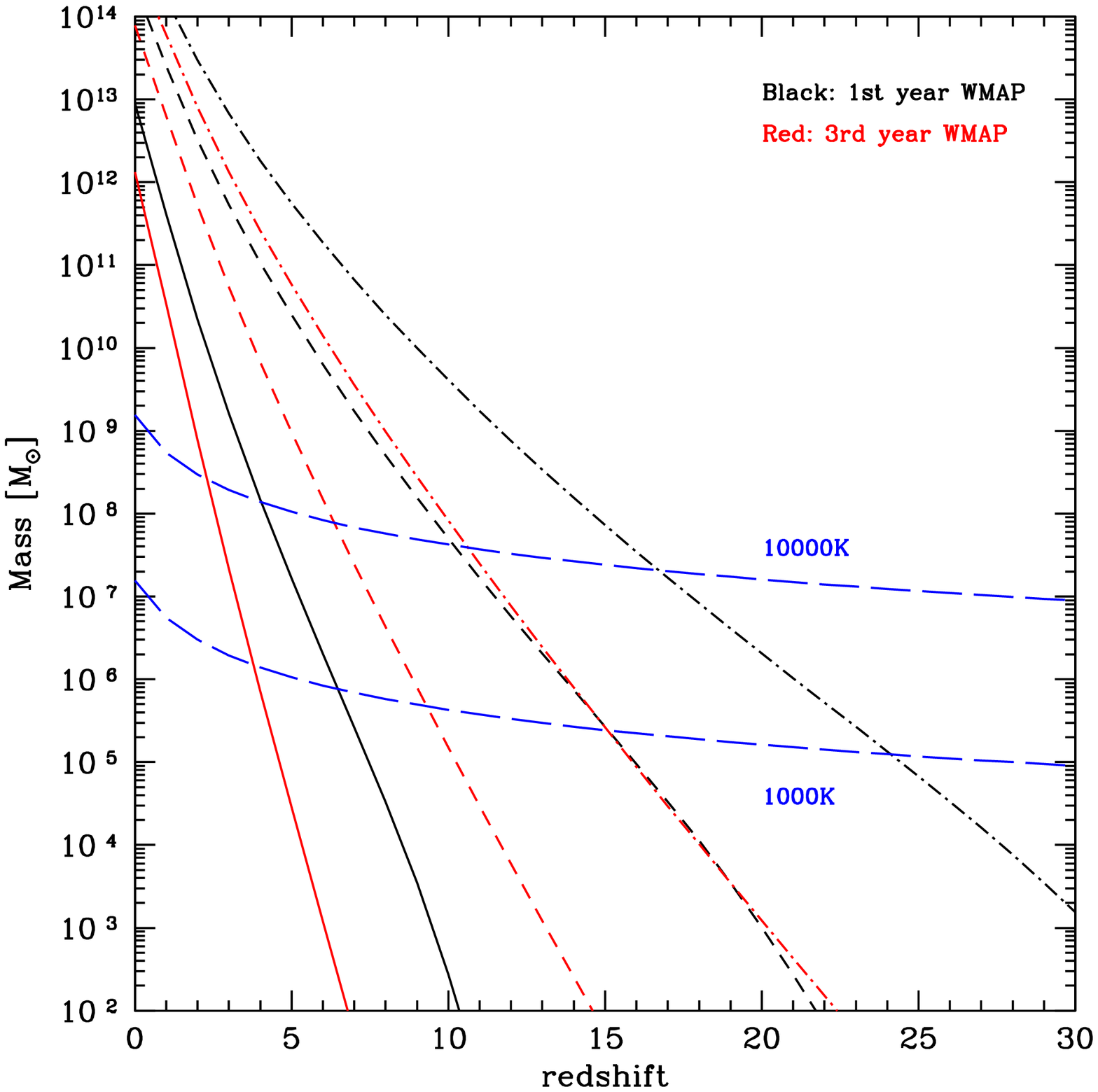}}
\hspace{0.13cm} \resizebox{8cm}{!}{\includegraphics{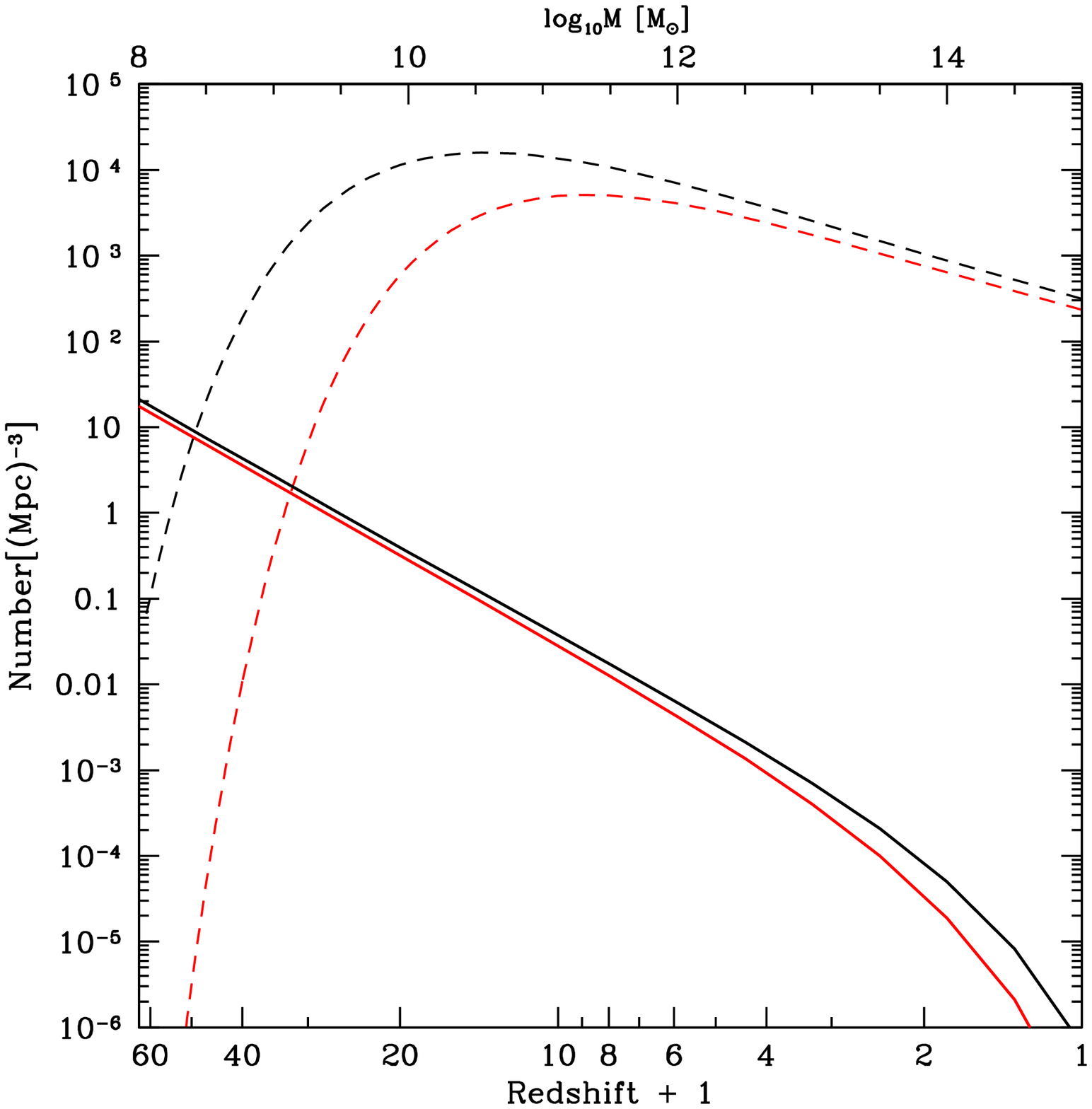}}
\caption{Left-hand panel: halo mass as a function of redshift. The
solid lines corresond to the mass ${\rm M_*}$, defined through
$\nu=\sigma(M)/\delta_c=1$; the dashed lines correspond to the mass of
$2\sigma$ halos ($\nu=2$); the dot-dashed lines correspond to the mass
of $3\sigma$ ($\nu=3$) halos. The black curves show results for
cosmological parameters derived from the 2dFGRS and the first year
{\small WMAP} data and the red curves for parameters derived from the
2dFGRS and three-year {\small WMAP} data (Sanchez et al. 2005, Spergel
et al 2006). The two long-dashed blue lines show the halo mass which,
at each redshift, has a virial temperature of more than $1000\,{\rm
K}$ and $10000\,{\rm K}$, respectively, as indicated by the labels.
Right-hand panel: the predicted halo abundance in the two
cosmologies. The solid lines show cumulative comoving number densities
of halos more massive than a given mass as a function of redshift
(upper axis). The dashed lines show the predicted comoving number
density of halos that have a virial temperature exceeding $1000\,{\rm
K}$ as a function of redshift (lower axis). This is roughly the
critical temperature at which gas can start efficient cooling by
molecular hydrogen.}
\label{fig:abundance}
\end{figure*}

\section{Abundance of star-forming halos}
So far, we have concentrated on the physical processes that lead to
the formation of the first protostars in a $\Lambda$CDM universe. We
now turn to a discussion of the abundance of early star-forming halos
and its dependance on cosmological parameters in a $\Lambda$CDM
universe.  As we have shown in Section~3, the basic principles that
determine the formation paths of the first protostars are simple:
once gas is heated up to a temperature of around $1000\,{\rm K}$, the
production of molecular hydrogen is boosted sufficiently for the gas
cloud to cool very rapidly. The ensuing collapse proceeds faster if it takes
place in a halo at higher redshift because in that case the gas starts
off with a higher density and temperature and a shorter dynamical time.

As shown in Figures~1 and~5, the virial temperature provides a
reasonable estimate of the initial temperature of the star-forming
material at the time when it begins to settle into the halo.  In
Figure~\ref{fig:abundance}, we show the abundance of objects with
virial temperature above $1000\,{\rm K}$. Note that the gas in a halo
with this temperature is predicted to {\em start} collapsing at this
time, but the collapse process itself, eventually yielding a
protostellar object, takes a further $\sim 10^6-10^8$ years. This
corresponds to is significant interval in redshift, $\Delta z \sim$ of
a few, depending on the exact redshift.

The first stars in the standard $\Lambda$CDM cosmology are often
assumed to form typically at redshifts between $z=20$ and $z=30$
because $3\sigma$ halos with a virial temperature of $\simeq
1000\,{\rm K}$ collapse in this redshift range. In the left-hand panel
of Figure~\ref{fig:abundance}, we plot the abundance of $1\sigma$,
$2\sigma$ and $3\sigma$ halos as a function of redshift. A similar
figure has been shown in a number of previous studies (e.g.~Barkana \&
Loeb 2001) but here we compare results obtained for two choices of
cosmological parameters, those inferred by combining the galaxy
power spectrum of the 2dFGRS with the first-year and three-year
{\small WMAP} data respectively. In the figure, $1\sigma$ halos are
shown with solid lines, $2\sigma$ halos with short-dashed lines and
$3\sigma$ with dot-dash lines. The two long-dashed blue lines show the
halo mass which, at each redshift, has a virial temperature of more
than $1000\,{\rm K}$ and $10000\,{\rm K}$, respectively.

As a consequence of the significant suppression of power on small
scales in the 2dFGRS+three-year {\small WMAP} cosmology (due to a
lower value of $\sigma_8$ and a red tilt in the primordial power
spectrum index), the curve for the $3\sigma$ halos in this cosmology
is shifted so much that it nearly coincides with that for the
$2\sigma$ halos in the older cosmology. The curve for halos which have
a virial temperature exceeding $1000\,{\rm K}$ crosses the curve for
$3\sigma$ halos in the 2dFGRS+three-year {\small WMAP} cosmology at $z
\sim 15$ instead of at $z \sim 25$ as in the older cosmology.

The dramatic changes in the number of halos that can host the first
stars arising from the relatively small change in the values of the
cosmological parameters in the 2 models we considering are illustrated
in the right-hand panel of Figure~\ref{fig:abundance}. Here, the
comoving abundance of potential first-star-bearing halos, i.e. those
with a virial temperature exceeding $1000\,{\rm K}$ is plotted 
as a function of redshift (labelled along the lower axis)
for the two sets of cosmological parameters (distinguished by the two
colours as in the left panel). For reference, we also plot the
comoving abundance of all halos at the present day, as a function of
halo mass (labelled along the upper axis), based on the fitting
formula of Sheth, Mo \& Tormen (2000), which works reasonably well
for large masses (Jenkins et al. 2001; Springel et al. 2005; Reed et
al. 2007).

The abundance of potentially star-forming halos differs dramatically
in the two cosmologies at redshifts greater than $z=10$. At redshift
$z=50$, the predicted abundance of halos capable of molecular cooling
is $\sim 7$ orders of magnitudes lower in the 2dFGRS+three-year
{\small WMAP} cosmology than in the older 2dFGRS+first-year {\small
WMAP} cosmology. As cosmic time increases, the predicted abundance of
star-forming halos increases very rapidly, as noted by G05. By
redshift $z \sim 35$, the abundance of halos with $T_{\rm vir} >
1000$K has increased by $6$ orders of magnitude, making it comparable
in the cosmology that uses the 3-year {\it WMAP} data to that of dwarf
galaxies at the present day. By redshift $z \sim 15$, the abundance of
such objects is $1000$ times higher than that of $10^8\,{\rm M_\odot}$
halos today. It is clear that the current uncertainty in the values of
the cosmological parameters, particularly $\sigma_8$, is a major
source of uncertainty for the predicted abundance and evolution of
Pop-III stars.

\section{Summary and discussion}

In this paper, we have presented a large set of cosmological {\small
SPH} simulations of the formation of the first stars. Our simulations
reach sub-solar mass resolution and include an accurate and detailed
chemistry network to treat radiative cooling by molecular
hydrogen. Our large suite of simulations allows us to investigate the
scatter in the properties of star-forming clouds and the redshift
dependence of their formation process.  Our simulations stop when the
central gas density has reached ${n_{\rm H}\sim 10^{10}{\rm
cm^{-3}}}$, the point at which our assumption that the gas is
optically thin begins to break down.

Despite the complexities in detailed formation paths, our simulations
suggest a relatively simple picture leading up to the development of
the first star-forming clouds. Once primordial gas is heated to $\sim
1000\, {\rm K}$, either by shocks or by adiabatic compression, the
production of molecular hydrogen is boosted.  When the molecular
fraction reaches a few times $10^{-4}$, the gas cools rapidly and
settles to the centre of the gravitational potential well of the
halo. Our simulations bear out the same sequence of events sketched in
earlier semi-analytical work by Tegmark et al.  (1997) and seen in the
detailed simulations of Fuller \& Couchman (2000), Machacek et
al. (2001) and Yoshida et al. (2003). Our main results may be
summarized as follows:
\begin{description}

\item[(1)] The formation path of Pop-III stars is broadly similar in all
  halos, independently of collapse redshift and environment. However,
  the timescale for the collapse differs systematically depending on
  the redshift at which the collapse takes place.  The gas condenses
  more rapidly into stars at high redshift and so the formation
  timescale can differ by factors as large as $20$ for halos that
  collapse between redshifts $z \sim 50$ and $z \sim 15$. For example,
  for the R5 object which starts to form at $z \sim 50$, the collpase
  takes as little as a few million years whereas for the Z2wt halo,
  which began to form at $z \sim 15$, the  collapse took $100$ million years.
\item[(2)] The spherically averaged radial gas density profiles of the
  star-forming clouds agree well with each other up to a radius of
  $\sim 10\,{\rm pc}$. The profiles exhibit a power-law distribution
  $\rho \propto r^{-2.2-2.3}$ which extends over 10 orders of
  magnitude in density and is independent of formation redshift and
  environment. The molecular hydrogen fractions and temperature
  structures are also very similar. The main reason why these
  similarities are present is that once the gas has started to
  collapse, it largely isolates itself from the rest of the evolving
  halo and so the final state of the star-forming cloud becomes
  largely insensitive to the global properties of the host halo such
  as its virial mass, temperature or spin.
\item[(3)] Despite the similarity in the radial profiles of gas density,
  temperature, and molecular hydrogen fraction, the velocity field of
  the gas can differ significantly from object to object.
\item[(5)] At the moment when our simulations stop, the star-forming
  clouds exhibit a variety of morphologies. Some of them are nearly
  rotationally supported disks, others resemble flattened spheroids,
  while others possess self-gravitating bars. The instantaneous
  accretion rate onto the central region distinguishes between two
  kinds of star-forming clouds. Those with a disk-like structure
  have the smaller accretion rates, while those that resemble
  flattened spheroids or possess bars have the larger accretion
  rates. The scatter in accretion rate can be more than a factor of
  $10$ and is related to the morphology of the cloud.
\item[(6)] A higher baryon fraction increases the gas density
  everywhere within a halo and speeds up the formation of
  the first stars.
\item[(7)] The predicted abundance of Pop-III halos in the
  $\Lambda$CDM model varies dramatically depending on whether one
  assumes the cosmological parameters derived using either the first
  year or the three-year {\small WMAP} data. For example, the
  abundance of objects similar to our R5 halo differs by $7$ orders of
  magnitudes at redshift $50$ in the two cosmologies. 
\end{description}

At the moment when our simulations stop, i.e. when $n_{\rm H}$ reaches
a value $\sim 10^{10}{\rm cm^{-3}}$, none of the central star-forming
clouds have fragmented. This is in agreement with earlier simulations
(ABN02; Bromm et al. 2002; O'Shea \& Norman 2006a,b; Yoshida et
al. 2006). At this time, all our simulated objects experience a much
higher instantaneous accretion rate onto their central region than is
common in protostars today. If fragmentation does not occur and the
accretion does not slow down significantly, these objects are likely
to bear a star of more than a few tens of solar masses which would
have accumulated on a timescale of $\sim 10^5$ years. However, the
velocity structure around the protostars, and thus the instantaneous
accretion rate, varies greatly from object to object, and this
suggests that the final masses of the first stars may not lie in a
narrow range. However, it is unclear whether the differences in the
instantaneous accretion rates will be directly proportional to the
final mass of the stars because the velocity structure of the gas in
the protostellar environment changes very rapidly with time.

The fate of the star-forming clouds resolved in our simulations is
unclear. In order to reach stellar densities, the gas clouds must
still collapse by more than four orders of magnitude in radius and
this requires efficient transport of angular momentum. This is a
complicated problem and the various morphologies of the star-forming
clouds we find suggest that its solution may involve more than a
single physical process. One way to transport angular momentum
outwards and facilitate further efficient accretion is through
gravitational torques generated by spiral density fluctuations. Only
$3$ of our $8$ objects exhibit spiral structure in the central regions
at the time when we halt our simulations and it is unclear if the
other objects will also develop spiral structure later on. A further
and critical complicating factor is the uncertain effect of feedback
generated by the protostar on the dynamics of the accreting gas.

Because of all of the complications just discussed, the masses of the
first stars remain uncertain. A greater level of complexity than is
possible with the current generation of simulations will be required
to answer this fundamental question. These simulations will need to be
able to model, in a self-consistent way, the dynamics of the accreting
gas, the evolution of the protostar and the various radiative,
chemical and mechanical feedback effects that are likely to be at play
in the formation of the first generation of stars.

\section*{Acknowledgements}
We would like to thank the referee, Volker Bromm, for very helpful
comments. We thank Shude Mao, Darren Reed and Yuexing Li for a careful
reading of an earlier version of the draft. LG is particularly
grateful to Simon White for extensive discussions. CSF ackowledges a
Royal Society Wolfson Research Merit Award. The work is supported in
part by the Grants-in-Aid for Young Scientists (A) 17684008 (NY) by
the Ministry of Education, Culture, Science and Technology in Japan.
The simulations described in this paper were carried out on the Blade
Center cluster of the Computing Center of the Max-Planck-Society in
Garching and on the Cosmology Machine of the Institute for
Computational Cosmology in Durham.  
\label{lastpage}

\begin{appendix}
\medskip

\section{Evolution of velocity profiles of the R5 object}
\begin{figure}
\resizebox{8cm}{!}{\includegraphics{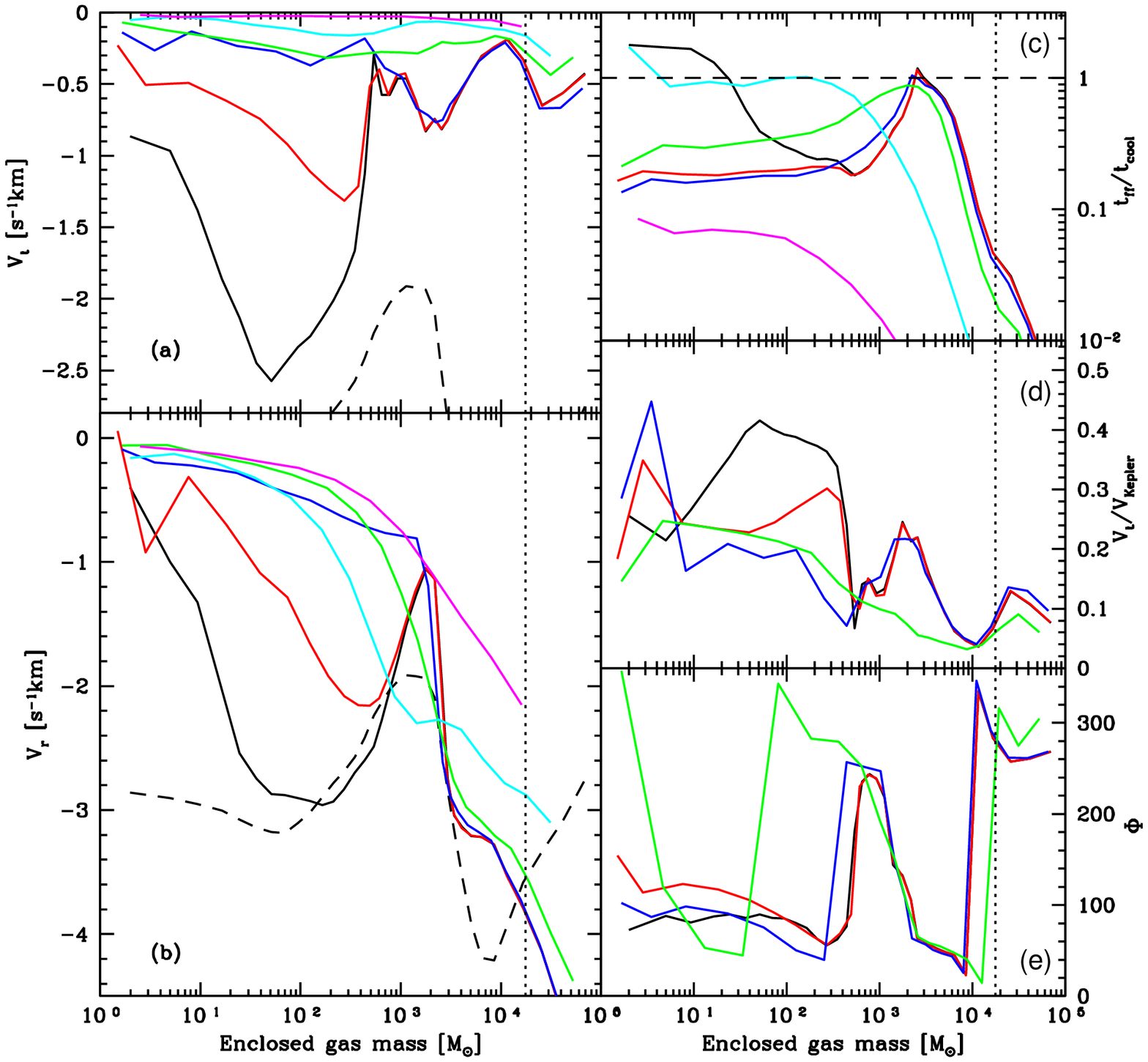}}
\caption{Radially averaged profiles of velocity and angular momentum
related physical quantities as a function of enclosed gas mass for $5$
output times of the R5 object. The lines correspond to the same times
as in Figure~3. Panel (a): Negative tangential velocity profiles. The
dashed lines in this panel and panel (b) refer to negative sound
speed. Panel (b): Radial velocity profiles. Panel (c): The ratio of
free-fall time to cooling time. Panel (d): The ratio of tangential
velocity to the required Keplerian velocity $v_{Kepler}=(GM/r)^{1/2}$.
Panel (e): The orientation of angular momentum. Note that only the
angle $\phi$ is shown. The vertical doted lines in all panels indicate
the enclosed gas mass within the virial radius. Note that only the
last four outputs in panel (d) and (e) are shown, for clarity.}
\label{fig:velr5}
\end{figure}

Figure~\ref{fig:velr5} is the evolution of velocity profiles of the
R5 object.

\section{Velocity profiles of $5$ objects}

\begin{figure}
\resizebox{8cm}{!}{\includegraphics{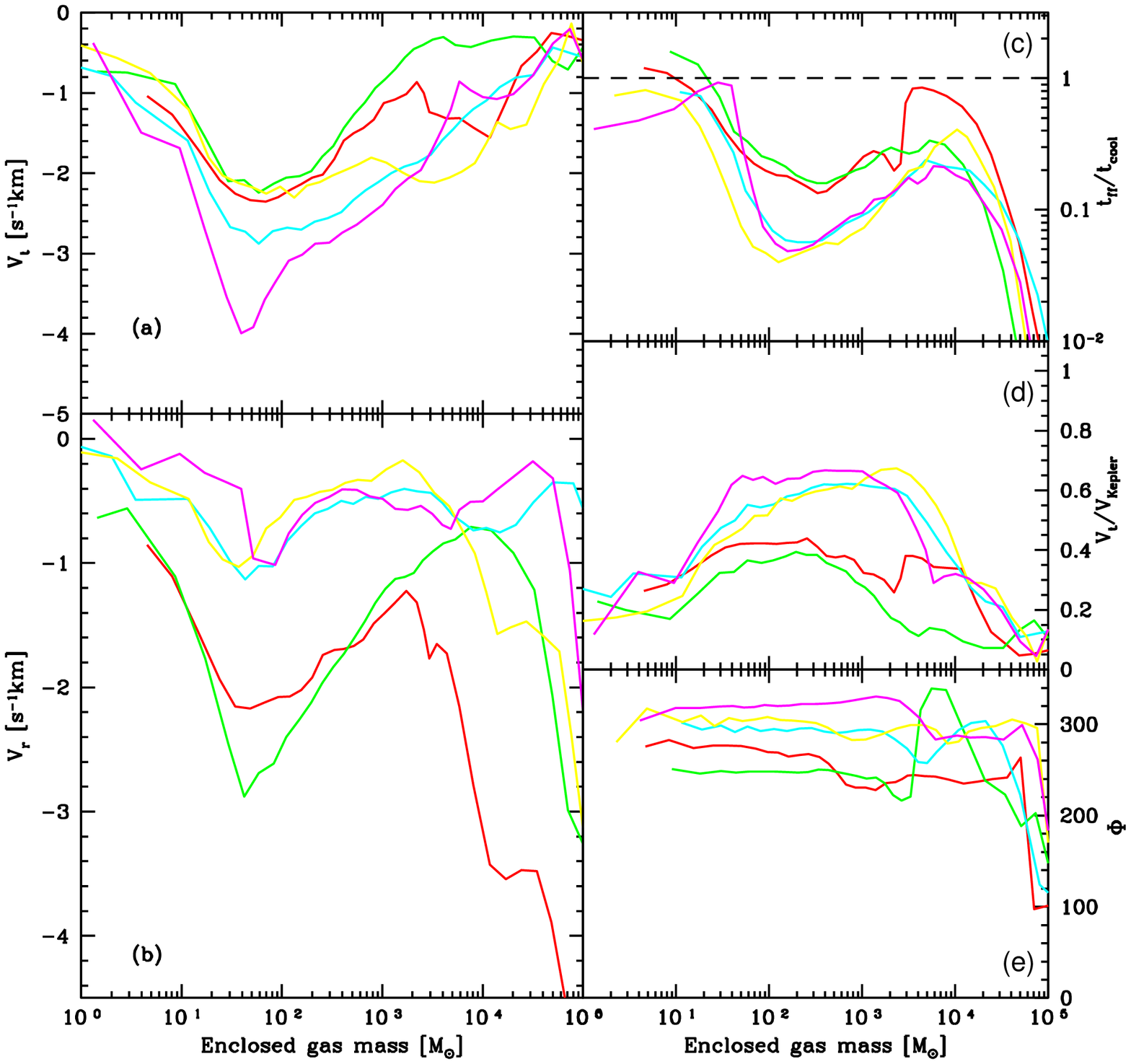}}
\caption{The same plots as in Figure~\ref{fig:allvel} but for a
  different set of $5$ simulations taken from our suite of runs: R5wt,
  Z1w, Z1wt, Z2 and Z2w. Panel (a): Negative tangential velocity
  profiles. Panel (b): Radial velocity profiles. Panel (c): The ratio
  of the free-fall time to the cooling time. Panel (d): The ratio of
  the tangential velocity to the Keplerian velocity
  $v_{Kepler}=(GM/r)^{1/2}$. Panel (e): The orientation of the angular
  momentum profiles. Note that only the angle $\phi$ is shown. The
  lines are the same as in Figure~\ref{fig:allden}.}
\label{fig:appvel}
\end{figure}

Figure~\ref{fig:appvel} is the same plot as Figure~\ref{fig:allvel}
but are for other $5$ simulations.

\end{appendix}
\end{document}